\documentclass[published]{JHEP3} 

\JHEP{}

\usepackage{epsfig, multicol}
\newcommand\fverb{\setbox\pippobox=\hbox\bgroup\verb}
\newcommand\fverbdo{\egroup\medskip\noindent%
			\fbox{\unhbox\pippobox}\ }
\newcommand\fverbit{\egroup\item[\fbox{\unhbox\pippobox}]}
\newbox\pippobox

\title{Effective Monopole Action at Finite Temperature in SU(2) Gluodynamics
}

\author{Katsuya Ishiguro and Tsuneo Suzuki \\
	Institute for Theoretical Physics,  Kanazawa University, \\
	Kanazawa 920-1192,  Japan.\\
	E-mail: \email{ishiguro@hep.s.kanazawa-u.ac.jp}\\
	E-mail: \email{suzuki@hep.s.kanazawa-u.ac.jp}
}
\author{Tateaki Yazawa \\
	Kinjo College,  Matto,  Ishikawa 924-8511,  Japan.\\
	E-mail: \email{yazawa@kinjo.ac.jp}
}
\received{\today} 		
\revised{\today}
\accepted{\today}		

\preprint{\heplat{20011216}}	

\abstract{
Effective monopole action at finite temperature in SU(2) gluodynamics
is studied on anisotropic lattices.
Using an inverse Monte-Carlo method and the blockspin transformation
for space directions,  we determine 4-dimensional effective
monopole action at finite temperature. We get an almost perfect action
in the continuum limit under the assumption that the action is composed of 
two-point interactions alone.
It depends on a physical scale $b_s$ and the temperature $T$. 
The temperature-dependence appears with respect to the spacelike
monopole couplings in the deconfinement phase,  whereas the timelike 
monopole couplings do not show any appreciable temperature-dependence.
The dimensional reduction of 
the 4-dimensional SU(2) gluodynamics ((SU(2))$_{4D}$) at
high temperature is the 3-dimensional Georgi-Glashow model ($(GG)_{3D}$).
The latter is  studied at the parameter region obtained from the dimensional reduction.
We compare the effective instanton  action of $(GG)_{3D}$ with
the timelike monopole action obtained from (SU(2))$_{4D}$.
We find that both agree very well for $T \ge 2.4T_c$ at large $b$ region.
The dimensional reduction works well also for the effective action.
}


\begin{document} 

\newcommand{\be}{\begin{eqnarray}}
\newcommand{\ee}{\end{eqnarray}}
\newcommand{\bc}{\begin{center}}
\newcommand{\ec}{\end{center}}
\newcommand{\sbra}[1] { \left( #1 \right)}
\newcommand{\mbra}[1] { \left\{ #1 \right\}}


\section{Introduction}

It is important to understand  nonperturbative 
effects of Quantum Chromodynamics (QCD) at finite temperature.
At zero temperature,  the typical nonperturbative phenomena are 
color confinement and the chiral symmetry breaking.
At high temperature,  QCD enters the Quark Gluon Plasma (QGP) phase 
in which colors are deconfined and chiral 
symmetry is restored.
It is known that not only perturbative but also nonperturbative 
effects such as the spatial string tension 
and the Debye-screening mass~\cite{kajantie97-2}
exist even in the deconfinement phase.

The  nonperturbative quantities have been studied also
using the 3-dimensional effective action 
obtained through the dimensional reduction.
The idea of the dimensional reduction for high temperature gauge theory 
was proposed in early 80's~\cite{pisarski81}.
The 3-dimensional effective action is derived perturbatively by the integration 
of non-zero modes for time direction of the fields.
After performing the dimensional reduction perturbatively 
in (SU(2))$_{4D}$,  the obtained effective action 
is $(GG)_{3D}$.
The effectiveness of the dimensional reduction at high temperature 
has been confirmed by numerical simulations 
on the lattice [1,3--7].
Quadratic and quartic interactions of the Higgs field 
are necessary for the infrared physics.
Spacelike Wilson loops 
and Polyakov loop correlators in $(GG)_{3D}$ agree well with those of 
(SU(2))$_{4D}$ for $T \ge 2T_c$~\cite{lacock92}.
The details of the relation between the phase diagram  
and the parameter region of the dimensional reduced 
$(GG)_{3D}$ in 2-loop perturbative calculation have been studied in ~\cite{kajantie97}.
Using the parameter  in 2-loop perturbative calculation, 
 the Debye-screening mass is shown to be a nonperturbative physical quantity in itself~\cite{kajantie97-2}.
The validity of the dimensional reduction for $T \ge 2T_c$ in 
(SU(2))$_{4D}$ have also been confirmed for 
the glueball spectrum~\cite{hart00} and 
the gauge-fixed propagator~\cite{cucchieri2001}.
In Ref.~\cite{karsch98} the parameters of the dimensional reduced effective action 
have been determined nonperturbatively.
However to the authors' knowledge,  there have been no nonperturbative studies using the  
dimensional reduction from the standpoint of topological quantity.

\DOUBLEFIGURE[b]{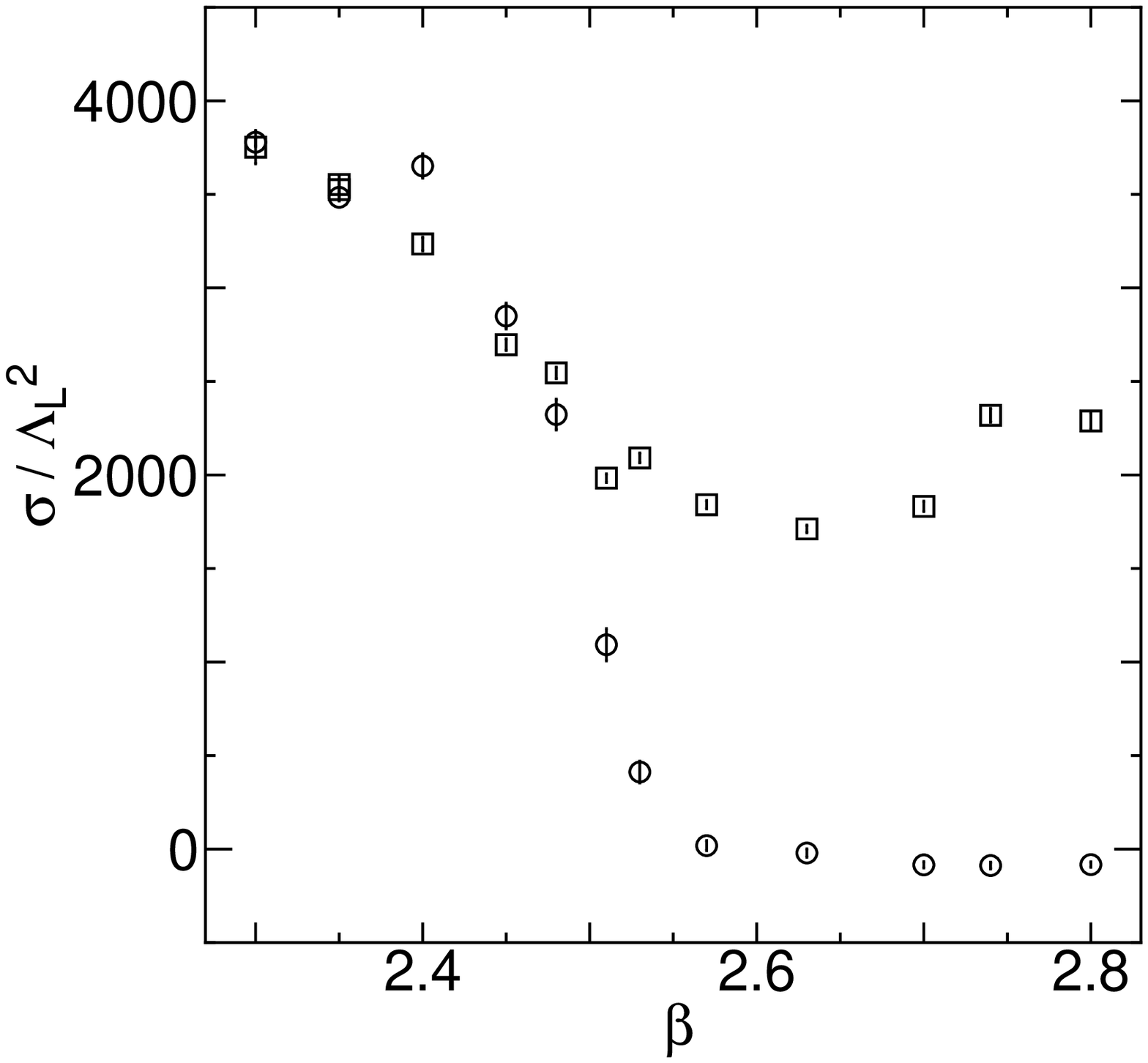, height=7cm, width=7cm}
{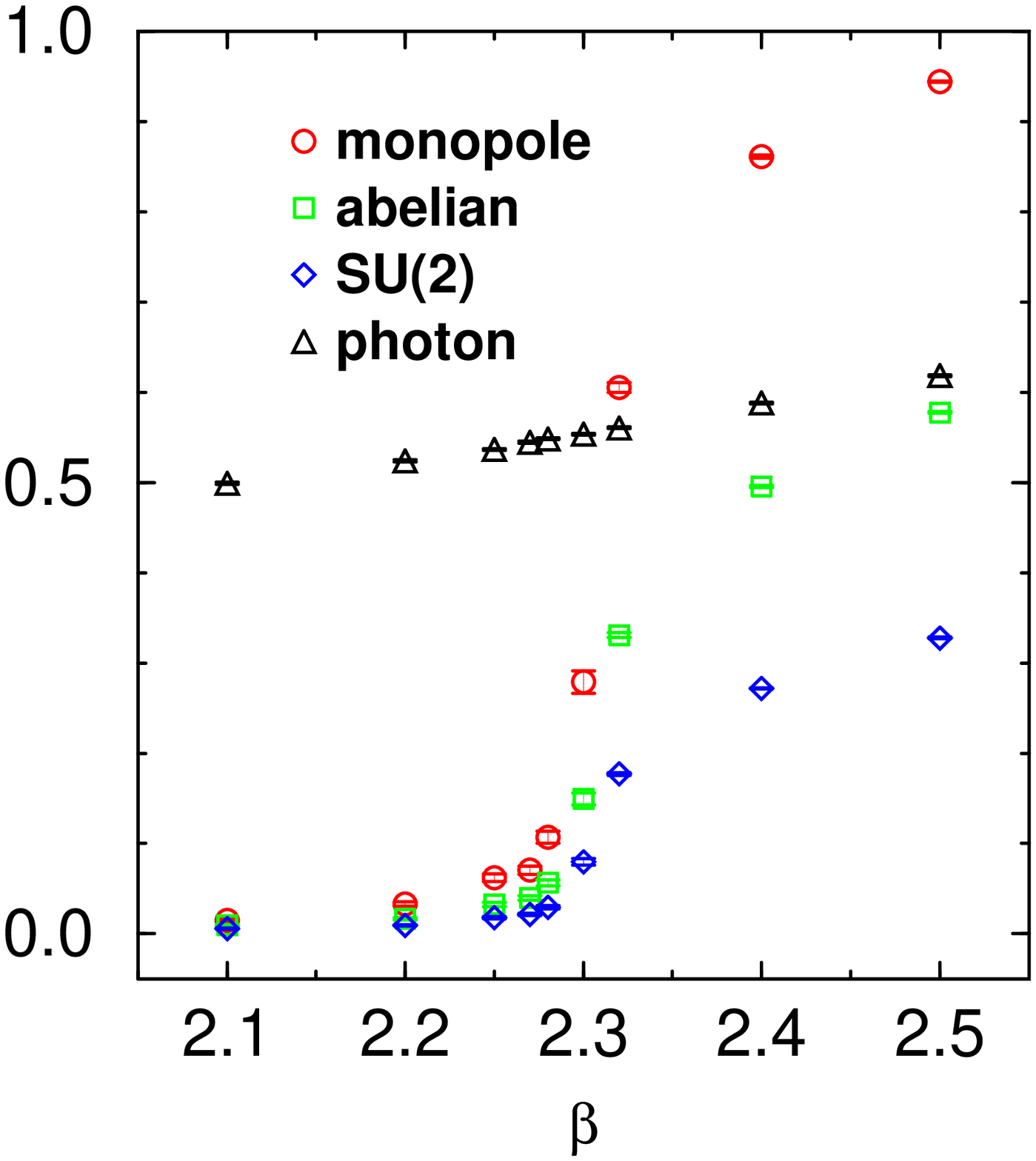, height=8cm, width=7cm}
{Physical string tensions (circle) and spatial string tensions 
(square) from monopoles in SU(2)QCD on $24^3 \times 8$ lattices. This figure is
from Ref.~\cite{ejiri95}.\label{sttensionref}
}
{Non-abelian and abelian Polyakov loops and 
monopole Dirac string and photon contributions 
to Polyakov loops in the MA gauge in SU(2) QCD on 
$24^3 \times 4$ lattice.
This figure is taken from Ref.~\cite{ejiri97}.\label{ploopref}
}

At zero temperature,  the dual superconductor picture of the QCD vacuum 
seems to be the color confinement mechanism in which 
magnetic monopoles condense and color-electric flux is squeezed 
(dual Meissner effect).
Monopoles are induced  by performing abelian projection 
 (partial gauge-fixing keeping $U(1)^2$).
In SU(2) and SU(3) gauge theory,  the string tension extracted from 
the monopole part reproduces the original one (monopole dominance).
This fact suggests that monopoles play an important role for confinement.
An effective monopole action described by monopole currents has been 
studied in detail and an almost perfect action (corresponding to the 
continuum limit) is derived successfully  in infrared region 
of QCD [9--11].

At finite temperature,  there have been interesting data suggesting 
the importance of monopoles [12--16]. 
The string tension from the monopole part of the Wilson loop 
almost agrees with that of the 
abelian Wilson loop in the confinement phase,  whereas it vanishes clearly 
in the deconfinement phase.
The data~\cite{ejiri95} for the temperature-dependence of the string tensions 
from monopoles and photons are shown in Fig.~\ref{sttensionref}.
The string tension from the photon part is negligibly small.

A non-abelian Polyakov loop is well known as an order parameter of the 
deconfinement phase transition.
Similarly an abelian Polyakov loop which is written in terms of 
abelian link variables alone is an order parameter of the deconfinement 
phase transition. It is given by a product of contributions from 
Dirac strings of monopoles and from photons.
The data~\cite{ejiri97} of SU(2) QCD in the MA gauge 
are shown in Fig.~\ref{ploopref}.
Here the confinement-deconfinement phase transition occurs 
at the critical coupling $\beta_c = 2.298$.
The abelian Polyakov loops vanish in the confinement phase whereas 
they have a finite value in the deconfinement phase.
The behaviors of the Dirac string contributions (monopole Polyakov loops) 
are  similar,  but more drastic than those of the abelian and the non-abelian 
Polyakov loops.
The photon part has a finite non-zero value in both phases.
So only the monopole Polyakov loops play a role as an order parameter 
of the deconfinement phase transition in the abelian Polyakov loops.

\FIGURE{
\epsfig{file=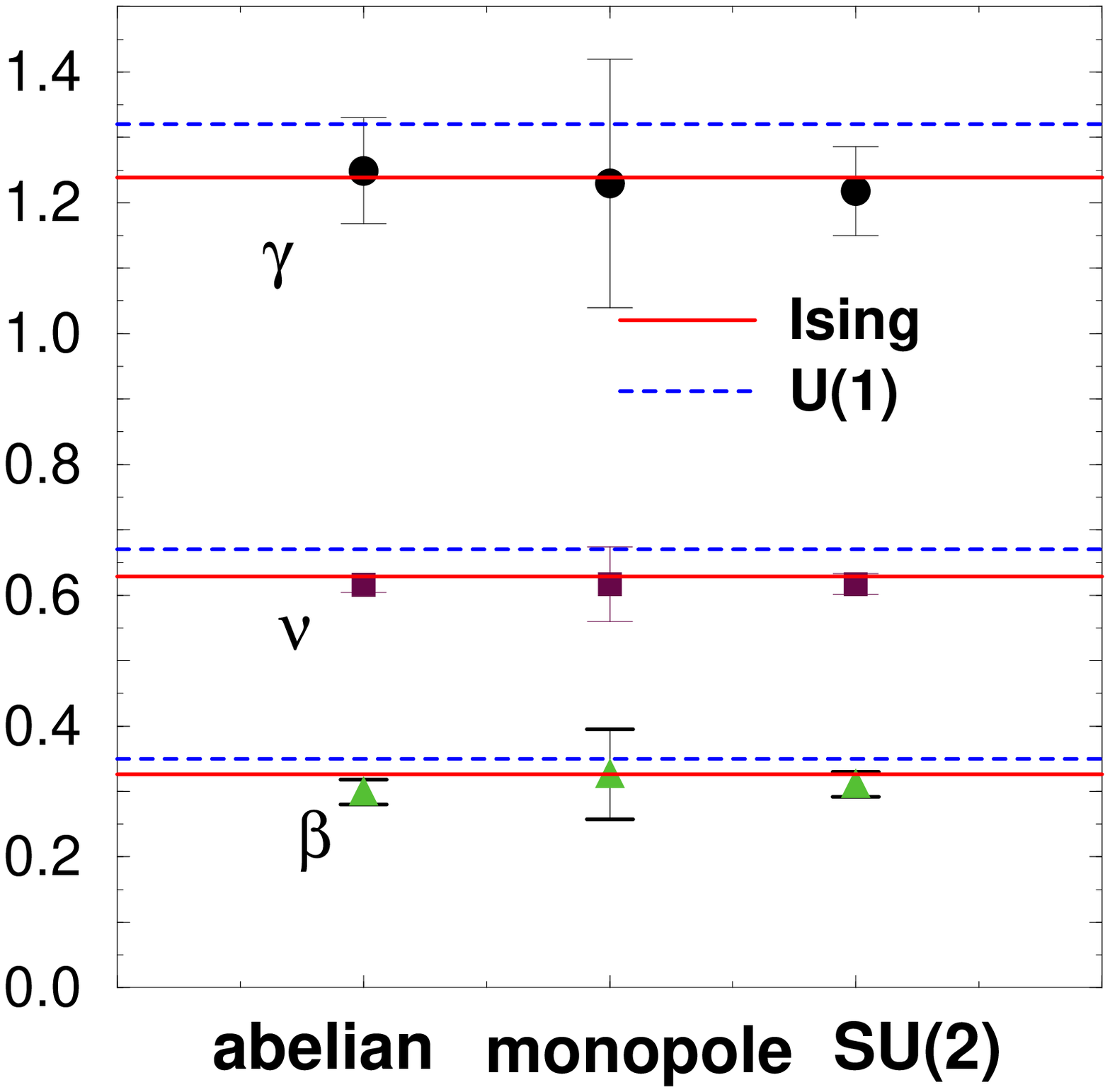, height=7cm, width=7cm}
\caption{Critical exponents of non-abelian, abelian and monopole 
Polyakov loops in SU(2) QCD.
This figure is taken from Ref.~\cite{ejiri97}.}
\label{criexpref}
}

The critical exponents have been determined from the behaviors of the Polyakov loops,  their susceptibility and the fourth cumulant.
The data~\cite{ejiri97} are shown in Fig.~\ref{criexpref}.
The critical exponents and the critical temperature 
determined  in the abelian and the monopole case are 
in agreement with those in the  non-abelian case within the statistical error.

What happens with respect to the nonperturbative effects at high temperature ?
There is also the monopole dominance for spatial string tension 
at high temperature ~\cite{ejiri95}. It is known that 
the timelike wrapped monopole loops are important 
which are closed through the periodic boundary condition~\cite{ejiri96}.
On the other hand,  $(GG)_{3D}$ has an instanton solution 
~\cite{thooft74, polyakov74}
and its Coulomb gas leads us to confinement~\cite{polyakov77, yazawa01}.
4D timelike monopoles tend to instantons  
in the high temperature limit.
These facts suggest that at high temperature nonperturbative effects  
are caused by timelike monopoles (when $T\ge T_c$) 
 and instantons (when $T\to\infty$). 

It is the purpose of this paper to confirm the above expectation.
We derive first infrared effective 
monopole actions numerically from finite temperature (SU(2))$_{4D}$ .
We adopt  anisotropic lattices and perform the blockspin transformations 
of the monopole currents to study the continuum limit.
The behaviors of spacelike monopole action and timelike monopole 
action in the confinement and in the deconfinement phases are 
discussed carefully. 
We then compare the timelike monopole effective 
action at high temperature in (SU(2))$_{4D}$ with the effective 
instanton action derived numerically from  $(GG)_{3D}$ to study if 
the dimensional reduction works also in the framework of effective monopole
 (instanton) action.

The paper is organized as follows.
In Section 2 we consider the effective monopole action at finite 
temperature in (SU(2))$_{4D}$ on anisotropic lattices.
In Section 3 we investigate the instanton action in $(GG)_{3D}$
 and compare it with the timelike monopole action in 
(SU(2))$_{4D}$ at high temperature. 
Section 4 is devoted to concluding remarks.

\section{The 4-Dimensional Effective Monopole Action }

\subsection{The Method}

In this section,  we review the method to determine the effective 
monopole action~\cite{shiba95, kato98}.
First we generate  thermalized non-abelian link fields 
$\{ U_\mu (s)\}$ using the  Wilson gauge action for pure SU(2) QCD.
Next,  we perform  abelian projection in the Maximally abelian (MA) 
gauge~\cite{kronfeld87-1, kronfeld87-2}.
 MA gauge fixing maximizes the following quantity under 
gauge transformations:
\\
\be
R = Tr \sum_{s, \mu} [ U_\mu(s) \sigma_3 
U_{\mu}^{\dagger}(s+\hat{\mu}) \sigma_3 ] .
\ee
\\
This means that
\\
\be
X(s) = \sum_\mu [ U_\mu(s) \sigma_3 U_{\mu}^{\dagger}(s) +
                  U_{\mu}^{\dagger}(s-\hat{\mu})
                  \sigma_3 U_{\mu}(s-\hat{\mu}) ] 
\ee
\\
is diagonalized.
After the gauge fixing,  we separate  abelian link fields 
$\{ u_\mu (s)\}$ from the gauge-fixed non-abelian ones 
$\{ \tilde{U}_\mu (s)\}$:
\\
\be
\tilde{U}_\mu (s) &=& C_\mu (s) u_\mu (s) ,  \\
 C_\mu (s) &=& \left( 
                 \begin{array}{cc}
                 \sqrt{1-|c_\mu (s)|^2} & -c_{\mu}^{\ast}(s) \\
                 c_{\mu}(s)             & \sqrt{1-|c_\mu (s)|^2}         
                 \end{array}
                 \right) , 
\\
u_\mu (s) &=& \left(
                 \begin{array}{cc}
                 e^{i \theta_\mu (s)} & 0 \\
                 0                    & e^{-i \theta_\mu (s)} 
                 \end{array}
                 \right) .
\ee
\\
Here $C_\mu (s)$ ($u_\mu (s)$) transforms like a charged matter 
 (a gauge field) under the residual 
U(1) symmetry.
Next we define a monopole current 
(DeGrand-Toussaint monopole)~\cite{degrand80}.
Abelian plaquette variables $\theta_{\mu\nu}(s)$ are written as 
\\
\be
\theta_{\mu\nu}(s) = \theta_\mu (s) + \theta_\nu (s+\hat{\mu}) 
                   - \theta_\mu (s+\hat{\nu}) - \theta_\nu (s) ,  
\hspace*{1cm} ( -4\pi < \theta_{\mu\nu}(s) \le 4\pi ) .
\ee
\\
It is decomposed into two terms using integer variables $n_{\mu\nu}(s)$ :
\\
\be
\theta_{\mu\nu}(s) \equiv \bar{\theta}_{\mu\nu}(s) + 2\pi n_{\mu\nu}(s) , 
\hspace*{1cm} ( -\pi < \bar{\theta}_{\mu\nu}(s) \le \pi ) .
\ee
\\
Here $\bar{\theta}_{\mu\nu}(s)$ is interpreted as an electromagnetic 
flux through the plaquette and $n_{\mu\nu}(s)$ corresponds to the 
number of Dirac string piercing the plaquette.
The monopole current is defined as 
\\
\be
k_\mu (s) = \frac{1}{2} \epsilon_{\mu\nu\rho\sigma}
            \partial_\nu n_{\rho\sigma}(s+\hat{\mu}) . 
\ee
\\
It satisfies the conservation law  
$\partial_{\mu}^{\prime} k_\mu (s) = 0 $.

The abelian dominance and the monopole dominance in the infrared region 
suggest the existence of an effective U(1) action and an 
effective monopole action respectively.
An effective U(1) action is described only by the abelian degree of 
freedom and it is related to 
the original non-abelian action $S[C, u]$ as follows : 
\\
\be
Z   &=& \int Du \bigl[ 
                    \int DC e^{-S[C, u]} \delta(X) \Delta_{FP}(U)
              \bigr] \\
  &=& \int Du e^{-S_{eff}[u]} .
\ee
\\
Here $X=0$ is the gauge-fixing condition and $\Delta_{FP}(U)$ is 
the Faddeev-Popov determinant.
Then an effective monopole action which is written only by monopole currents 
$\{ k_\mu (s) \}$ is derived from the effective U(1) action:
\\
\be
Z  &=& \int Du e^{-S_{eff}[u]} \\
   &=& \Bigl( \prod_{s, \mu} \sum_{k_\mu(s) = -\infty}^{\infty} \Bigr)
        \int Du \delta \bigl( k_\mu(s) - \frac{1}{2} \epsilon_{\mu\nu\rho\sigma}
            \partial_\nu n_{\rho\sigma}(s+\hat{\mu}) \bigr)
        e^{-S_{eff}[u]} \\
   &=& \Bigl( \prod_{s, \mu} \sum_{k_\mu(s) = -\infty}^{\infty} \Bigr)
        \Bigl( \prod_s \delta_{ \partial_{\mu}^{\prime} k_\mu (s), 0} \Bigr)
        e^{-S_{eff}[k]} .
\ee
\\
We derive the effective monopole action  using an inverse Monte-Carlo Method
 from  monopole current configurations $\{ k_\mu (s) \}$ 
  generated by usual Monte-Carlo simulations of SU(2) gluodynamics. 
For more details, see Appendix~\ref{inversemonte}.

\subsection{Anisotropic Lattice}

In zero temperature case,  an almost perfect monopole action has been obtained 
by Kanazawa group [9--11,25].
In the infrared region they get an effective monopole 
action which depends only on a physical scale $b$ alone and is free from 
the lattice spacing $a$.
They take the following steps.
(1) First thermalized monopole current configurations $\{ k_\mu (s) \}$ 
are generated from the  Wilson action at some $\beta$.
These configurations depend on lattice spacing $a(\beta)$.
(2) In order to consider the infrared region of QCD, 
they perform a blockspin transformation in terms of the monopole currents 
and define the extended monopoles. After the blockspin transformation,  
renormalized lattice spacing is $b=na(\beta)$,  where $n$ is the number of steps of 
blockspin transformations.
(3) Using the renormalized monopole current configurations,  they determine 
an effective monopole action numerically on the renormalized lattice $b$.
(4) The continuum limit is taken as the limit $a \to 0$ and $n \to \infty$ 
for a fixed physical scale $b$.
They have found that scaling looks good for $b \ge 1$ in unit of  
the physical string tension $\sqrt{\sigma_{phys}}$ under the assumption 
that the action is composed of 2,  4 
and 6 point monopole interactions.

Now let us  consider the finite temperature case.
A special feature of this system is a periodic boundary condition for 
time direction and the physical size of the  time direction is finite.
The physical length in the time direction is limited to less than $1/T$.
In this case it is useful to introduce anisotropic lattices 
[26-28].
In the space directions,  we perform the blockspin transformation as done in  
the zero temperature case. The continuum limit is taken as  $a_s \to 0$ and 
$n_s \to \infty$ for a fixed physical scale $b_s = n_s a_s$.
Here $a_s$ is the lattice spacing in the space directions and $n_s$ 
is the blockspin factor.
In the time direction,  the continuum limit is taken as $a_t \to 0$ 
and $N_t \to \infty$ for a fixed temperature $T= 1/(N_t a_t)$.
Here $a_t$ is the lattice spacing in the  time direction and $N_t$ is 
the number of lattice site for the time direction.
We finally get an effective monopole action 
which depends on the physical scale $b_s$ and the temperature $T$,  if 
the scaling is satisfied.

\subsection{Determination Of The Lattice Spacings $(a_s,  a_t)$}

The Wilson action on anisotropic lattice for $SU(2)$ gauge theory 
is written as 
\\
\be
S = \beta \Bigl\{ \frac{1}{\gamma} \sum_{s, i > j \ne 4}
                   P_{ij} (s) + \gamma \sum_{s, i \ne 4}
                   P_{i4} (s)
          \Bigr\} , 
\ee
\be
P_{\mu \nu} (s) \equiv \frac{1}{4}
    Tr \bigl[ {\bf 1} - U_\mu (s) U_\nu (s + \hat{\mu})
                        U_{\mu}^{\dagger} (s + \hat{\nu})
                        U_{\nu}^{\dagger} (s)
       \bigr]
    + h.c .
\ee
\\
If $\gamma=1$,   the lattice is isotropic $(a_s = a_t)$.
The procedure to determine the lattice spacing $(a_s,  a_t)$ 
from the above action is the following~\cite{klassen98}.

First we determine an anisotropy $\xi \equiv a_s/a_t$ 
for various values $(\beta,  \gamma)$ 
considering the zero-temperature case.
We calculate $V(I, J)$ from  Wilson loops $W(I, J)$ as 
\\
\be
V(I, J) = \log \frac{W(I, J-1)}{W(I, J)} . \label{eq:vrt}
\ee 
\\
This is the static potential if we take the limit $J \to \infty$.
Using (\ref{eq:vrt}),  we define $V_s (R_s,  J)$ and $V_t (R_t,  J)$ as
\\
\be
V_s (R_s,  J) \equiv \log \frac{W(R_s,  J-1)}{W(R_s,  J)} ,  \\
V_t (R_t,  J) \equiv \log \frac{W(R_t,  J-1)}{W(R_t,  J)} .
\ee
\\
Here $R_s$ and $J$ are taken to be lattice sizes 
of the Wilson loop in space directions 
and $R_t$ is the size for time direction.
In other words,  $V_s (R_s , J)$ and $V_t (R_t, J)$ are calculated 
from spacelike and timelike Wilson loops respectively.
Then we define the ratio $R(R_s,  R_t,  J)$ as
\\
\be
R(R_s,  R_t ,  J) \equiv \frac{V_s (R_s,  J)}{V_t (R_t,  J)} .
\ee
\\
We vary $R_t$ for fixed $R_s$ and $J$ and look for the value 
$R_t$ for $R(R_s,  R_t,  J)=1$.
It is impossible to vary $R_t$ continuously,  so that we use an interpolation.
If $R(R_s,  R_t,  J)=1$,  then $a_s R_s = a_t R_t$ and 
$\xi = a_s / a_t = R_t / R_s$.
In the classical level an anisotropy $\xi = \gamma$,  
but that is not the case  
in the quantum level.
So we define $\eta$ using the parameter $\gamma$ as 
$\xi \equiv \eta \gamma$.

Next to determine the lattice spacings $(a_s,  a_t)$ in unit of the 
physical string tension at zero temperature,  we calculate the string tension 
for $(\beta,  \gamma)$ on the lattice.
From the timelike Wilson loop,  the static potential is calculated by 
\\
\be
V(R_s) = \lim_{R_t \to \infty} \log \frac{W(R_s, R_t-1)}{W(R_s, R_t)}.\label{eq:potential}
\ee
\\
We fit it with the form $linear + Coulomb + constant$.
We use the smearing procedure~\cite{ape87} for spacelike link variables.
The relation between the lattice string tension $\sqrt{\sigma_{lat}}$ 
and the physical string tension $\sqrt{\sigma_{phys}}$ is 
\\
\be
\frac{\sigma_{lat}}{a_s a_t} = \sigma_{phys} .
\ee
\\
So we can determine the lattice spacing $(a_s,  a_t)$ as follows:
\\
\be
a_s = \sqrt{ \frac{ \xi  \sigma_{lat} }{ \sigma_{phys} } }  \ \ \  ,  \ \  
a_t = \sqrt{ \frac{ \sigma_{lat}  }{ \xi \sigma_{phys} } } .
\ee
\\

The values of $(\beta,  \gamma)$ and the lattice sizes and the number of 
configurations used in simulations are summarized in Table~\ref{gabesico}.
The results of $\eta$ for each $(\beta,  \gamma)$ 
are given in Fig.~\ref{etagabe}.
The lattice spacing $(a_s,  a_t)$ obtained from $\eta$,  $\gamma$ and 
$\sigma_{lat}$ are in Fig.~\ref{asatgabe}.
Using these results we determine the parameter $(\beta,  \gamma)$ for 
arbitrary $(a_s,  a_t)$ by the interpolation.

\TABLE[t]{
{\small
  \begin{tabular}{|c|c|c|c||c|c|c|c| |c|c|c|c|}                   \hline
    $\gamma$ & $\beta$ &    Lattice size    &   conf. &
    $\gamma$ & $\beta$ &    Lattice size    &   conf. &
    $\gamma$ & $\beta$ &    Lattice size    &   conf. \\ \hline\hline
      $1.0$  &  $2.0$  & $16^3 \times 48$   &       $100$       
    & $2.0$  &  $2.0$  & $16^3 \times 48$   &      $1500$        
    & $3.0$  &  $2.0$  & $16^3 \times 64$   &      $2550$       \\ 
             &  $2.1$  & $16^3 \times 48$   &       $100$       
    &        &  $2.1$  & $16^3 \times 48$   &      $1500$        
    &        &  $2.1$  & $16^3 \times 64$   &      $2550$       \\ 
             &  $2.2$  & $16^3 \times 48$   &       $100$       
    &        &  $2.2$  & $16^3 \times 48$   &      $1500$        
    &        &  $2.2$  & $16^3 \times 64$   &      $2550$       \\ 
             &  $2.3$  & $16^3 \times 48$   &       $100$       
    &        &  $2.3$  & $16^3 \times 48$   &      $1000$        
    &        &  $2.3$  & $16^3 \times 64$   &      $1150$       \\ 
             &  $2.4$  & $16^3 \times 48$   &       $100$       
    &        &  $2.4$  & $16^3 \times 48$   &      $1000$        
    &        &  $2.4$  & $16^3 \times 64$   &      $1150$       \\ 
             &  $2.5$  & $24^3 \times 48$   &       $100$       
    &        &  $2.5$  & $20^3 \times 60$   &      $1050$        
    &        &  $2.5$  & $20^3 \times 72$   &      $1090$       \\ 
             &  $2.6$  & $24^3 \times 48$   &       $100$        
    &        &  $2.6$  & $20^3 \times 60$   &      $1050$        
    &        &  $2.6$  & $20^3 \times 72$   &      $1090$       \\ \hline 
      $1.2$  &  $2.0$  & $16^3 \times 48$   &      $1500$       
    & $2.5$  &  $2.0$  & $16^3 \times 64$   &      $1950$        
    & $3.5$  &  $2.0$  & $16^3 \times 80$   &      $3200$       \\ 
             &  $2.1$  & $16^3 \times 48$   &      $1500$       
    &        &  $2.1$  & $16^3 \times 64$   &      $1950$        
    &        &  $2.1$  & $16^3 \times 80$   &      $3200$       \\ 
             &  $2.2$  & $16^3 \times 48$   &      $1500$       
    &        &  $2.2$  & $16^3 \times 64$   &      $1950$        
    &        &  $2.2$  & $16^3 \times 80$   &      $3200$       \\ 
             &  $2.3$  & $16^3 \times 48$   &      $1000$       
    &        &  $2.3$  & $16^3 \times 64$   &      $1050$        
    &        &  $2.3$  & $16^3 \times 80$   &      $3200$       \\ 
             &  $2.4$  & $16^3 \times 48$   &      $1000$       
    &        &  $2.4$  & $16^3 \times 64$   &      $1050$        
    &        &  $2.4$  & $16^3 \times 80$   &      $3200$       \\ 
             &  $2.5$  & $24^3 \times 48$   &      $1100$       
    &        &  $2.5$  & $20^3 \times 72$   &      $1090$        
    &        &  $2.5$  & $20^3 \times 72$   &      $1730$       \\ 
             &  $2.6$  & $24^3 \times 48$   &      $1100$        
    &        &  $2.6$  & $20^3 \times 72$   &      $1090$        
    &        &  $2.6$  & $20^3 \times 72$   &      $1730$       \\ \hline 
      $1.5$  &  $2.0$  & $16^3 \times 48$   &      $1500$       \\
             &  $2.1$  & $16^3 \times 48$   &      $1500$       \\
             &  $2.2$  & $16^3 \times 48$   &      $1500$       \\
             &  $2.3$  & $16^3 \times 48$   &      $1000$       \\
             &  $2.4$  & $16^3 \times 48$   &      $1000$       \\
             &  $2.5$  & $24^3 \times 48$   &      $1100$       \\
             &  $2.6$  & $24^3 \times 48$   &      $1100$       \\ \cline{1-4}  
  \end{tabular}
\caption{$(\gamma,  \beta)$, lattice size and the number of configurations used in 
simulations to determine $(a_s,  a_t)$.}
}
\label{gabesico}
}

\FIGURE{
\epsfig{file=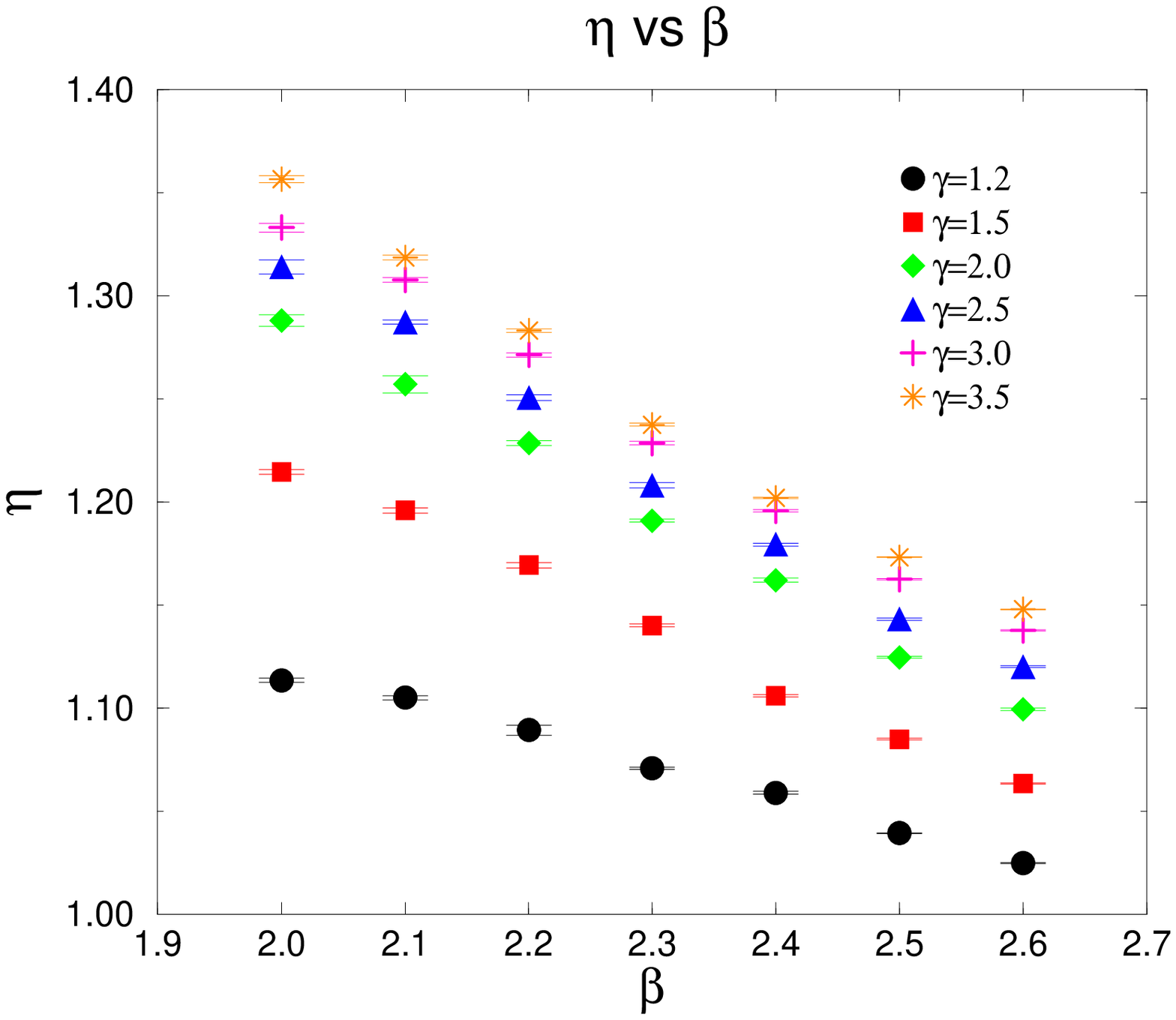, height=7cm, width=7cm}
\epsfig{file=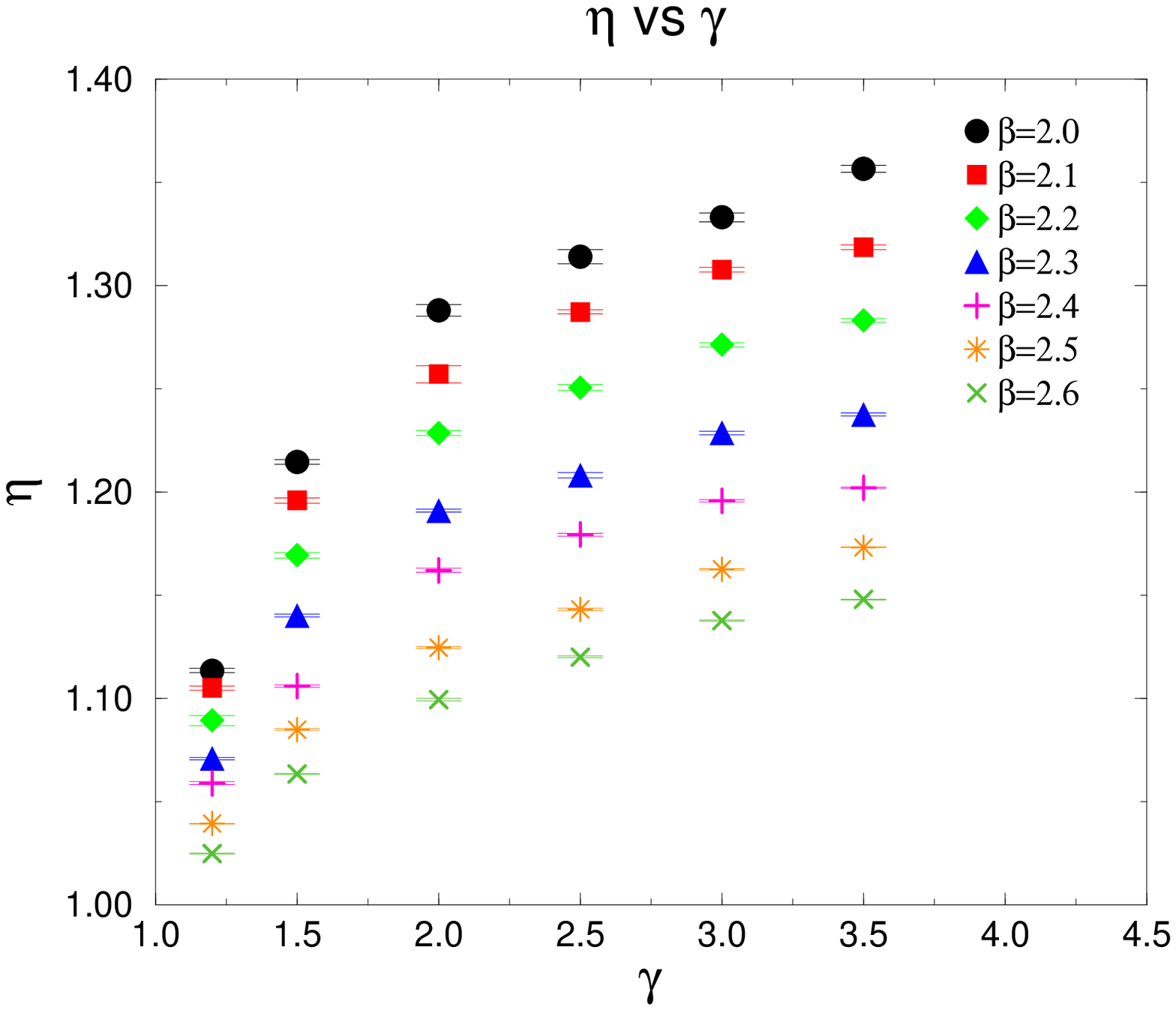, height=7cm, width=7cm}
\caption{The relations $\eta$ vs $\beta$ (left)
and $\eta$ vs $\gamma$ (right).}
\label{etagabe}
}

\FIGURE{
\epsfig{file=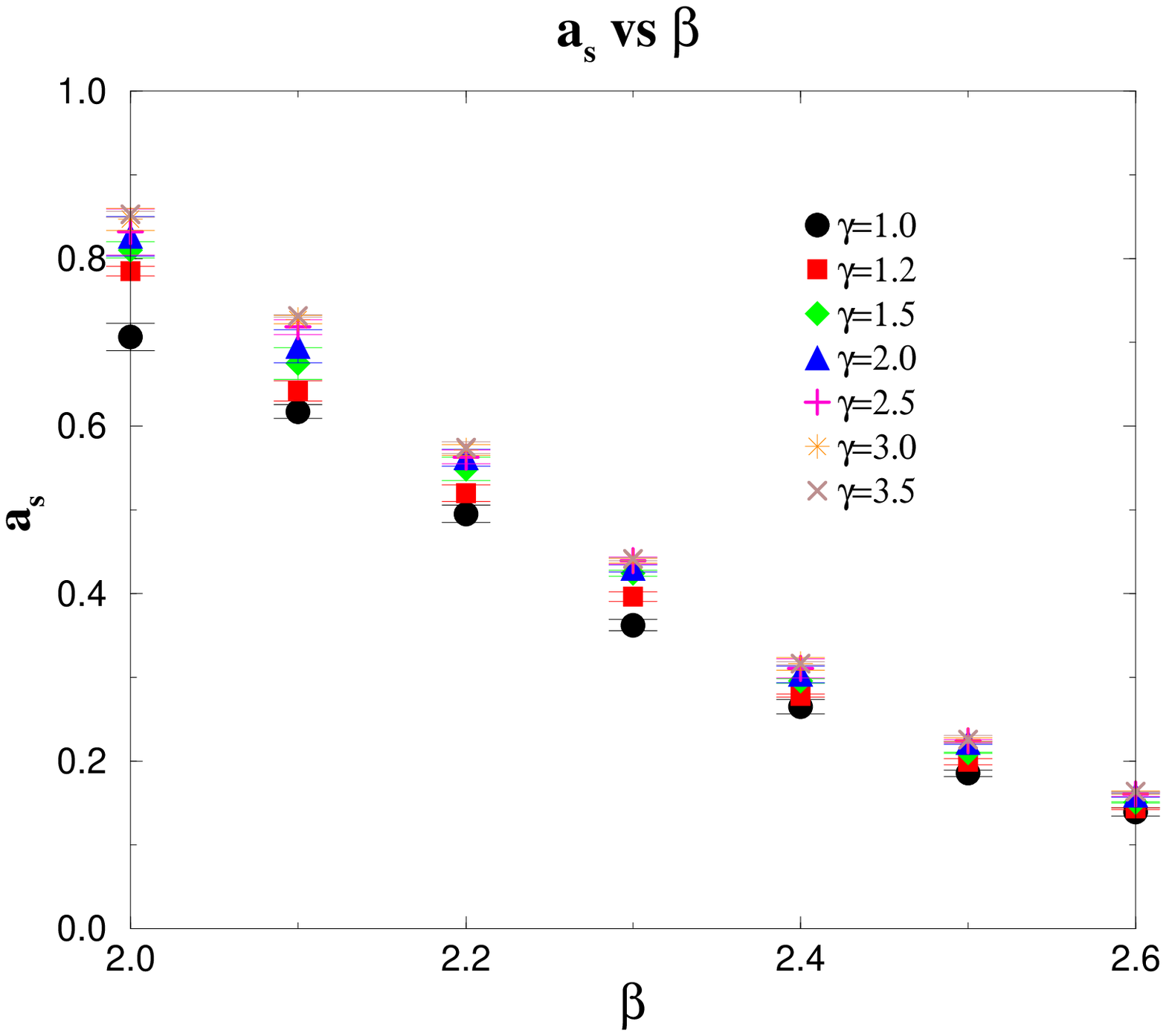, height=7cm, width=7cm}
\epsfig{file=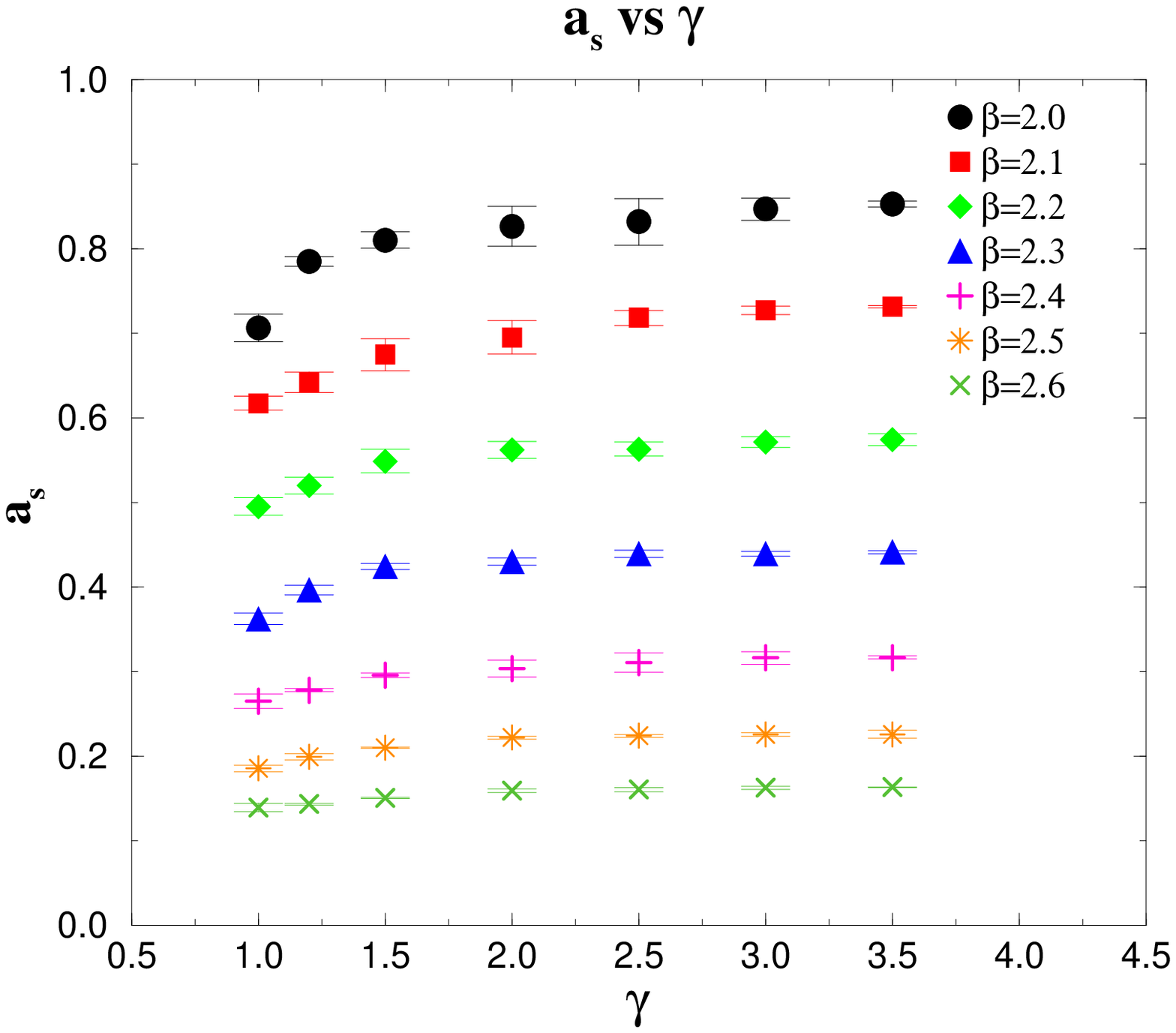, height=7cm, width=7cm}
\epsfig{file=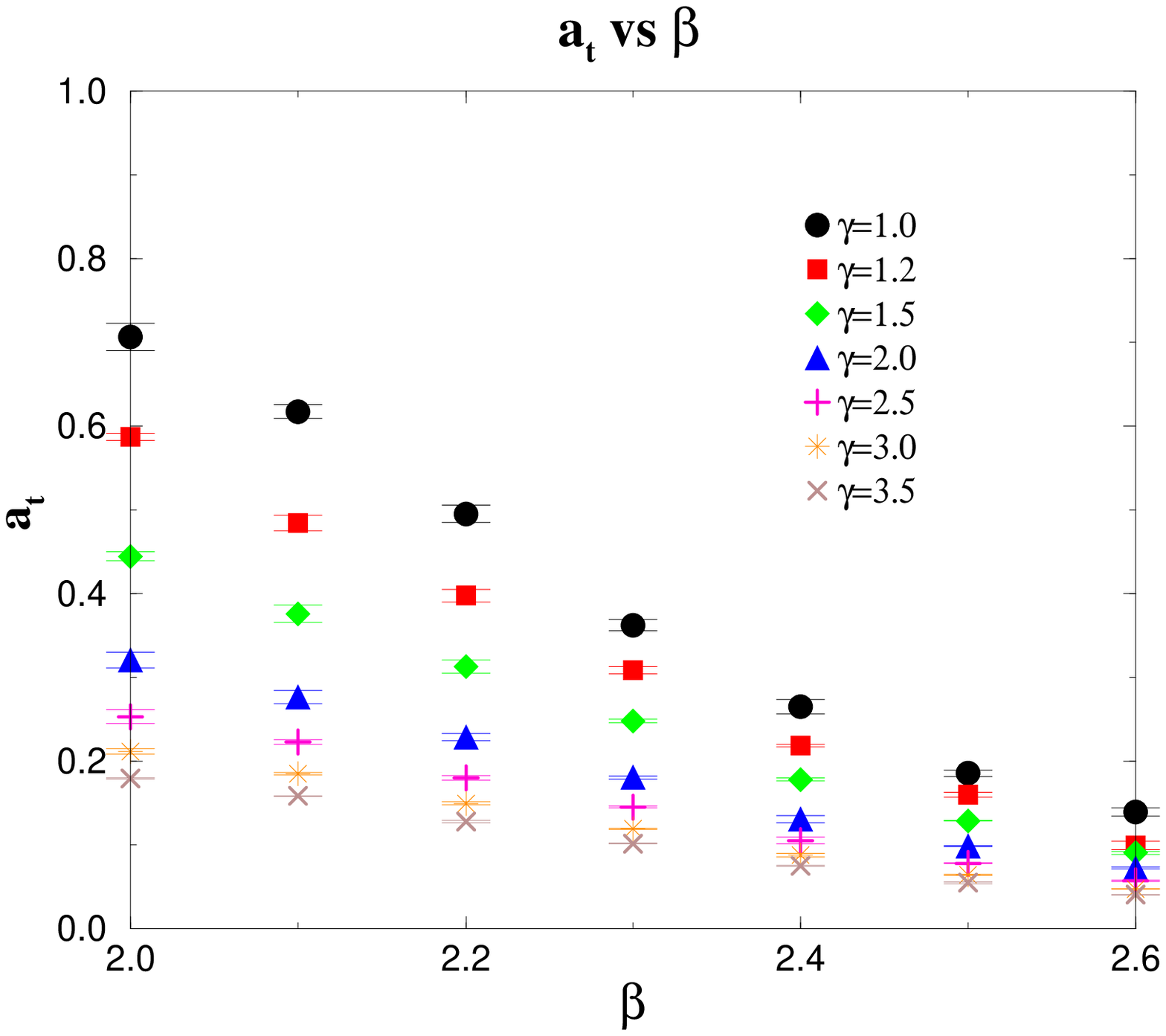, height=7cm, width=7cm}
\epsfig{file=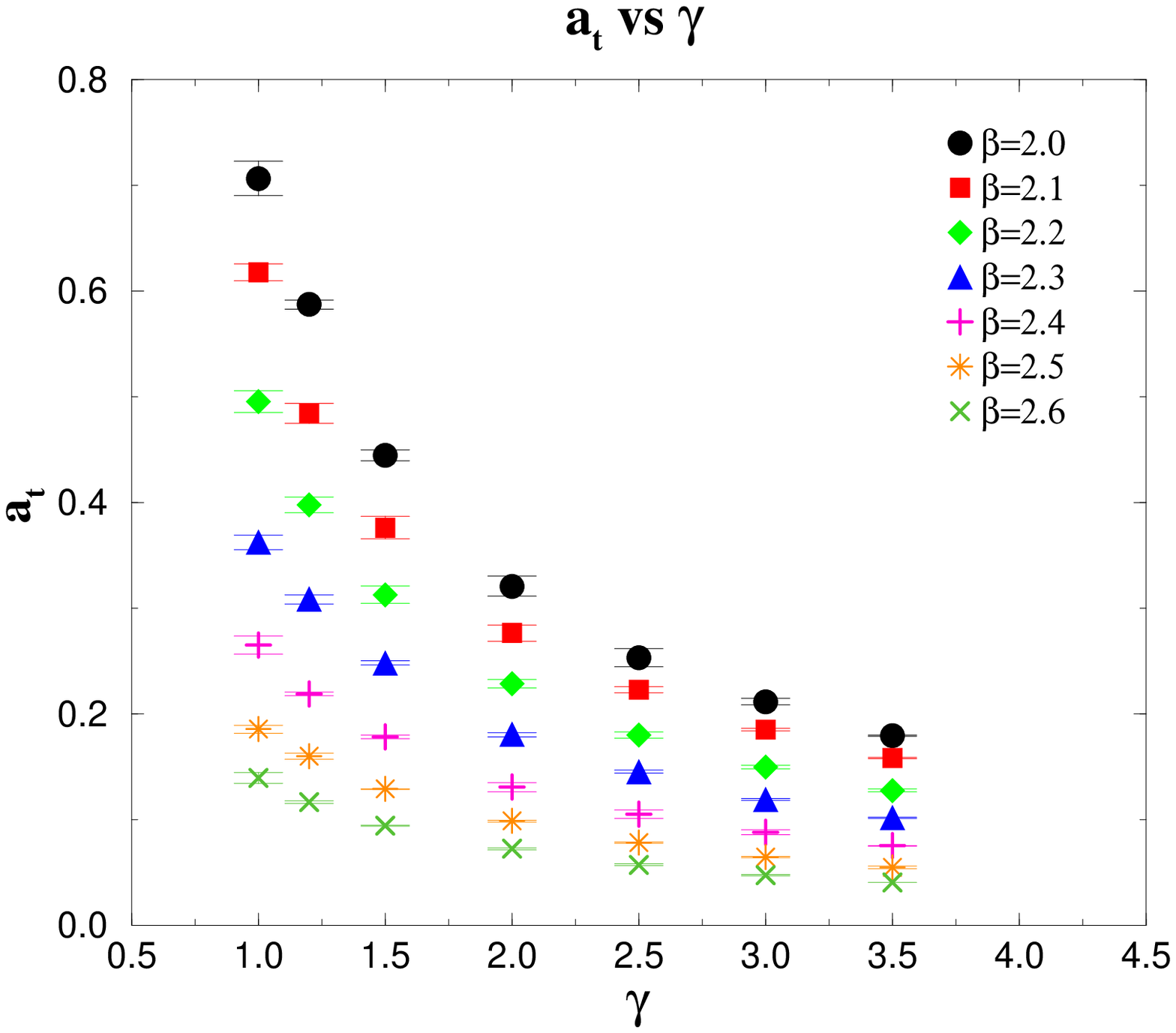, height=7cm, width=7cm}
\caption{$a_s$ vs $\beta$,  $a_s$ vs $\gamma$, 
$a_t$ vs $\beta$ and $a_t$ vs $\gamma$}
\label{asatgabe}
}

\subsection{Monopole Action At Finite Temperature}

Now let us construct the 4D effective monopole action 
at finite temperature adopting $N_s^3 \times N_t$ ($N_s \gg N_t$) lattices. 
Here we have to consider 
 spacelike monopole currents $k_i (i=1, 2, 3)$
and timelike monopole current $k_4$ separately.
An abelian Wilson loop operator is expressed as 
\\
\be
W_a = \exp \bigl\{ i \sum_s J_\mu(s) \theta_\mu(s) \bigr\} , 
\ee
\\
where $J_\mu(s)$ is an external current taking $\pm 1$ along the 
Wilson loop. Since $J_\mu(s)$ is conserved,  it is rewritten for a simple 
flat Wilson loop in terms of an antisymmetric variable $M_{\mu\nu}(s)$ 
as $J_\nu(s) = \partial_{\mu}^{\prime} M_{\mu\nu}(s)$.
$M_{\mu\nu}(s)$ takes $\pm 1$ on the surface with the Wilson loop boundary.
Then we get 
\\
\be
W_a = \exp \bigl\{ -\frac{i}{2} \sum_s M_{\mu\nu}(s) \theta_{\mu\nu}(s) \bigr\} , 
\ee
\\
where 
$\theta_{\mu\nu}(s) = \partial_\mu \theta_\nu (s) -
                      \partial_\nu \theta_\mu (s) $.
Using the decomposition 
$\theta_{\mu\nu}(s) = \bar{\theta}_{\mu\nu}(s) + 2\pi n_{\mu\nu}(s)$, 
we get 
\\
\be
W_a &=& W_p \cdot W_m ,  \\
W_p &=& \exp \bigl\{ -i \sum_{s, s^\prime} \partial_{\mu}^{\prime} 
                    \bar{\theta}_{\mu\nu}(s) D(s-s^\prime) J_\nu(s^\prime)
             \bigr\} ,  \\
W_m &=& \exp \bigl\{ 2 \pi i \sum_{s , s^\prime} k_\beta (s)
                   D( s - s^\prime ) \frac{1}{2} 
                   \epsilon_{\alpha \beta \rho \sigma}
                   \partial_\alpha M_{\rho \sigma} (s^\prime)   
           \bigr\} , 
\ee
\\
where $D(s)$ is the lattice Coulomb propagator~\cite{shiba94}.
Since $\partial_{\mu}^{\prime}\bar{\theta}_{\mu\nu}(s)$ contains 
only the photon fields,  $W_p$ ($W_m$) is the photon (monopole) 
contribution to the Wilson loop.
An ordinary space-time Wilson loop has a contribution only from spacelike 
monopoles,  whereas both space and timelike monopoles 
contribute to a spacelike Wilson loop.
The physical string tension has a finite value in the  confinement phase 
but it is zero in the deconfinement phase. 
On the other hand,  the spatial string tension determined by the spacelike 
Wilson loop has a finite value in both phases.
Another special feature of the monopole action at finite temperature 
comes from the finite size in the time direction.
We define a blockspin transformation of monopole currents  \cite{ivanenko90} as
\\
\be
  K_{\mu \ne 4} (s_s ,  s_4) &=& \sum_{i ,  j = 0}^{n_s - 1}
                             \sum_{l = 0}^{n_t -1}
           k_{\mu \ne 4} \bigl( n_s s_s + ( n_s -1 ) \hat{\mu}
                                + i \hat{\nu} + j \hat{\rho} ,  
                                n_t s_4 + l 
                         \bigr) , 
\\
  K_{4} (s_s ,  s_4) &=& \sum_{i ,  j ,  l = 0}^{n_s - 1}
           k_{4} \bigl( n_s s_s + i \hat{\mu}
                                + j \hat{\nu} + l \hat{\rho} ,  
                                n_t s_4 + ( n_t -1 ) 
                         \bigr) , 
\ee 
\\
where $n_s$ ($n_t$) is a blockspin factor for space (time) direction. 
Actually,  we consider mostly the $n_t=1$ case.

\subsection{Results}

The parameters used in the simulations and the corresponding lattice spacing 
($a_s,  a_t$) are summarized in Table~\ref{bgaa}.
The lattice sizes and the temperatures are written in Table~\ref{tsintat}.
We perform 6000 thermalization sweeps and take 40 configurations totally at  
every 100 sweeps.
The inverse Monte-Carlo method used  here is the modified Swendsen's method 
extended to monopole currents with the conservation law 
(see Appendix \ref{modswed})~\cite{shiba95, shiba94}.
For simplicity,  
we assume that the effective monopole action is composed of  only quadratic 
interactions. We adopts 84 interactions (For the explicit definition
of each interaction,  see Appendix \ref{4dcoupling}).

\DOUBLETABLE[t]{
\begin{tabular}{|c||c||c||c|} \hline
  $\beta$  &  $\gamma$  &   $a_s$  &   $a_t$    \\ \hline
   2.470   &   2.841    &   0.250  &   0.075    \\ \hline
   2.500   &   2.615    &   0.225  &   0.075    \\ \hline
   2.533   &   2.354    &   0.200  &   0.075    \\ \hline
   2.548   &   2.256    &   0.190  &   0.075    \\ \hline
   2.565   &   2.152    &   0.180  &   0.075    \\ \hline
   2.573   &   2.098    &   0.175  &   0.075    \\ \hline
   2.581   &   2.042    &   0.170  &   0.075    \\ \hline
   2.598   &   1.927    &   0.160  &   0.075    \\ \hline
\end{tabular}
}
{
\begin{tabular}{|c||c||c|} \hline
      T       &   Lattice size     &   $N_t a_t (= \frac{1}{T})$     \\ \hline
  $ 0.6T_c$   &  $72^3 \times 32$  &        $2.4$                    \\ \hline
  $ 0.8T_c$   &  $72^3 \times 24$  &        $1.8$                    \\ \hline
  $0.96T_c$   &  $72^3 \times 20$  &        $1.5$                    \\ \hline
  $ 1.2T_c$   &  $72^3 \times 16$  &        $1.2$                    \\ \hline
  $ 1.6T_c$   &  $72^3 \times 12$  &        $0.9$                    \\ \hline
  $ 2.4T_c$   &  $72^3 \times  8$  &        $0.6$                    \\ \hline
\end{tabular}
}
{Parameter ($\beta$ ,  $\gamma$) and lattice spacing ($a_s$,  $a_t$).
\label{bgaa}}
{Temperature, lattice size and $N_t a_t$.
\label{tsintat}}

\FIGURE[t]{
\epsfig{file=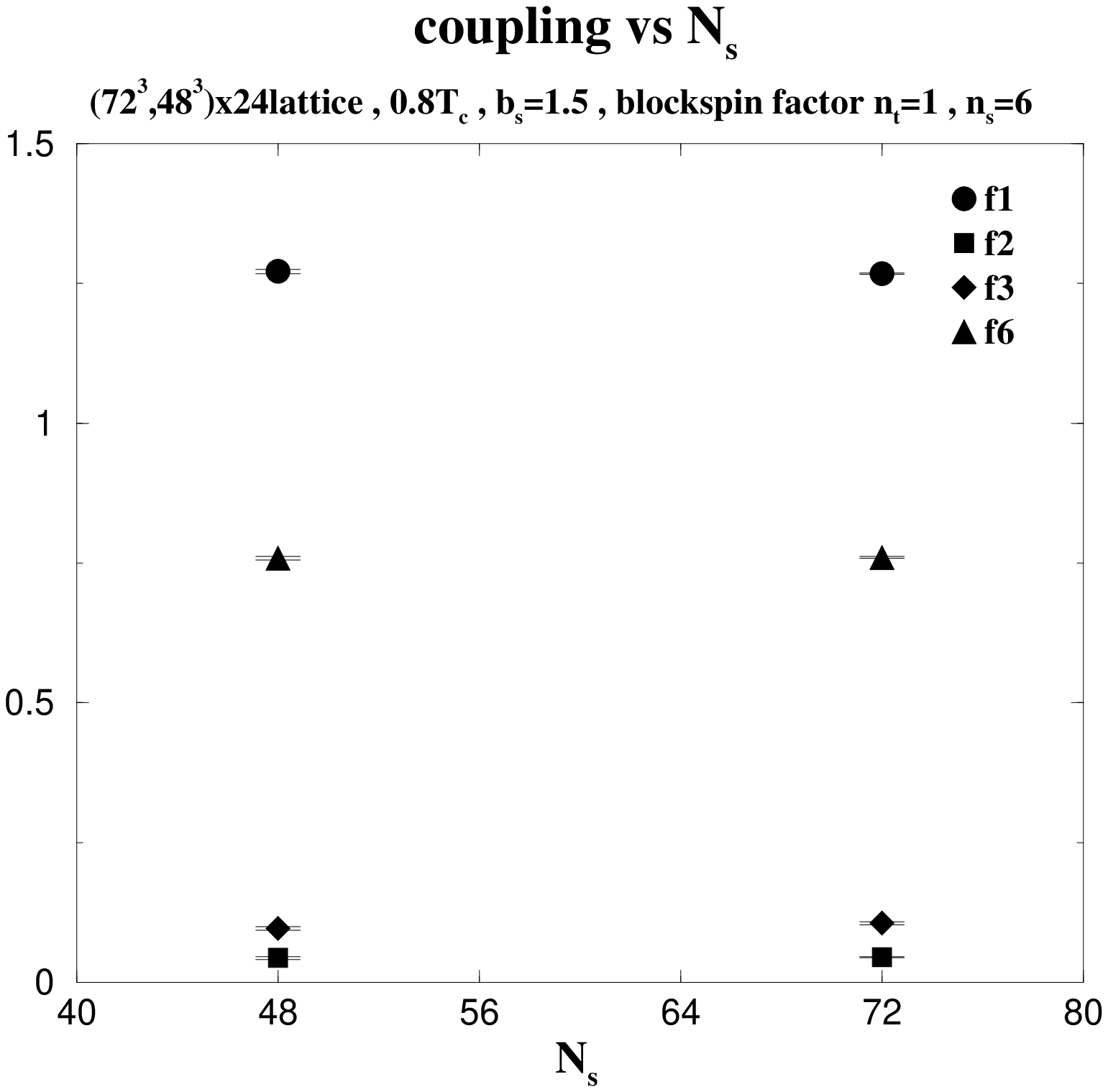, height=7cm, width=7cm}
\epsfig{file=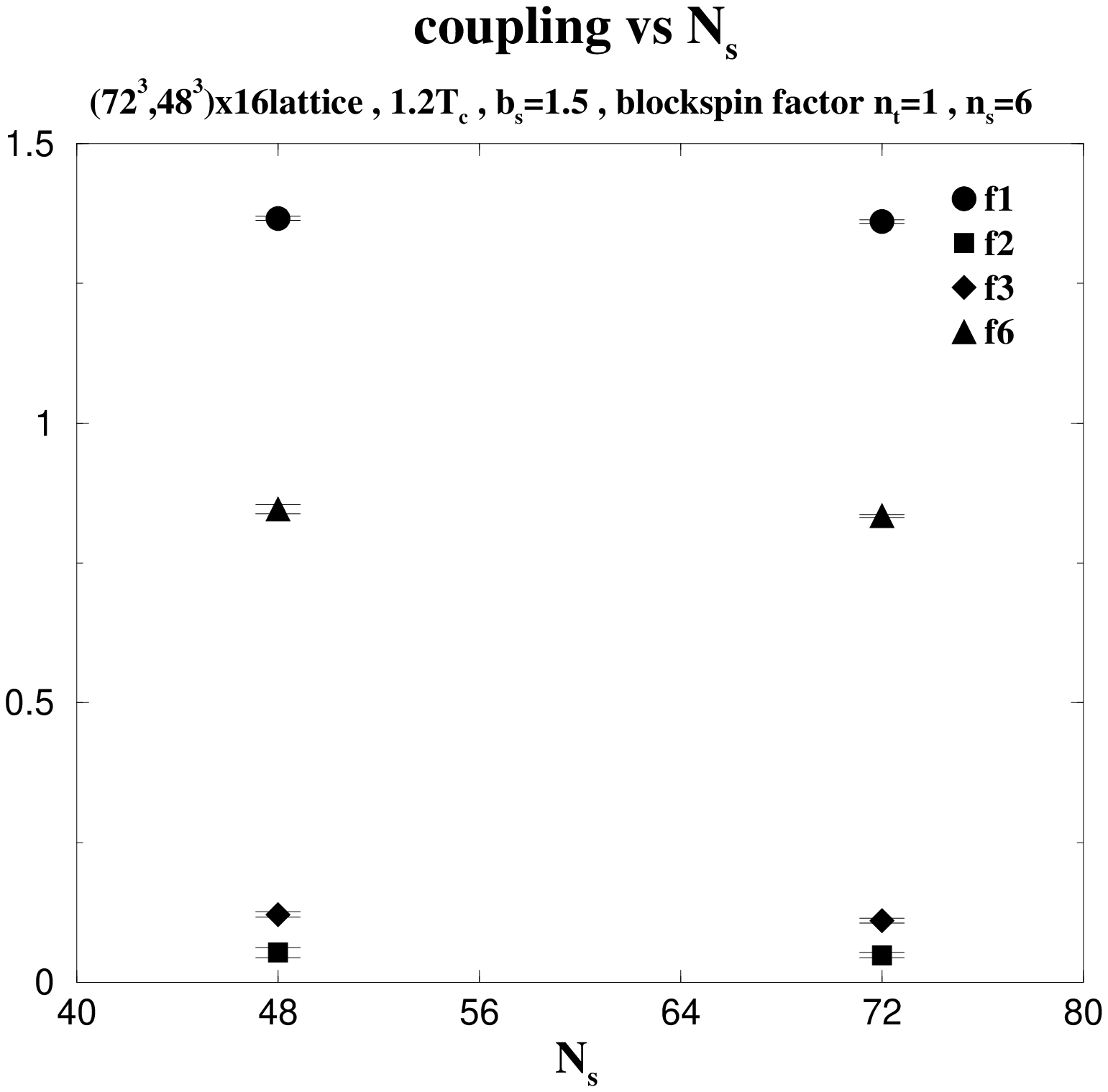, height=7cm, width=7cm}
\caption{The couplings vs $N_s$ for $b_s = 1.5$. 
$T=0.8T_c$(left) and $T=1.2T_c$(right).}
\label{Ns0812tcnt1f1236bs15}
}

First to get the infinite-volume limit,  we determine the actions 
for different lattice sizes at each ($\beta$,  $\gamma$) and 
temperatures.
We  consider two different lattice sizes. The data show that 
the volume dependence is hardly seen. The examples for $b_s = 1.5$ and 
$T=0.8T_c$,  $1.2T_c$ are shown in Fig.~\ref{Ns0812tcnt1f1236bs15}.

\FIGURE[t]{
\epsfig{file=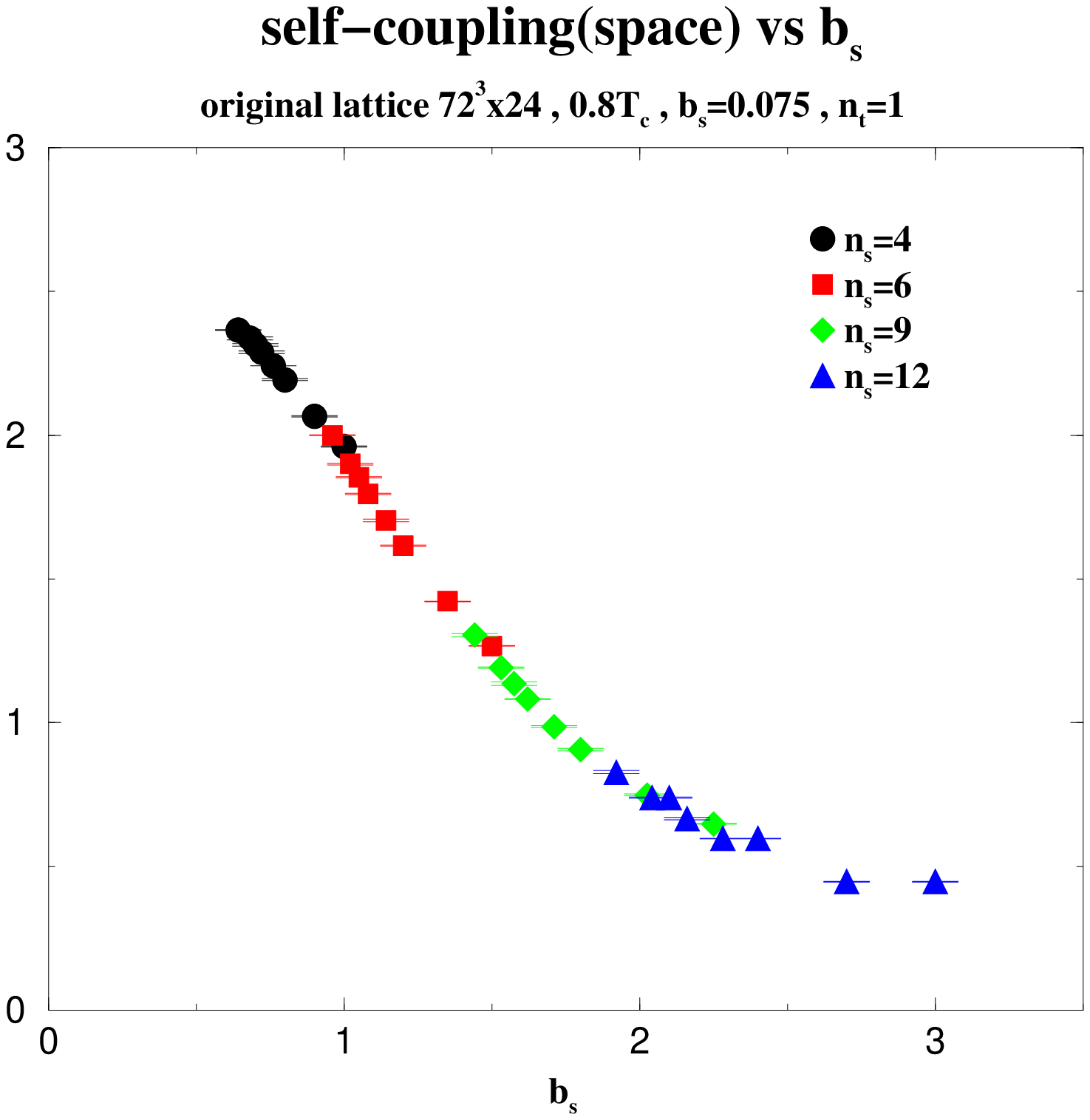, height=7cm, width=7cm}
\epsfig{file=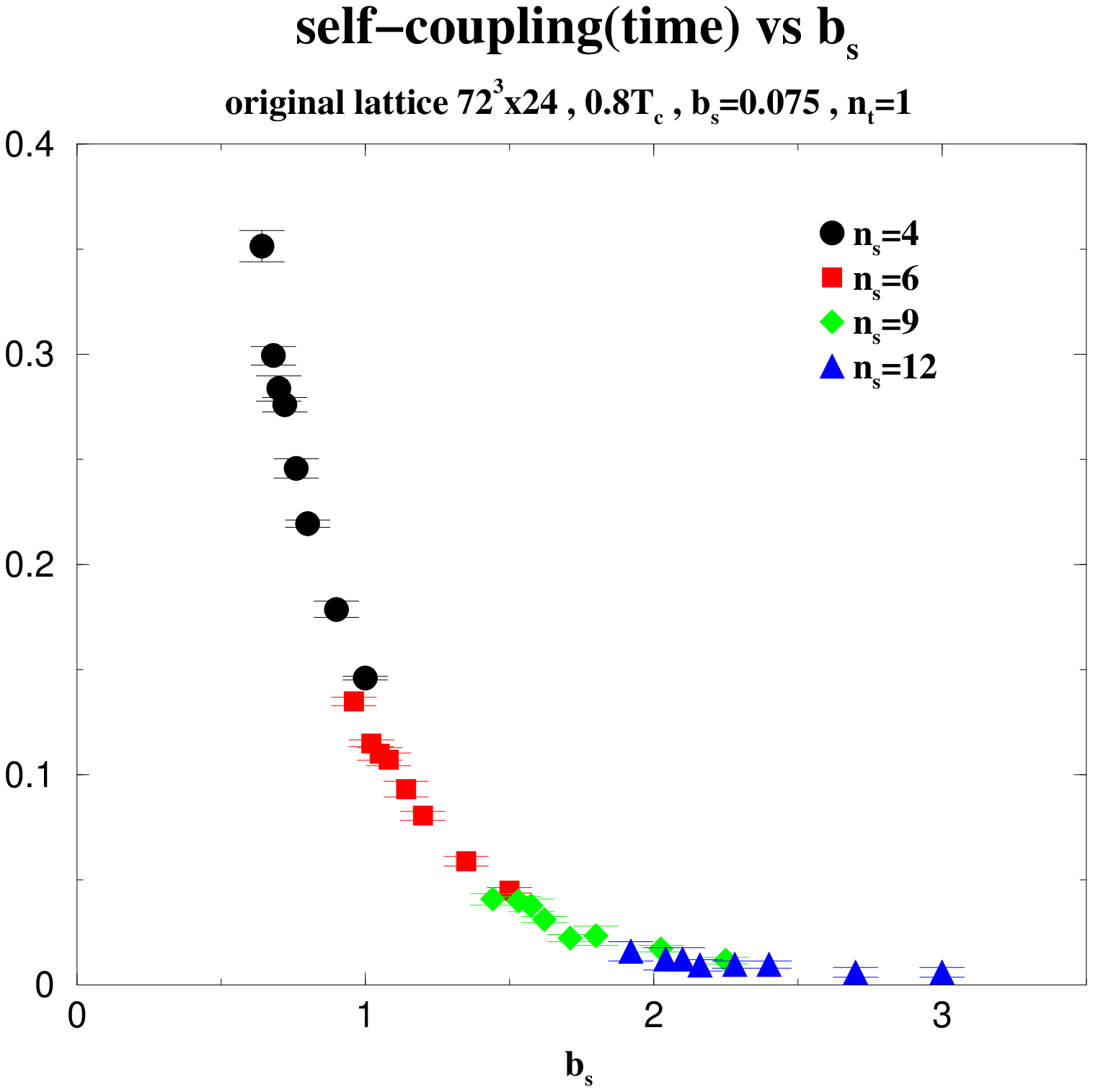, height=7cm, width=7cm}
\caption{$n_s$-dependence of the couplings $f_1$(left) 
and $f_2$(right) at $0.8T_c$.}
\label{nsdep08tc}
}

\FIGURE[t]{
\epsfig{file=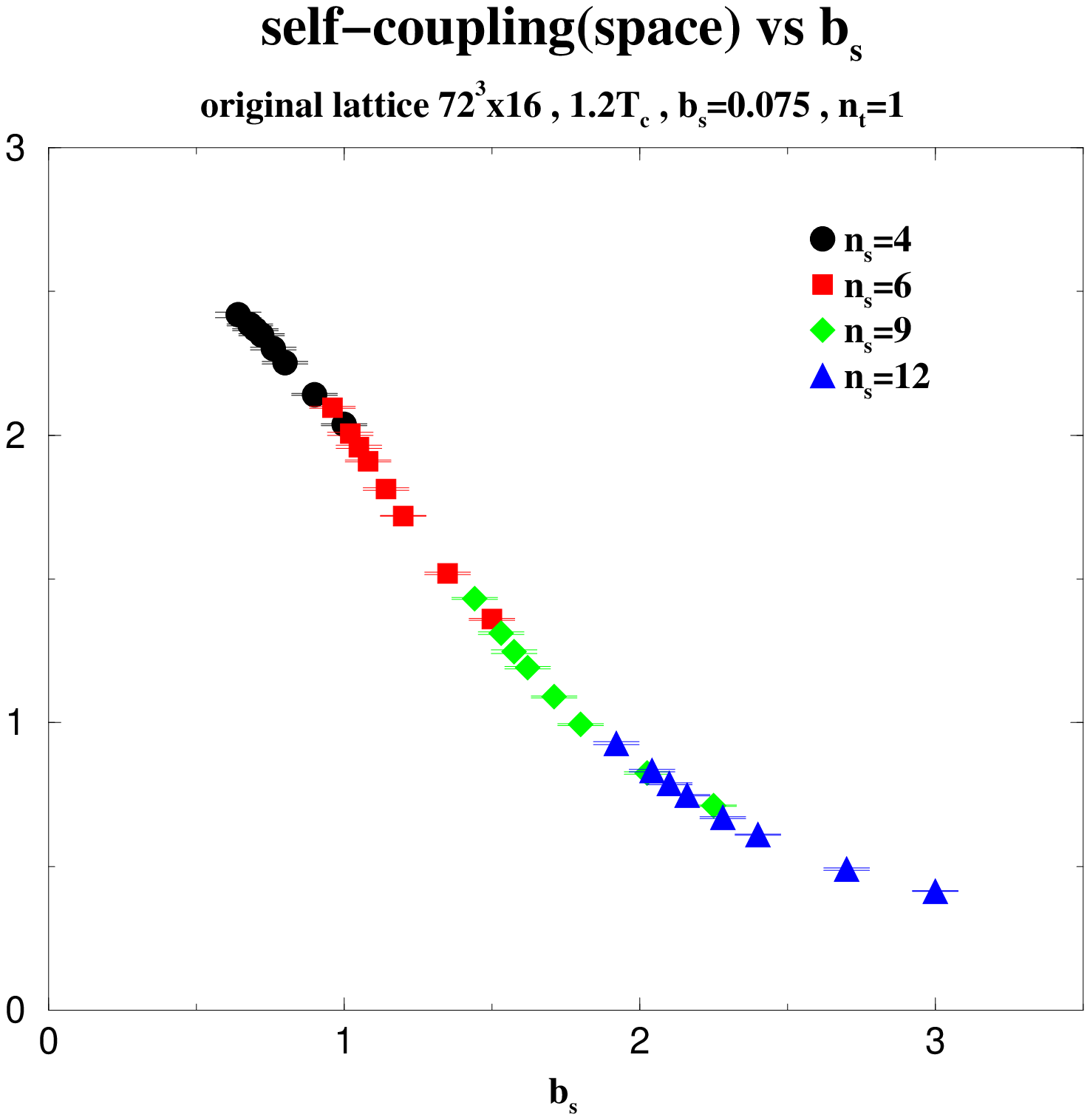, height=7cm, width=7cm}
\epsfig{file=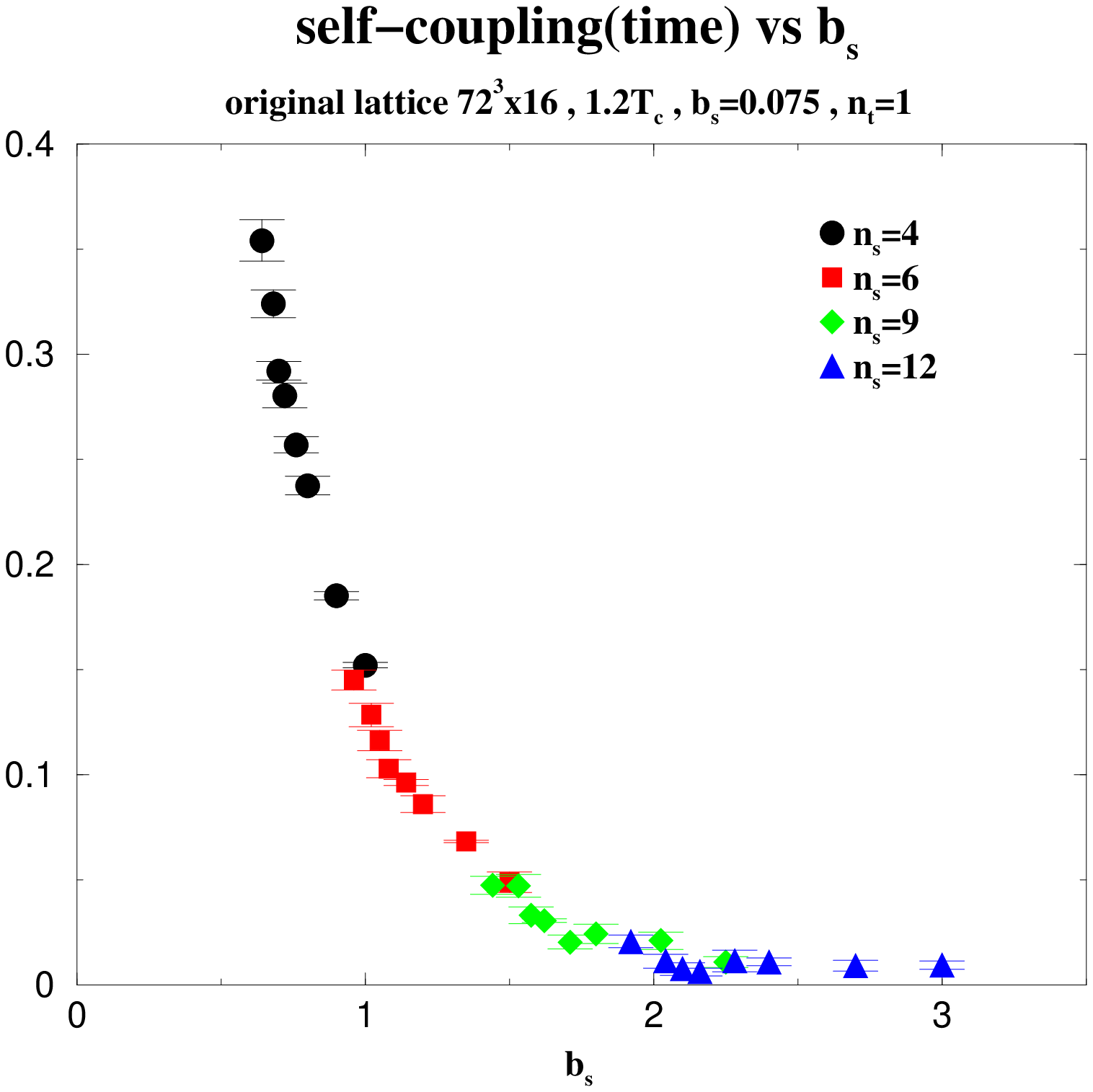, height=7cm, width=7cm}
\caption{$n_s$-dependence of the couplings $f_1$(left) 
and $f_2$(right) at $1.2T_c$.}
\label{nsdep12tc}
}

To get the continuum limit for space directions,  we perform the blockspin 
transformation ($n_s = 4,  6,  9, 12$) for each temperature.
The $n_s$-dependences of the couplings $f_1$ and $f_2$ for 
$0.8T_c$ and $1.2T_c$ are shown 
in Fig.~\ref{nsdep08tc} and Fig.~\ref{nsdep12tc}.
These figures indicate $n_s$-independence.
The data of the couplings f1 and f2 for all temperatures are seen 
in Fig.~\ref{bsnt1f12}. We can see the nice scaling behaviors 
 at each temperature.

\FIGURE{
\epsfig{file=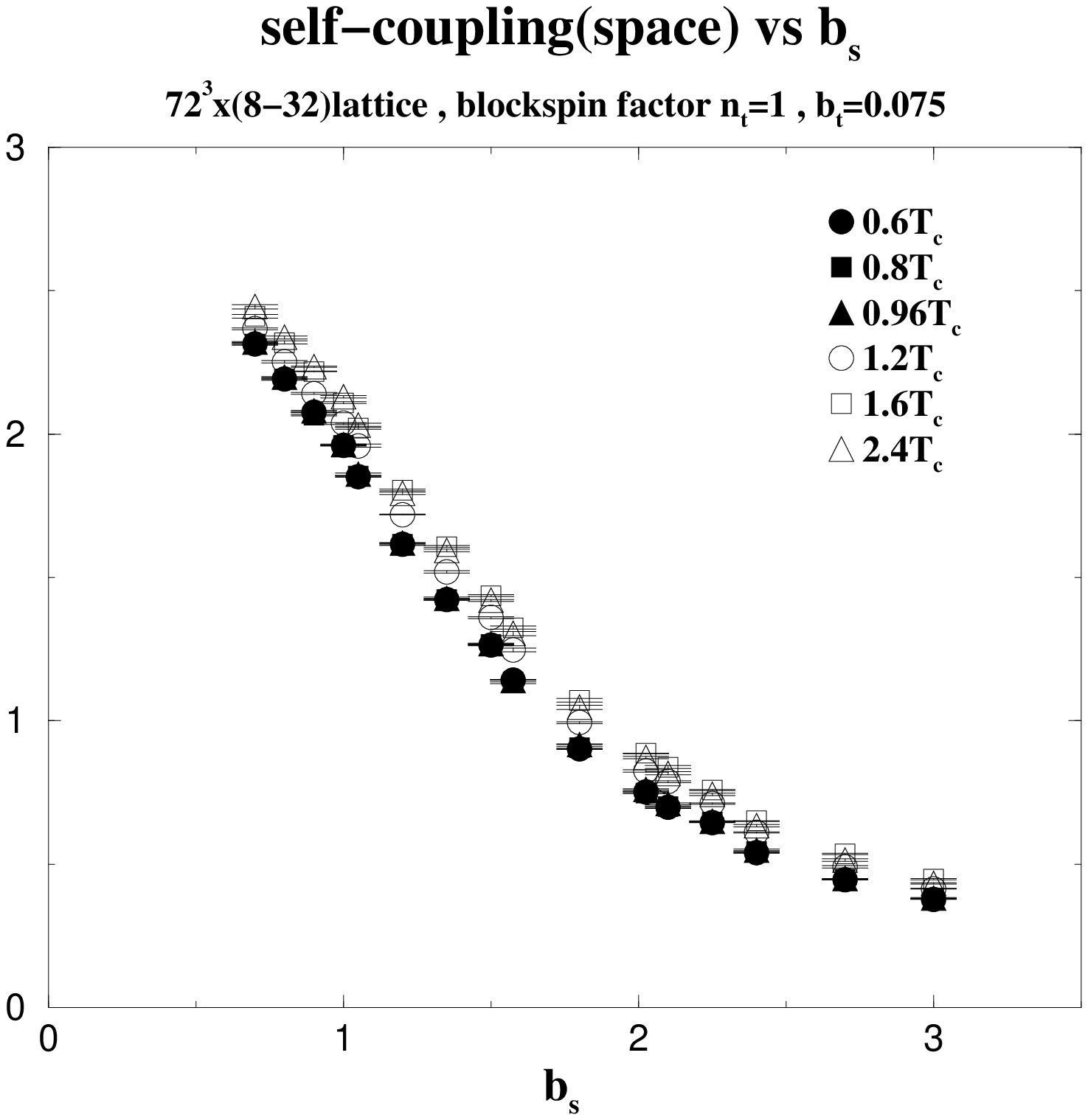, height=7cm, width=7cm}
\epsfig{file=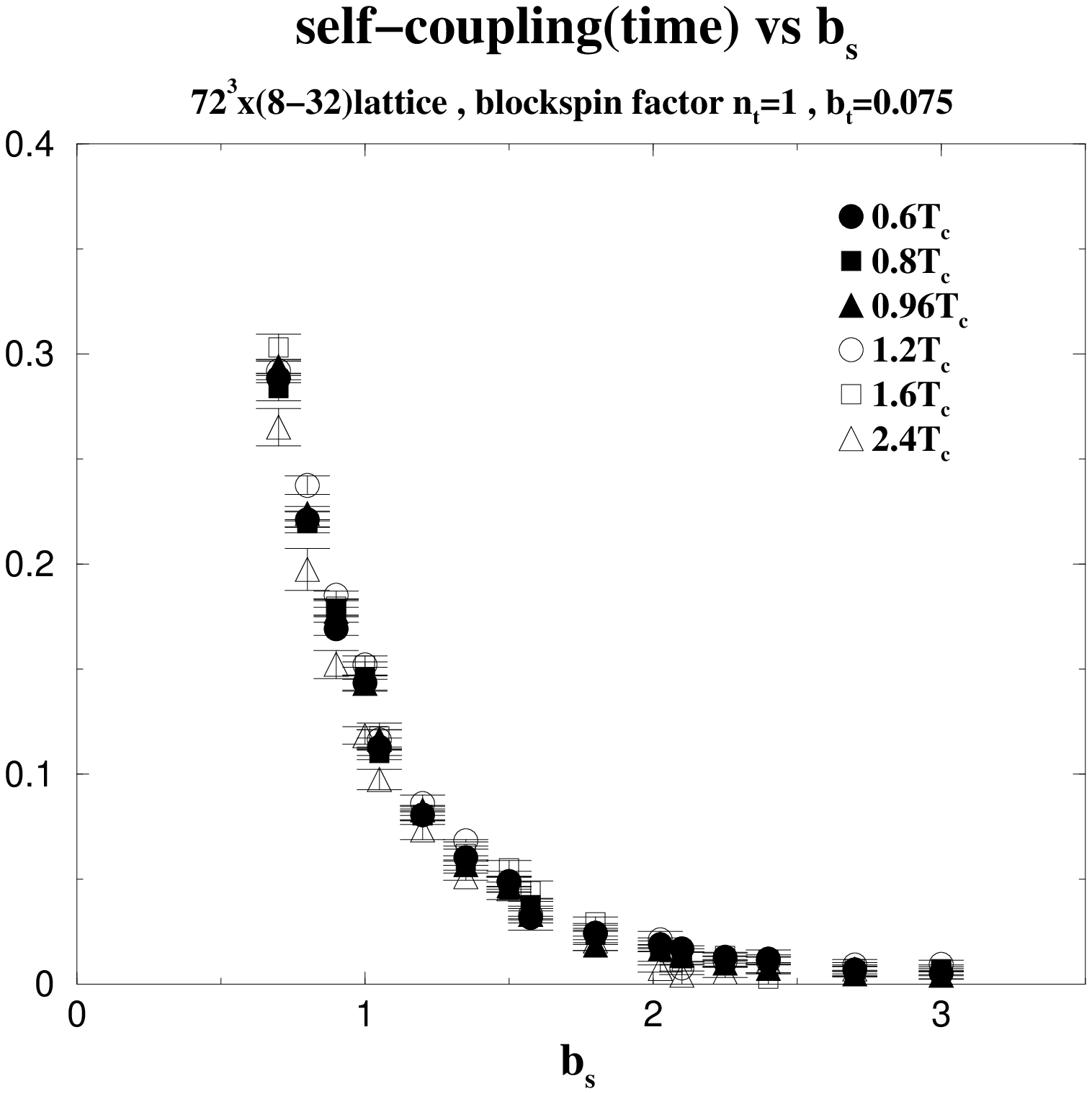, height=7cm, width=7cm}
\caption{Temperature and $b_s$ dependence of self-couplings 
for spacelike monopole (left) and timelike monopole (right).
}
\label{bsnt1f12}
}

\DOUBLETABLE{
\begin{tabular}{|c|c|c|c|c|}      \hline 
   $N_t$         & $\beta$ & $\gamma$ &  $a_s$  &  $a_t$   \\ \hline
    20           &  2.446  &  2.400   &  0.250  &  0.090   \\  
                 &  2.497  &  2.200   &  0.225  &  0.090   \\  
                 &  2.532  &  1.981   &  0.200  &  0.090   \\ 
                 &  2.564  &  1.750   &  0.175  &  0.090   \\ \hline 
   $N_t$         & $\beta$ & $\gamma$ &  $a_s$  &  $a_t$    \\ \hline
    16           &  2.462  &  1.942   &  0.250  &  0.113   \\  
                 &  2.490  &  1.767   &  0.225  &  0.113   \\  
                 &  2.519  &  1.607   &  0.200  &  0.113   \\  
                 &  2.552  &  1.450   &  0.175  &  0.113   \\ \hline 
\end{tabular} 
}
{
\begin{tabular}{|c|c|c|c|c|}      \hline 
   $N_t$         & $\beta$ & $\gamma$ &  $a_s$  &  $a_t$   \\ \hline
    12           &  2.465  &  2.178   &  0.250  &  0.100   \\  
                 &  2.496  &  1.985   &  0.225  &  0.100   \\  
                 &  2.525  &  1.781   &  0.200  &  0.100   \\
                 &  2.558  &  1.598   &  0.175  &  0.100   \\ \hline 
   $N_t$         & $\beta$ & $\gamma$ &  $a_s$  &  $a_t$    \\ \hline
     8           &  2.450  &  1.509   &  0.250  &  0.151   \\  
                 &  2.476  &  1.386   &  0.225  &  0.151   \\  
                 &  2.504  &  1.262   &  0.200  &  0.151   \\  
                 &  2.534  &  1.131   &  0.175  &  0.151   \\ \hline 
\end{tabular} 
}
{Parameters to see the $N_t$-dependence at $0.8T_c$.
\label{08tcNtbgaa}}
{Parameters to see the $N_t$-dependence at $1.2T_c$.
\label{12tcNtbgaa}}

\DOUBLEFIGURE
{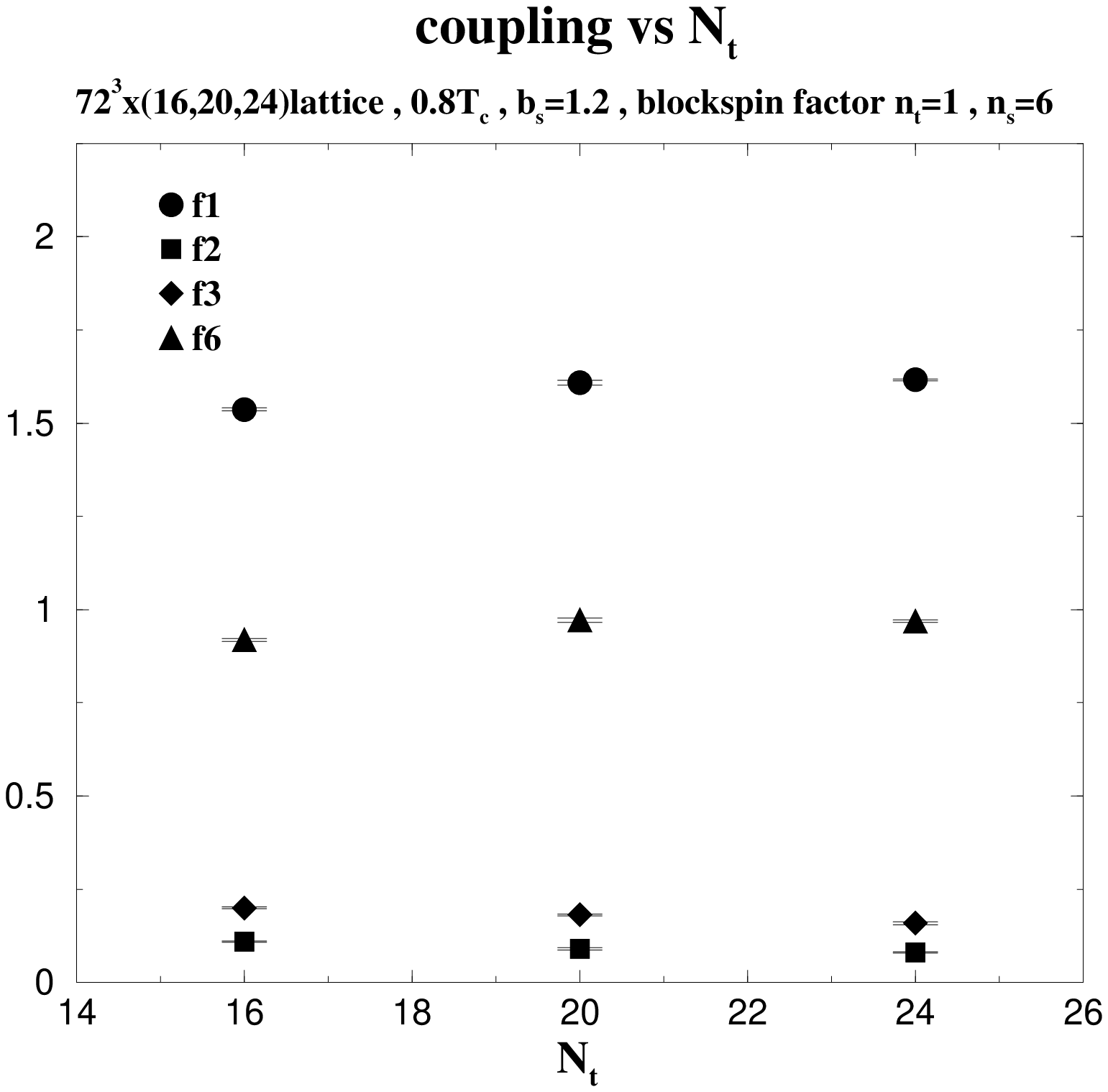, width=7cm, height=7cm}
{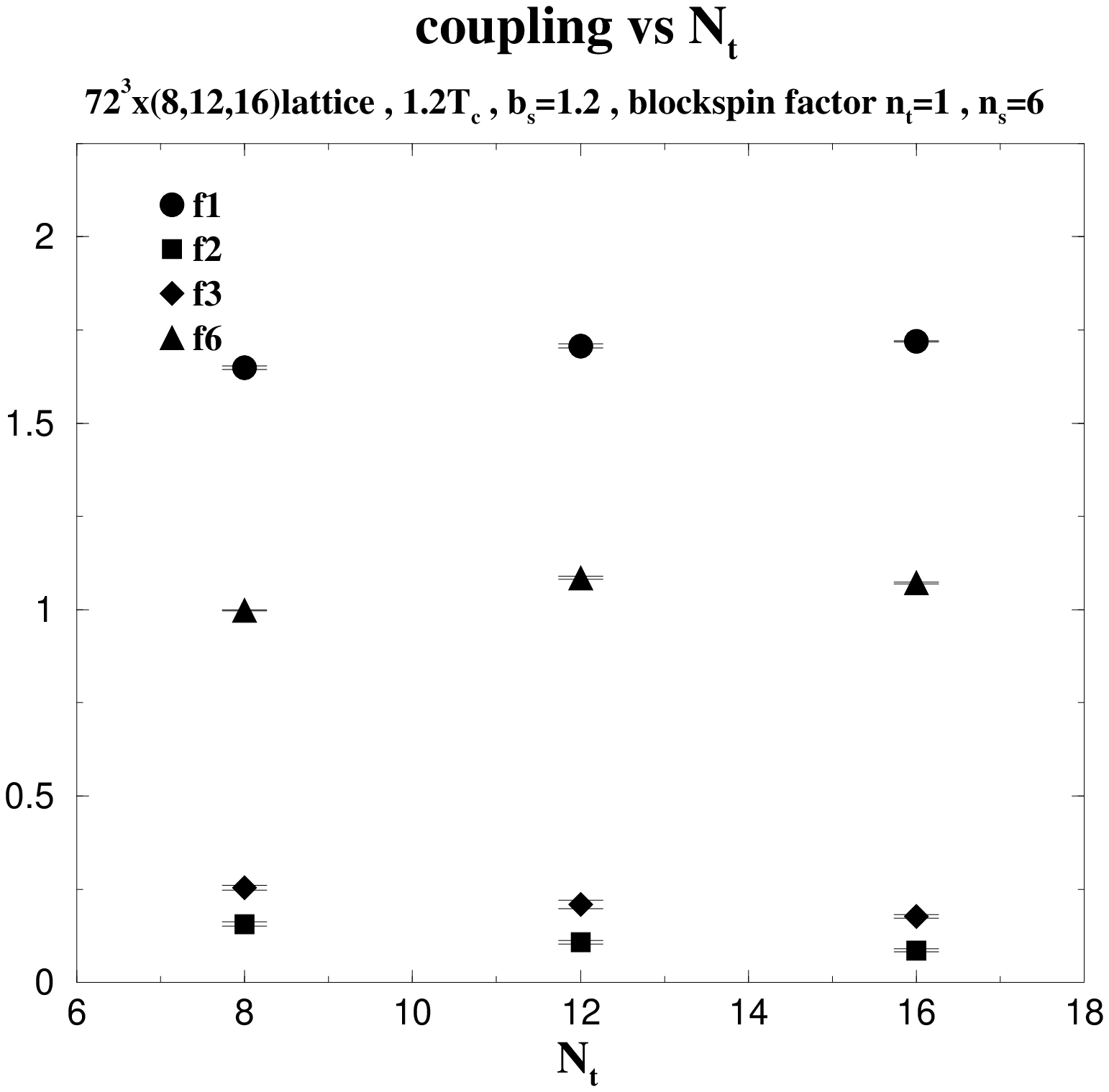, width=7cm, height=7cm}
{$N_t$-dependence for some couplings at 
$b_s=1.2$, $T=0.8Tc$.\label{08Ntnt1f1236b12}}
{$N_t$-dependence for some couplings at 
$b_s=1.2$, $T=1.2T_c$.\label{12Ntnt1f1236b12}}

\FIGURE{
\epsfig{file=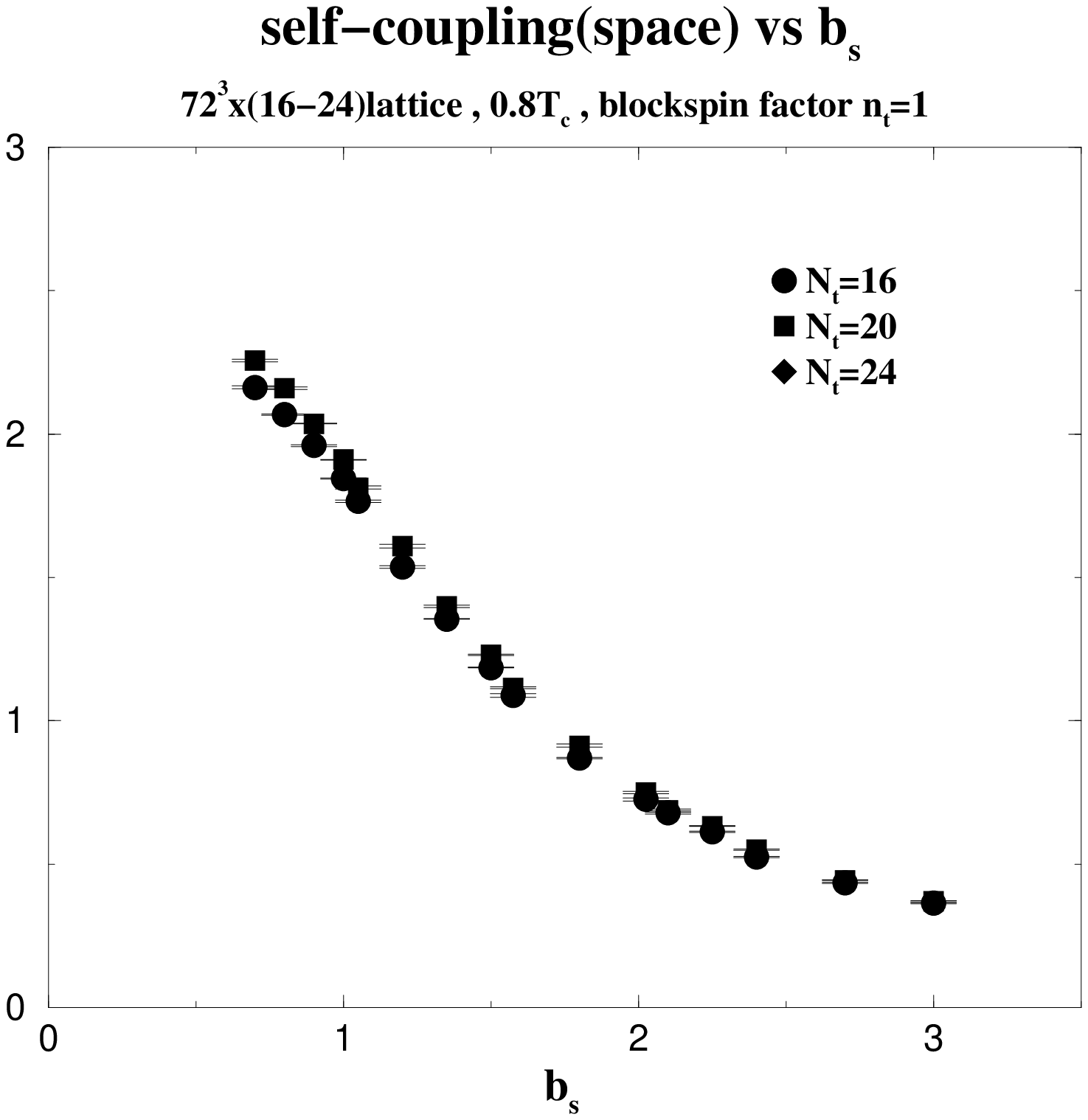, width=7cm, height=7cm}
\epsfig{file=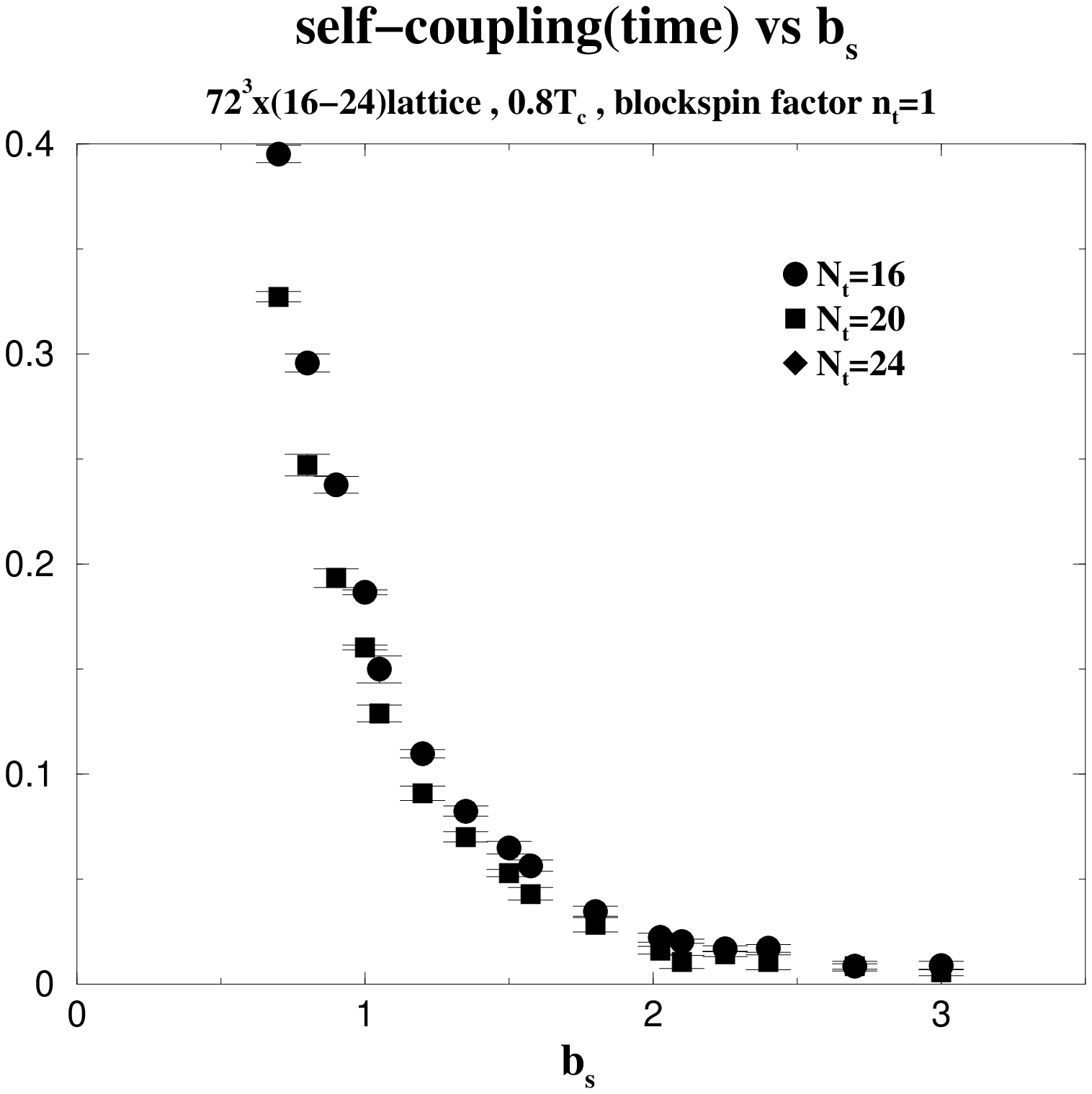, width=7cm, height=7cm}
\caption{$N_t$ and $b_s$ dependence of self-couplings for 
spacelike monopole (left) and timelike monopole (right) at $0.8Tc$.}
\label{08tcNtnt1f12}
}

\FIGURE{
\epsfig{file=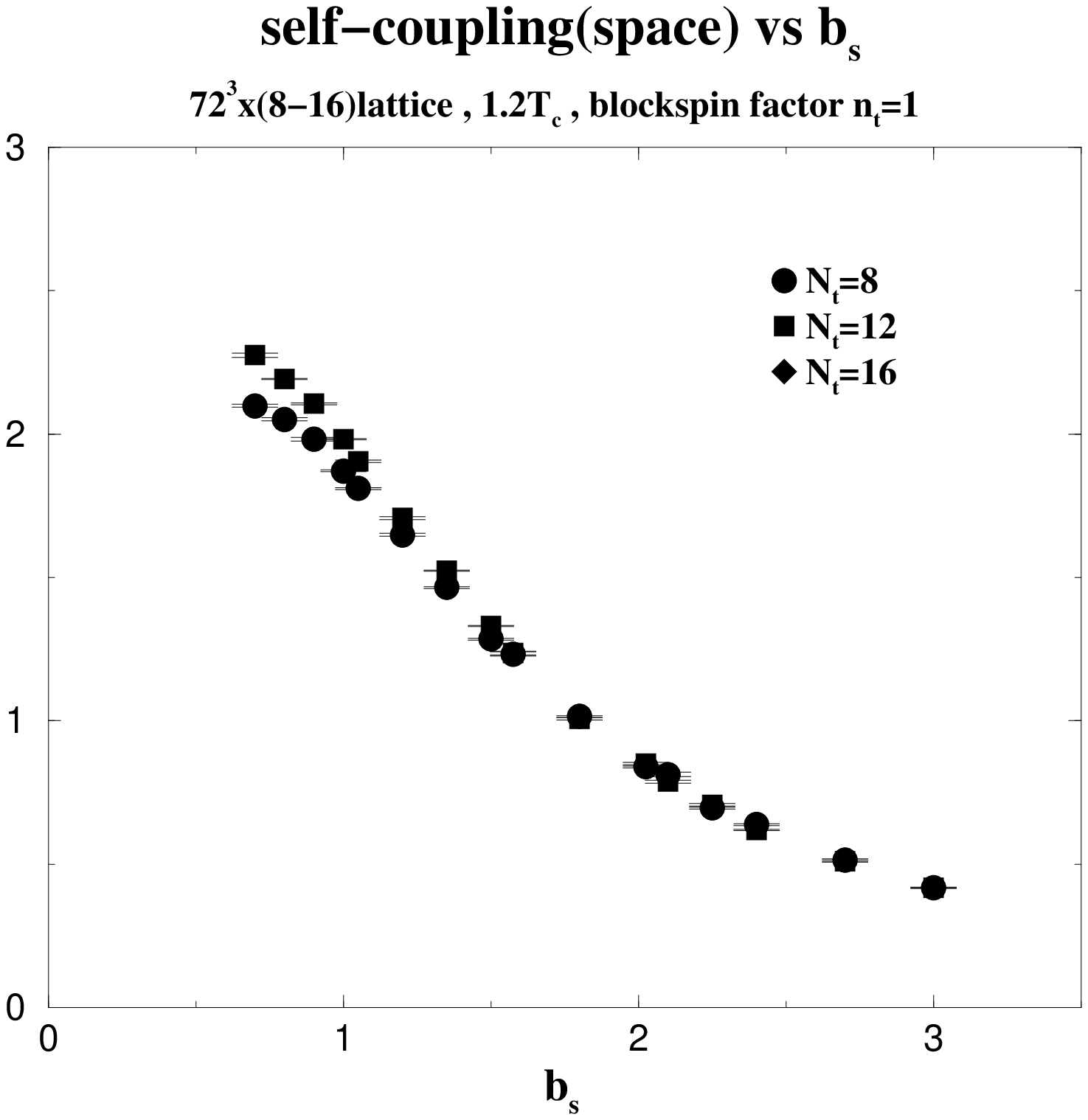, width=7cm, height=7cm}
\epsfig{file=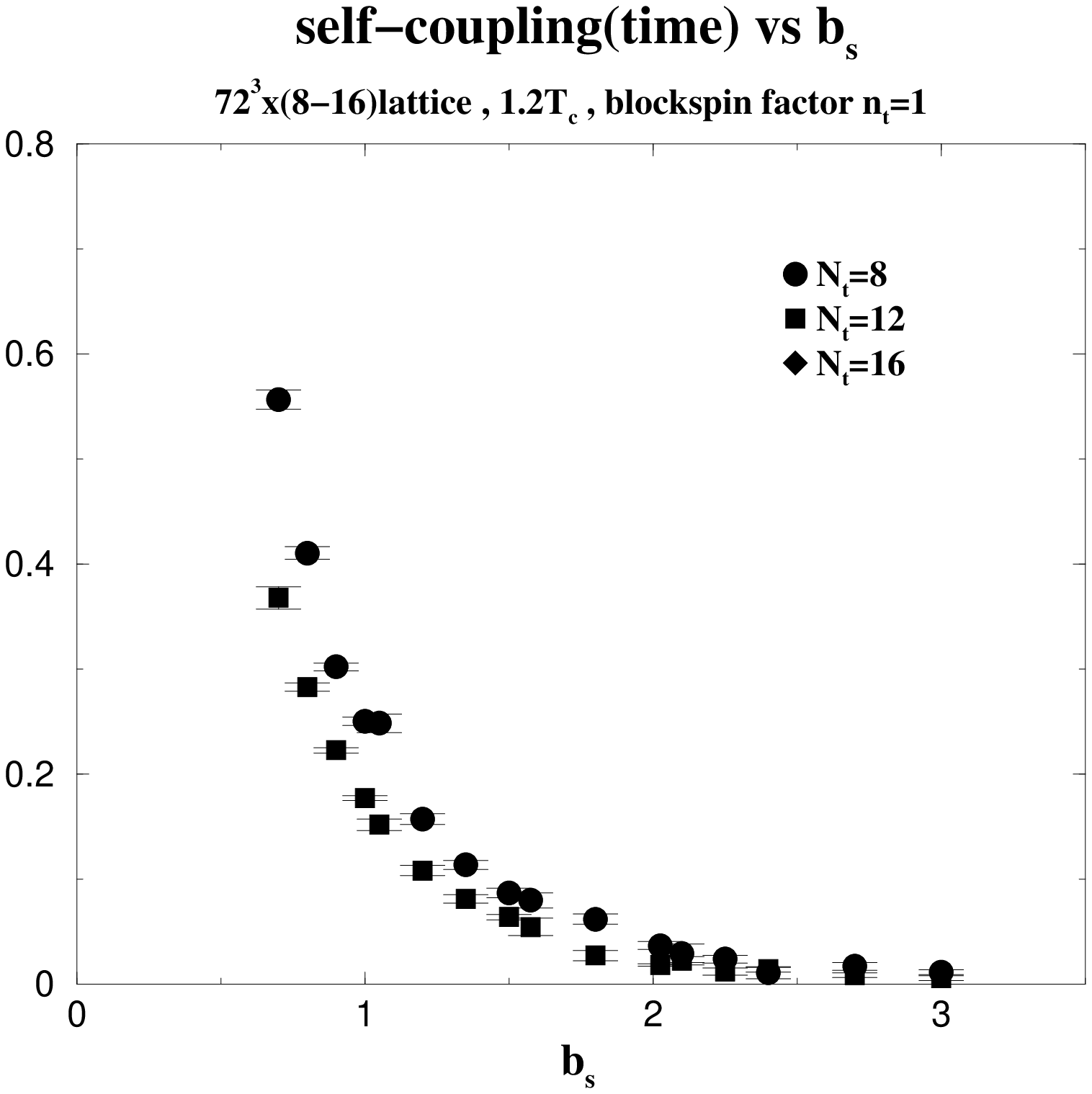, width=7cm, height=7cm}
\caption{$N_t$ and $b_s$ dependence of self-couplings for 
spacelike monopole (left) and timelike monopole (right) at $1.2Tc$.}
\label{12tcNtnt1f12}
}

Next let us discuss the continuum limit in the  time direction,  studying
$N_t$-dependence of the actions.
The parameters used in different $N_t$ are 
in Table~\ref{08tcNtbgaa} ($0.8T_c$) 
and in Table~\ref{12tcNtbgaa} ($1.2T_c$). 
Figures~\ref{08Ntnt1f1236b12} and \ref{12Ntnt1f1236b12} 
show $N_t$-independence of the actions 
for $N_t \ge 20$ (at $T=0.8T_c$) and 
$N_t \ge 12$ (at $T=1.2T_c$).
The data for all $b_s$ are plotted in Fig.~\ref{08tcNtnt1f12} ($0.8T_c$) and 
in Fig.~\ref{12tcNtnt1f12} ($1.2T_c$).
Because the temperatures are fixed,  this means $a_t$-independence also.

\FIGURE{
\epsfig{file=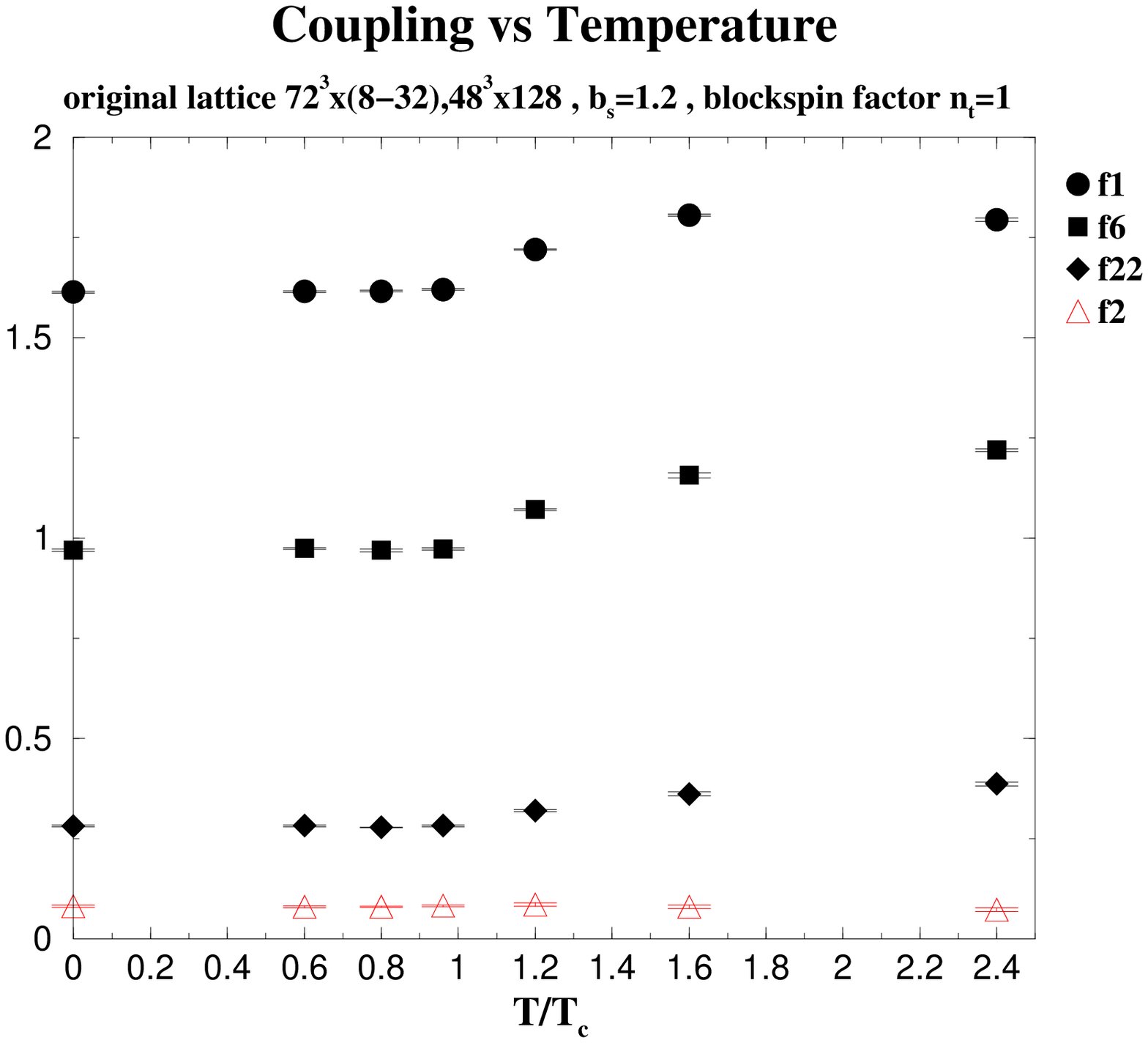, width=10cm, height=10cm}
\caption{Temperature dependence for 
f1, f6, f22 (space) and f2 (time) at $b_s = 1.2$}
\label{Tf1622bs12}
}

\FIGURE{
\epsfig{file=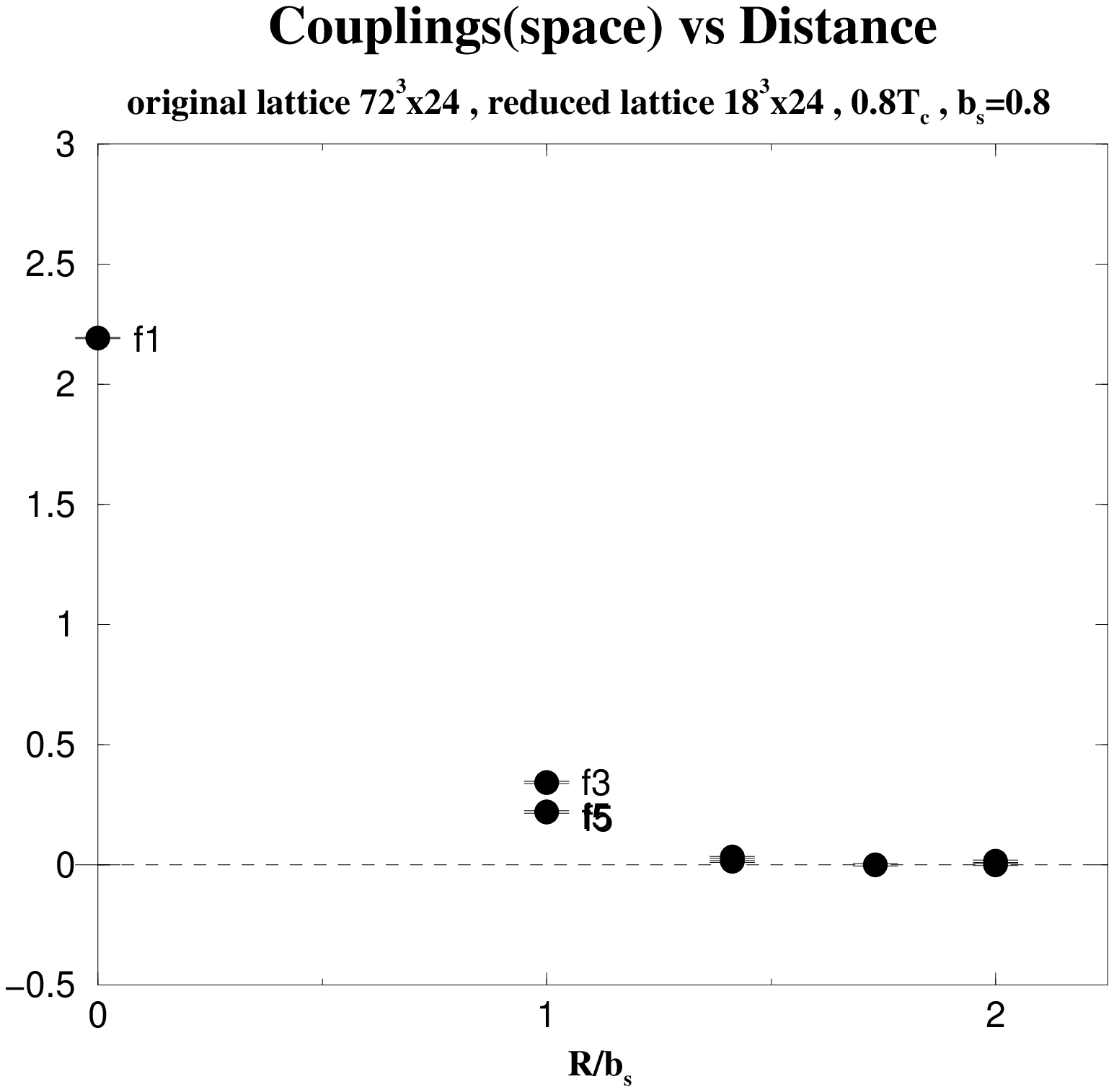, height=7cm, width=7.1cm}
\epsfig{file=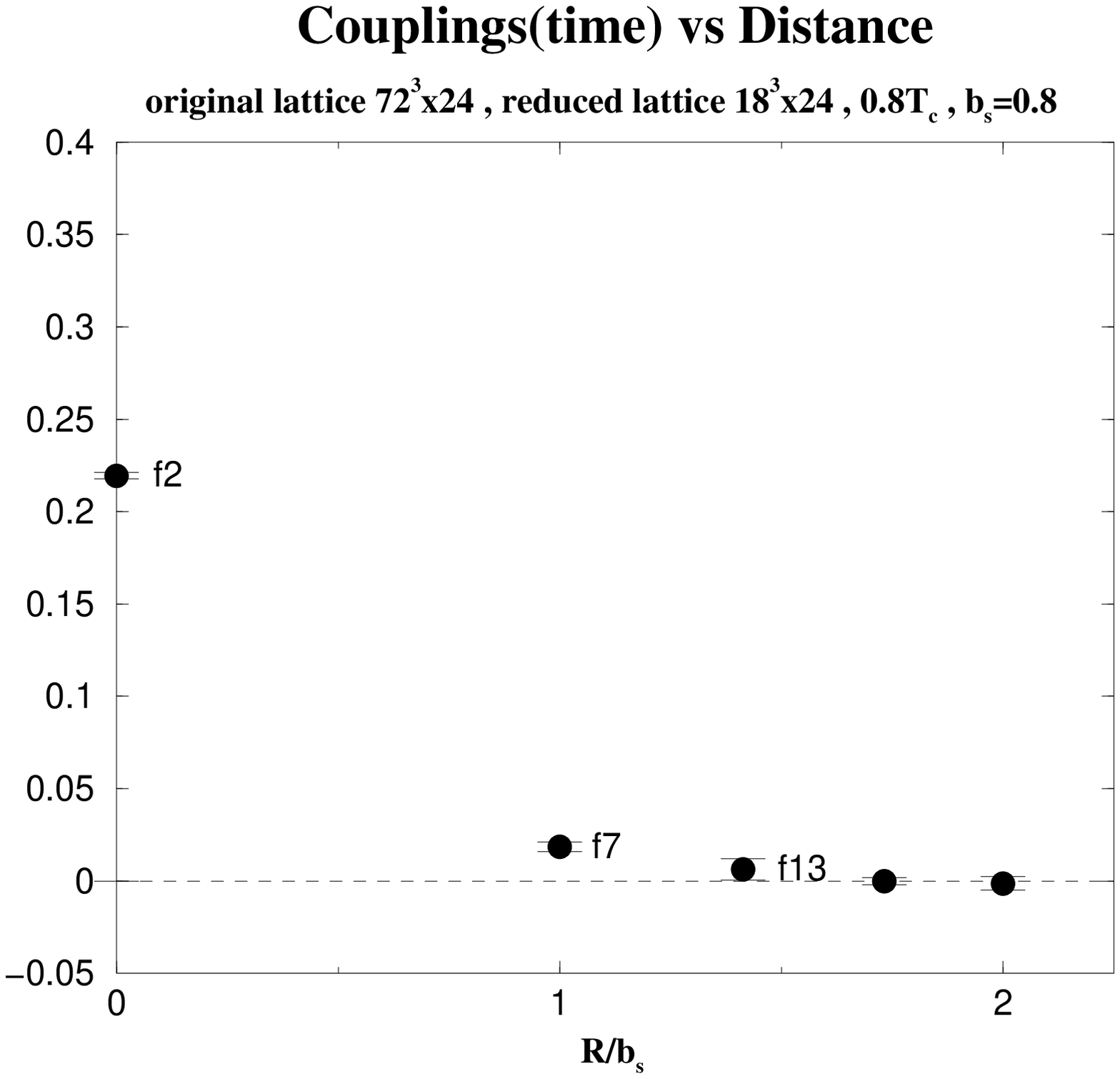, height=7cm, width=7.1cm}
\caption{Distance-dependence of the couplings apart in the space direction. 
Left is the spacelike monopole case and right is the timelike monopole
case at $0.8Tc$, $b_s=0.8$.}
\label{Rdepsdirection}
}

\FIGURE{
\epsfig{file=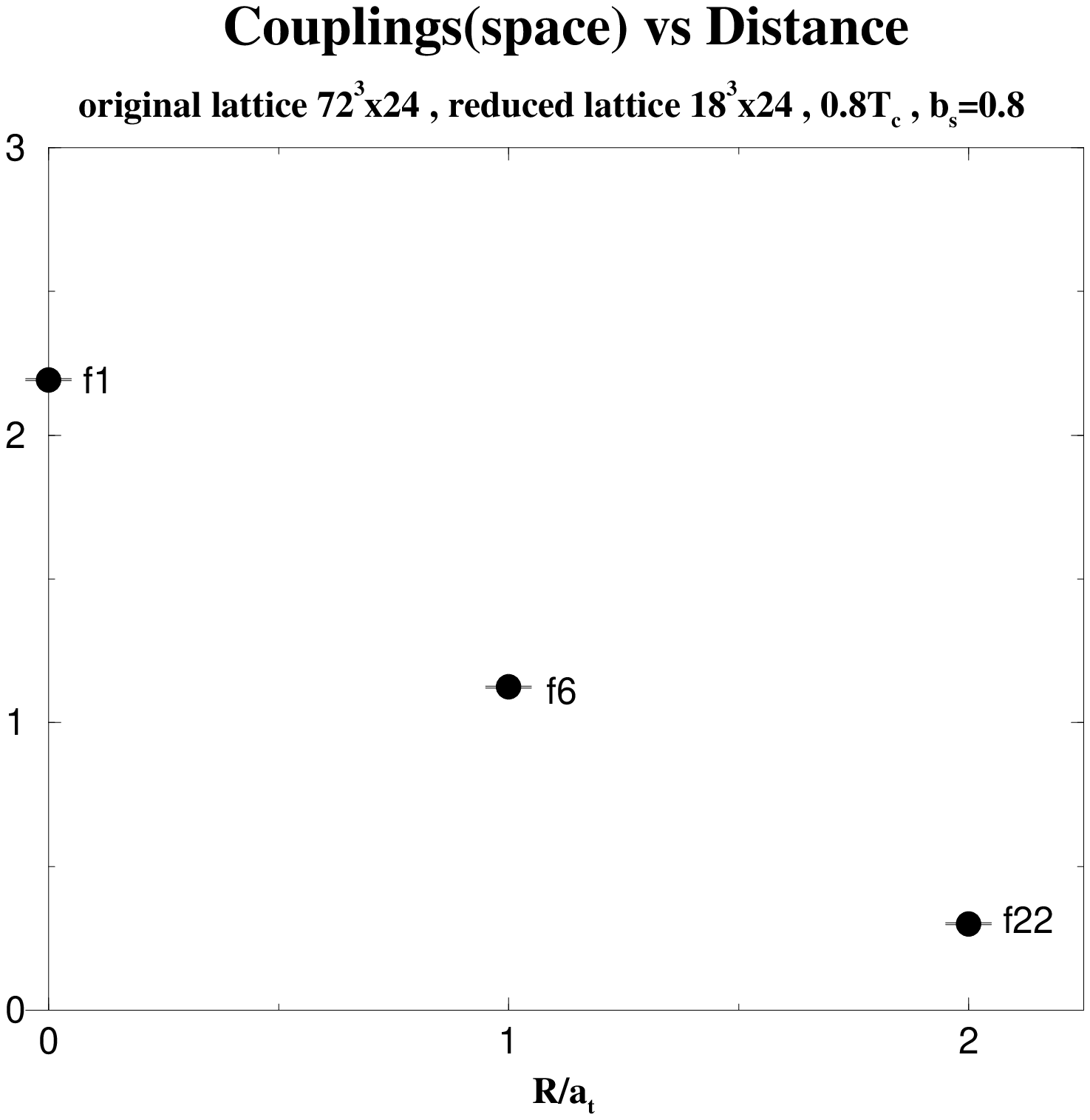, height=7cm, width=7cm}
\epsfig{file=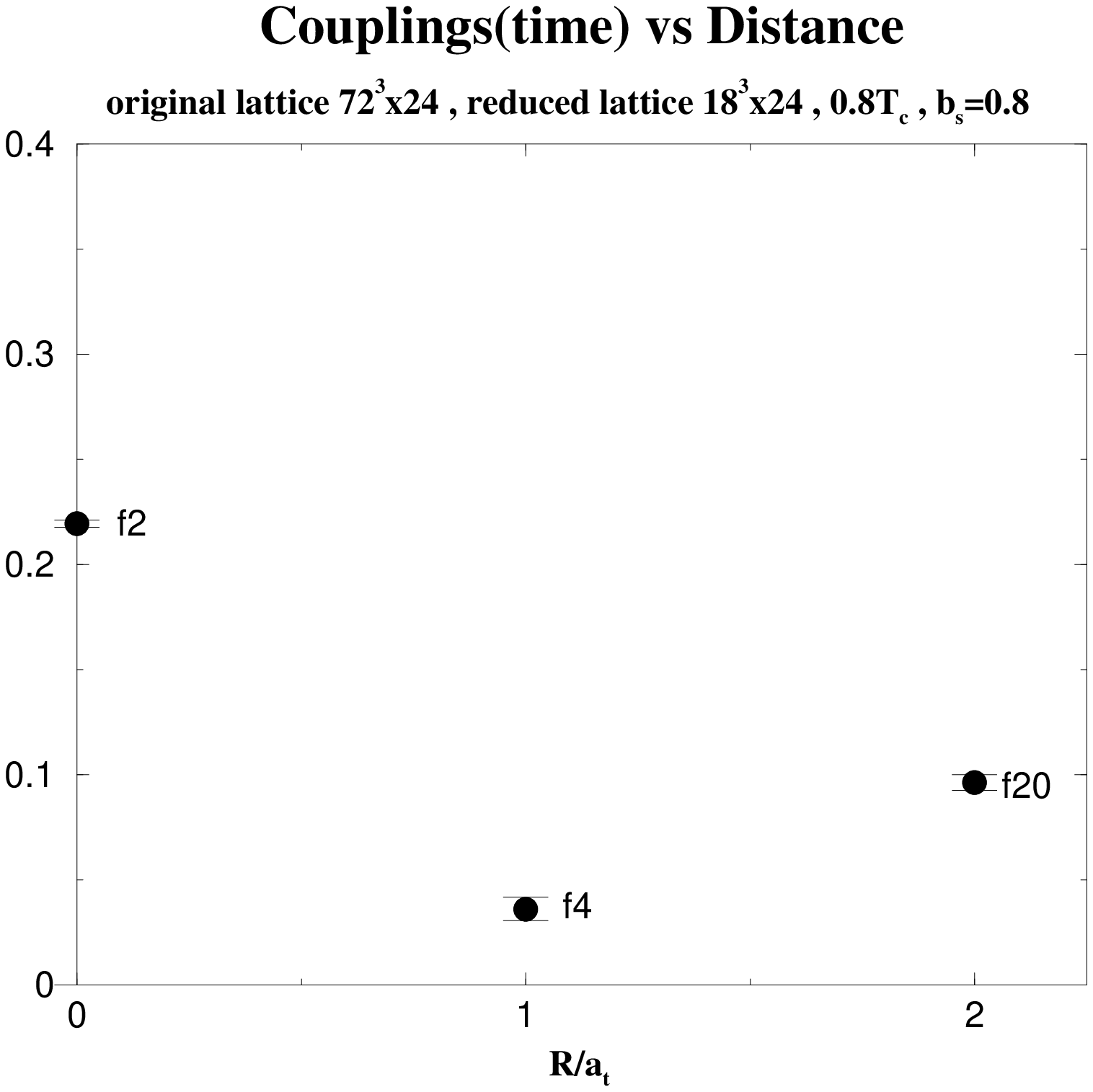, height=7cm, width=7cm}
\caption{Distance-dependence of the couplings apart in the time direction. 
Left is the spacelike monopole case and right is the timelike monopole
case at $0.8Tc$, $b_s=0.8$.}
\label{Rdeptdirection}
}

The features of the almost perfect 
monopole action at finite temperature are the
following:
(1)
Perpendicular interactions are 
 found to be negligible.
We can discuss spacelike and timelike monopole actions 
separately.
(2) Fig.~\ref{bsnt1f12} and Fig.~\ref{Tf1622bs12}  show that 
interactions of spacelike monopoles have no temperature-dependence 
in the confinement phase but have an obvious dependence in the deconfinement phase.
On the other hand,  interactions of timelike monopoles have no 
temperature-dependence in both phases. 
(3) We can examine the critical temperature $T_c$ of 
the confinement-deconfinement 
phase transition from the change of spacelike monopole interactions
(Fig.~\ref{Tf1622bs12}).
(4) The distance-dependence of the couplings is shown 
in Fig.~\ref{Rdepsdirection} and Fig.~\ref{Rdeptdirection}.
In both type of monopole actions,  
the self-coupling $f1$ (in the spacelike case) and $f2$ (in the timelike case)
 are dominant. The interactions between distant currents and perpendicular
 currents  are very small 
except $f_{20}$. 
The coupling $f_{20}$ may get any truncation error.
The couplings apart in the time direction (Fig.~\ref{Rdeptdirection}) 
are larger than the ones apart in the space direction (Fig.~\ref{Rdepsdirection}), 
because the lattice is anisotropic and the lattice distance in the 
 space direction ($b_s$) 
is larger than the one in the time direction ($a_t$). Moreover, 
the extended timelike monopole is defined on the $b_s^3$ cube,  whereas
the extended spacelike monopole is defined on the $b_s^2 a_t$ volume.
If we consider both monopoles using the same scale,  both couplings are of 
the same order~\cite{kitahara95}.

In the confinement phase,  the monopole currents form a long connected loop, 
 but there appear only  small loops in the deconfinement phase~\cite{kitahara95}.
It seems that the temperature-dependence of the spacelike 
monopoles corresponds to the change of monopole current configurations.
However,  we can not yet find a key explanation of 
the confinement-deconfinement 
mechanism due to the spacelike monopoles,  since the change of the spacelike 
monopole actions is not so drastic.

\clearpage

\section{Monopole Action At High Temperature}

\subsection{The Dimensional Reduction}

In this section we consider the effective monopole action 
beyond the critical  temperature 
and investigate the origin of the nonperturbative effect in the 
deconfinement phase.

\FIGURE{
\epsfig{file=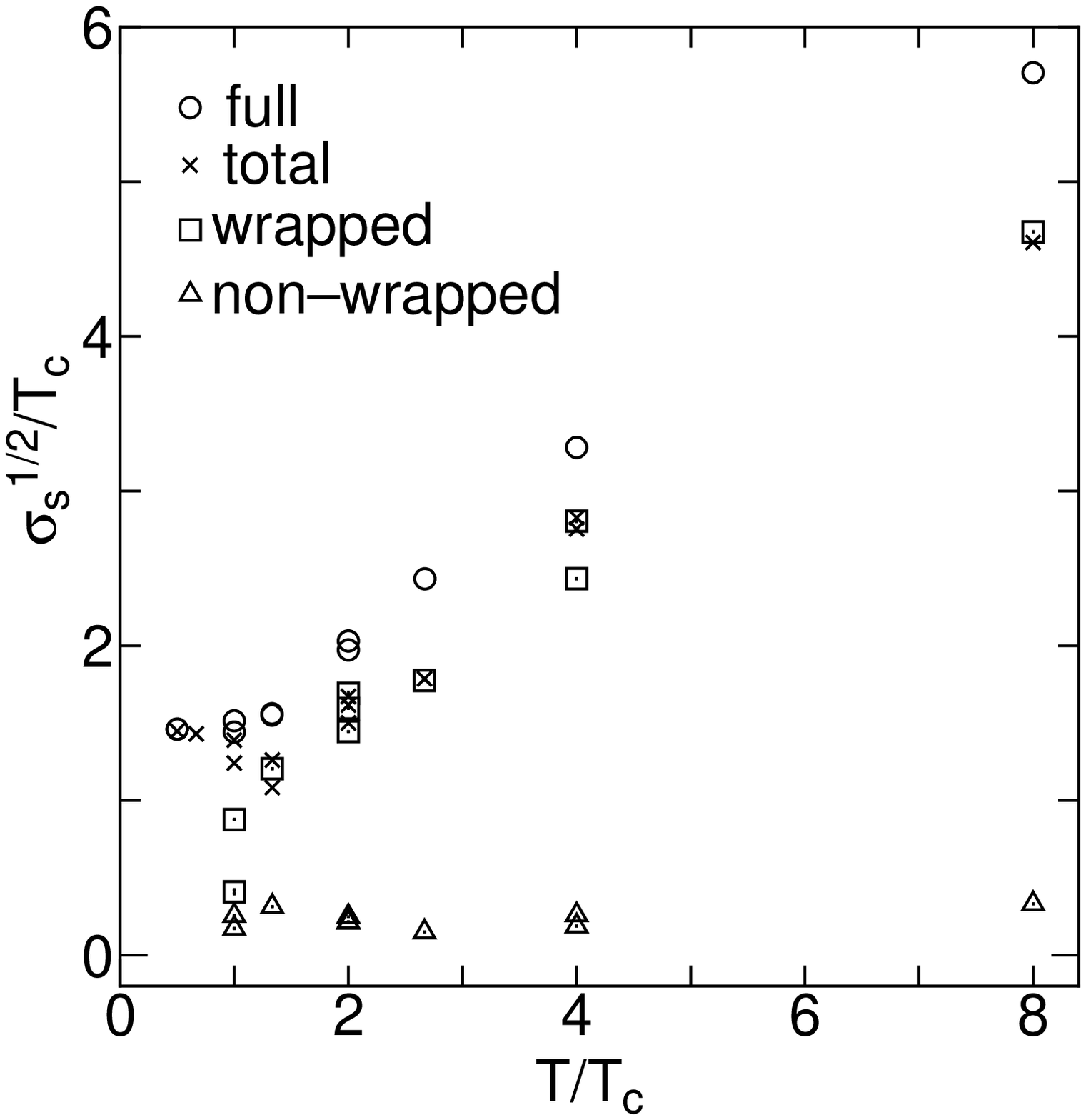, width=7cm, height=7cm}
\leavevmode
\caption{The full (non-abelian) spatial string tension (circle), 
the total monopole contribution (cross), the wrapped monopole contribution 
(square) and the non-wrapped monopole contribution (triangle).
This figure is taken from Ref.~\cite{ejiri96}. }
\label{spatialstringtension}
}

The relations between the monopoles and the spatial string tension in 
(SU(2))$_{4D}$ have been studied and the interesting 
features are observed~\cite{ejiri95,ejiri96}.
The data of  the spatial string tensions in Ref.~\cite{ejiri96} is 
shown in Fig.~\ref{spatialstringtension}. 
These data suggest that we can understand the nonperturbative effects 
in the deconfinement phase by the dynamics of the timelike monopoles.

To study the roles of the timelike mono\-poles,  we consider the dimensional 
reduction.  
4D timelike mono\-poles become instantons in $(GG)_{3D}$. 
It  has a classical solution 
with a magnetic charge --- \mbox{'t Hooft}-Polyakov 
monopole
 (instanton)  \cite{thooft74, polyakov74}. \\
Polyakov showed analytically that under the dilute Coulomb gas approximation 
of the \mbox{'t Hooft}-Polyakov instantons,  the string
tension  has 
a finite value~\cite{polyakov77}.
The validity of the approximation has been proved by numerical 
simulations in the London limit~\cite{yazawa01}.
The instantons in $(GG)_{3D}$ play a very important role for the nonperturbative effects like the string tension.
It is expected that the mechanism reproducing the 
spatial string tension in (SU(2))$_{4D}$ at high temperature 
is the same as that in $(GG)_{3D}$.

The starting point of the dimensional reduction is the action of (SU(2))$_{4D}$ at finite temperature.
\\
\be
S &=& \int_{0}^{\beta} dx \int d^3x 
    \frac{1}{4} F_{\mu\nu}^{a} F_{\mu\nu}^{a} ,  \label{eq:su2c}\\
F_{\mu\nu}^{a} &=& \partial_\mu A_{\nu}^{a} - \partial_\nu A_{\mu}^{a}
                 - g f_{abc} A_{\mu}^{b} A_{\nu}^{c} .
\ee
\\
At high temperature region after performing the dimensional reduction, 
the action (\ref{eq:su2c}) is described 
by $(GG)_{3D}$ with the following action~\cite{kajantie97} :
\\
\be
S_{eff} = \int d^3x \Bigl\{ 
                     \frac{1}{4} F_{ij}^{a} F_{ij}^{a}
                   + \frac{1}{2} \bigl( D_i A_0^a \bigr)^2
                   + \frac{1}{2} m_{D0}^{2} A_0^a A_0^a 
                   + \lambda_A \bigl( A_0^a A_0^a \bigr)^2
                    \Bigr\} . 
\label{eq:gg3dc}
\ee
\\
The 2-loop calculations give us the relations between the parameters 
appearing in (\ref{eq:gg3dc}) 
and those of the original action (\ref{eq:su2c})~\cite{kajantie97} :  
\\
\be
g_3^2 &=& g^2 (\mu) T \bigl[ 1 + \frac{g^2}{16 \pi^2} ( \frac{44}{3} 
          \frac{\mu}{\mu_T} + \frac{2}{3} ) \bigr] ,  \\
m_{D0}^{2} &=& \frac{2}{3} g^2 (\mu) T^2
               \bigl[ 1 + \frac{g^2}{16 \pi^2} ( \frac{44}{3} 
          \frac{\mu}{\mu_T} + \frac{10}{3} ) \bigr] ,  \\ 
\lambda_A &=& \frac{g^4 ( \mu ) T}{3 \pi^2}
               \bigl[ 1 + 2 \frac{g^2}{16 \pi^2} (\frac{44}{3} 
          \frac{\mu}{\mu_T} + \frac{7}{3}  ) \bigr] , 
\ee
\\
where $g^2 (\mu)$ is the 4D gauge coupling and $T$ is 
the temperature in (SU(2))$_{4D}$  
and $\mu_T \approx 7.0555T$.
For convenience,  we redefine the parameters~\cite{kajantie97} as 
\\
\be
g_{3}^{2} \ \  ,  \ \ x \equiv \frac{\lambda_A}{g_{3}^{2}} \ \  ,  \ \  
y \equiv \frac{m_{D0}^{2}}{g_{3}^{4}} .
\ee
\\
After the redefinition the dimensionful parameter is the 3D gauge coupling 
$g_{3}^{2}$ only.

\subsection{The 3-Dimensional SU(2) Georgi-Glashow Model On The Lattice}

The lattice action for $(GG)_{3D}$ is expressed as 
\\
\be
S_{(GG)_{3D}} &=& S_G + S_A ,  \\
S_G &=& \beta_3 \sum_{x, i>j} 
        \Bigl\{ 1 - \frac{1}{2} 
 Tr \bigl[ U_i(x) U_j(x+\hat{i}) U_{i}^{\dagger}(x+\hat{j}) U_{j}^{\dagger}(x)
    \bigr] \Bigr\} , 
\\
S_A &=& \sum_{x, i} 2
  \bigl[ Tr a A_{0}^{2}(x) 
       - Tr a A_0(x) U_i(x) A_{0}(x+\hat{i}) U_{i}^{\dagger}(x)
    \bigr]
\nonumber \\
     &+& \sum_x 
       \bigl[  (\tilde{m}_{D0}a)^2 Tr aA_{0}^{2}(x)  
               + a \lambda_A \bigl( \frac{1}{2} Tr a A_{0}^{2}(x) \bigr)^2
       \bigr] , 
\\
\beta_3 &=& \frac{4}{g_{3}^{2} a} ,  
\ee
\\
where $a$ is the lattice spacing and $\tilde{m}_{D0}$ is 
the bare mass in the lattice scheme.
In order to relate the results of lattice calculation in $(GG)_{3D}$ 
to the physics of the original (SU(2))$_{4D}$ at high temperature, 
it is necessary to consider the relation between the bare mass 
$\tilde{m}_{D0}$ and the renormalized mass in the continuum theory.
The bare mass 
$\tilde{m}_{D0}$ is rewritten in terms of $\beta$,  $x$ and $y$ as shown  
in Refs.~\cite{kajantie97, laine95} 
from the requirement that the renormalized mass in the lattice scheme 
is the same as the one in the $\overline{MS}$ scheme.  
The lattice action  is finally expressed as follows:
\\
\be
S_{(GG)_{3D}} &=& S_G + S_A ,  \\
S_G &=& \beta_3 \sum_{x, i>j} 
        \Bigl\{ 1 - \frac{1}{2} 
 Tr \bigl[ U_i(x) U_j(x+\hat{i}) U_{i}^{\dagger}(x+\hat{j}) U_{j}^{\dagger}(x)
    \bigr] \Bigr\} , 
\\
S_A &=& \beta_3 \sum_{x, i} \frac{1}{2}
 Tr \bigl[ \tilde{A_0}(x) U_i(x) \tilde{A_{0}}(x+\hat{i}) U_{i}^{\dagger}(x)
    \bigr]
\nonumber \\
     &+& \sum_x 
       \Bigl\{ - \beta_3 ( 3 + \frac{1}{2} h ) 
                         \frac{1}{2} Tr\tilde{A_{0}}^{2}(x)  
               + \beta_3 x \bigl( \frac{1}{2} Tr \tilde{A_{0}}^{2}(x) \bigr)^2
       \Bigr\} , 
\\
\beta_3 &=& \frac{4}{g_{3}^{2} a} ,  
\\
h &\equiv& \frac{16}{\beta_{3}^{2}} y 
                   - \frac{\Sigma ( 4 + 5x )}{\pi \beta_3}
\nonumber \\
   &-& \frac{1}{\pi^2 \beta_{3}^{2}} 
   \Bigl\{ (20x - 10x^2)(\ln \frac{3}{2} \beta_3 + 0.09) + 8.7 + 11.6x
   \Bigr\} , 
\ee
\\
where $\Sigma=3.1759114$ and $\tilde{A_0}$ is defined by 
$a A_{0}^{2} = \beta_3 \tilde{A_{0}}^{2} / 4$ .

To compare the effective monopole action of (SU(2))$_{4D}$ with 
that of $(GG)_{3D}$,  we should take the same scale in both theories.
A lattice spacing in $(GG)_{3D}$ is controlled by a parameter $\beta_3$ and 
is given in unit of $g_3^2$ as 
\\
\be
a = \frac{4}{g_{3}^{2} \beta_3} .
\ee
\\
The relation between the 3D gauge coupling $g_3$ and the 4D gauge coupling 
$g(T)$ which depends on temperature $T$ in the 1-loop calculation is 
\\
\be
g_{3}^{2} = g^2 (T) T .
\ee
\\
The 4D gauge coupling $g(T)$ have been determined from the temperature-dependence 
of the spatial string tension in (SU(2))$_{4D}$ 
in the 1-loop calculation~\cite{bali93}:
\\
\be
\sqrt{\sigma_s (T) } &=& (0.334 \pm 0.014 ) g^2 (T) T ,   \\
g^{-2} (T) &=& \frac{11}{12\pi^2} \ln \frac{T}{\Lambda_T} ,   \\
\Lambda_T &=& 0.050(10) T_c .
\ee
\\
The string tension of the dimensional reduced $(GG)_{3D}$ have been measured  
in Ref.~\cite{hart00} and 
the value is fitted well only in terms of the gauge coupling as  
$\sqrt{ \sigma_{(GG)_{3D}}} = 0.326(7) g_{3}^{2}$.
This means that $\sqrt{\sigma_s (T) }$ is almost the same as 
$\sqrt{ \sigma_{(GG)_{3D}}}$ numerically.
Using the 4D gauge coupling,  the lattice spacing is rewritten 
in unit of the temperature $T$ as 
\\
\be
a = \frac{4}{g_{3}^{2} \beta_3}
  = \frac{4}{g^{2}(T) T \beta_3} .
\ee
\\
We also use the relation between the critical temperature $T_c$ and 
the (zero temperature) 4D physical string tension $\sigma_{phys}$
~\cite{fingberg93} :
\\
\be
\frac{T_c}{\sqrt{\sigma_{phys}}} = 0.69 \pm 0.02 .
\ee
\\
Hence we can determine the lattice spacing $a$ in $(GG)_{3D}$ for each $T$  
in unit of the square root of the (zero temperature) 4D physical string tension.

\subsection{Results}

Based on the method in Ref.~\cite{nadkarni90}, 
we perform  Monte-Carlo simulations of 
$(GG)_{3D}$.
Before the comparison of  both actions, we measure the string tension.
To evaluate the contribution of the instantons to the string tension, 
we define the instantons in $(GG)_{3D}$.
The methods for the abelian projection and the decomposition of the U(1) 
plaquette variables are the same as 
in (SU(2))$_{4D}$~\cite{yazawa01}.
After the decomposition we can define an instanton as 
\\
\be
k(s) = - \frac{1}{2} \epsilon_{ijk} \partial_i n_{jk}(s) ,  (i, j, k=1, 2, 3) , 
\ee
\\
and the instanton part of the Wilson loop in 3D is expressed as 
\\
\be
W_{3d-m} = \exp \Bigl\{
                       2 \pi i \sum_{s, s^{\prime}} k(s)D(s-s^\prime)
                       \frac{1}{2}\epsilon_{ijk} \partial_i M_{jk}(s^\prime)
                \Bigr\} .
\ee
\\

\DOUBLETABLE{
\begin{tabular}{|c||c||c||c|} \hline
    $a$    & $\beta_3$  &   $x$   &   $h$     \\ \hline
   0.160   &  6.394     &  0.010  &  -0.658   \\ \hline
   0.170   &  6.018     &  0.010  &  -0.696   \\ \hline
   0.175   &  5.846     &  0.010  &  -0.714   \\ \hline
   0.180   &  5.683     &  0.010  &  -0.732   \\ \hline
   0.190   &  5.384     &  0.010  &  -0.769   \\ \hline
   0.200   &  5.115     &  0.010  &  -0.805   \\ \hline
   0.225   &  4.547     &  0.010  &  -0.892   \\ \hline
   0.250   &  4.092     &  0.010  &  -0.977   \\ \hline
\end{tabular}
}
{\begin{tabular}{|c||c||c||c|} \hline
    $a$    & $\beta_3$  &   $x$   &   $h$     \\ \hline
   0.160   &   5.428    &  0.094  &  -0.749   \\ \hline
   0.170   &   5.109    &  0.094  &  -0.790   \\ \hline
   0.175   &   4.963    &  0.094  &  -0.810   \\ \hline
   0.180   &   4.825    &  0.094  &  -0.830   \\ \hline
   0.190   &   4.571    &  0.094  &  -0.870   \\ \hline
   0.200   &   4.342    &  0.094  &  -0.909   \\ \hline
   0.225   &   3.860    &  0.094  &  -1.002   \\ \hline
   0.250   &   3.474    &  0.094  &  -1.091   \\ \hline
\end{tabular}
}
{The parameters in $(GG)_{3D}$ 
corresponding to the lattice spacing $a$ at $1.92T_c$ in 
(SU(2))$_{4D}$.
\label{3dparameter192tc}}
{The parameters in $(GG)_{3D}$ 
corresponding to the lattice spacing $a$ at $2.4T_c$ in 
(SU(2))$_{4D}$.
\label{3dparameter24tc}}

\DOUBLETABLE{
\begin{tabular}{|c||c||c||c|} \hline
    $a$    & $\beta_3$  &   $x$   &   $h$     \\ \hline
   0.160   &   3.200    &  0.079  &  -1.068   \\ \hline
   0.170   &   3.012    &  0.079  &  -1.113   \\ \hline
   0.175   &   2.926    &  0.079  &  -1.134   \\ \hline
   0.180   &   2.844    &  0.079  &  -1.154   \\ \hline
   0.190   &   2.695    &  0.079  &  -1.193   \\ \hline
   0.200   &   2.560    &  0.079  &  -1.230   \\ \hline
   0.225   &   2.276    &  0.079  &  -1.308   \\ \hline
   0.250   &   2.048    &  0.079  &  -1.370   \\ \hline
\end{tabular}
}
{
\begin{tabular}{|c||c||c|} \hline
      T       &   Lattice size  &   Lattice size     \\
              &   (4DSU(2))    &   ($(GG)_{3D}$)     \\ \hline
  $1.92T_c$   &  $48^3 \times 10$    &  $48^3$       \\ \hline
  $ 2.4T_c$   &  $48^3 \times  8$    &  $48^3$       \\ \hline
  $ 4.8T_c$   &  $48^3 \times  4$    &  $48^3$       \\ \hline
\end{tabular}
}
{The parameters in $(GG)_{3D}$ 
corresponding to the lattice spacing $a$ at $4.8T_c$ in 
(SU(2))$_{4D}$.
\label{3dparameter48tc}
}
{Temperature and Lattice size for (SU(2))$_{4D}$ and 
$(GG)_{3D}$
\label{3dlatticesize}
}

\DOUBLEFIGURE[t]
{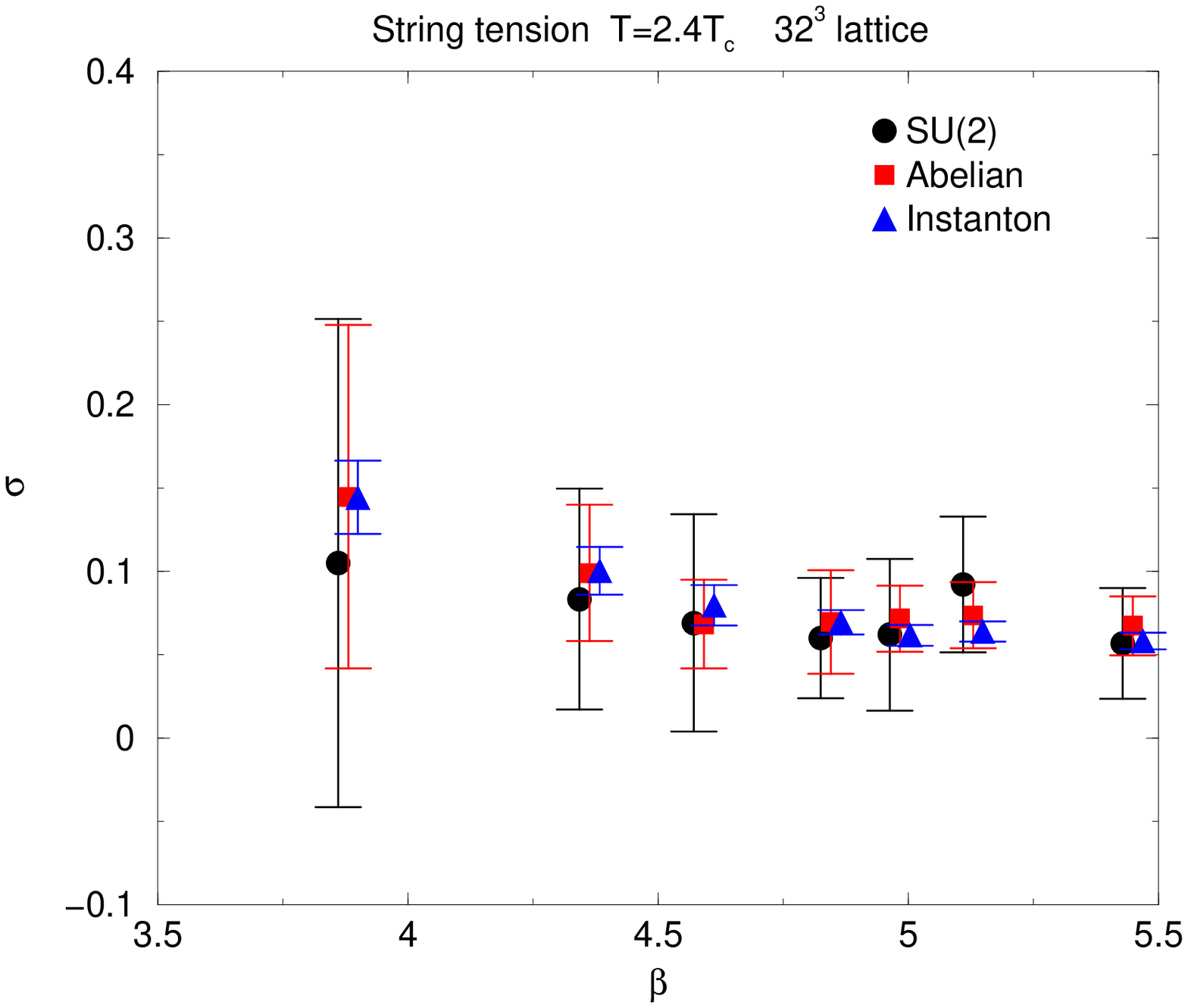, width=7cm, height=7cm}
{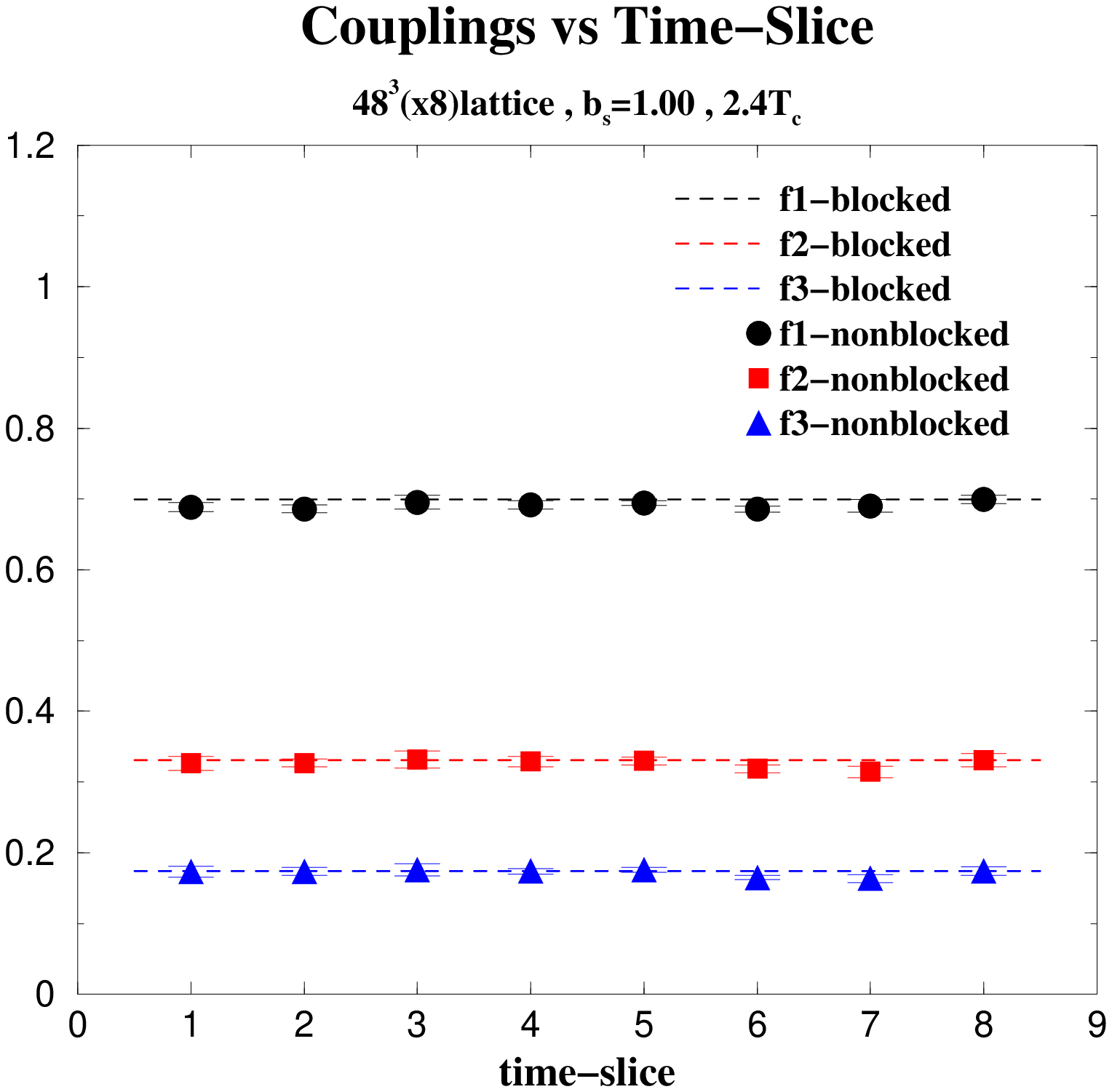, width=7cm, height=7cm}
{The string tension of $(GG)_{3D}$ at $2.4T_c$.\label{3dstringtension}
}
{Comparing the timelike monopole action at $2.4T_c$.\label{comparetimeaction}
}

The parameters used in the measurements of the string tension are determined by 
the above-mentioned procedure and are summarized 
in Tables~\ref{3dparameter192tc}--\ref{3dparameter48tc}.
The lattice sizes are summarized in Table~\ref{3dlatticesize}.
To get the string tensions we  fit 
the static potential (\ref{eq:potential}) with the function 
$\sigma R + \alpha \log R + c$ 
(where $\alpha$ and $c$ are constants).
The results in Fig.~\ref{3dstringtension} show that the abelian dominance and the 
instanton dominance for the string tension hold good.

Since the instanton dominance is observed, 
we try to derive  effective instanton actions in $(GG)_{3D}$ and 
compare those actions with the timelike monopole actions 
in (SU(2))$_{4D}$ in the deconfinement phase.
For the comparison,  we have to choose the time-slice in the 4D case.
\FIGURE{
\epsfig{file=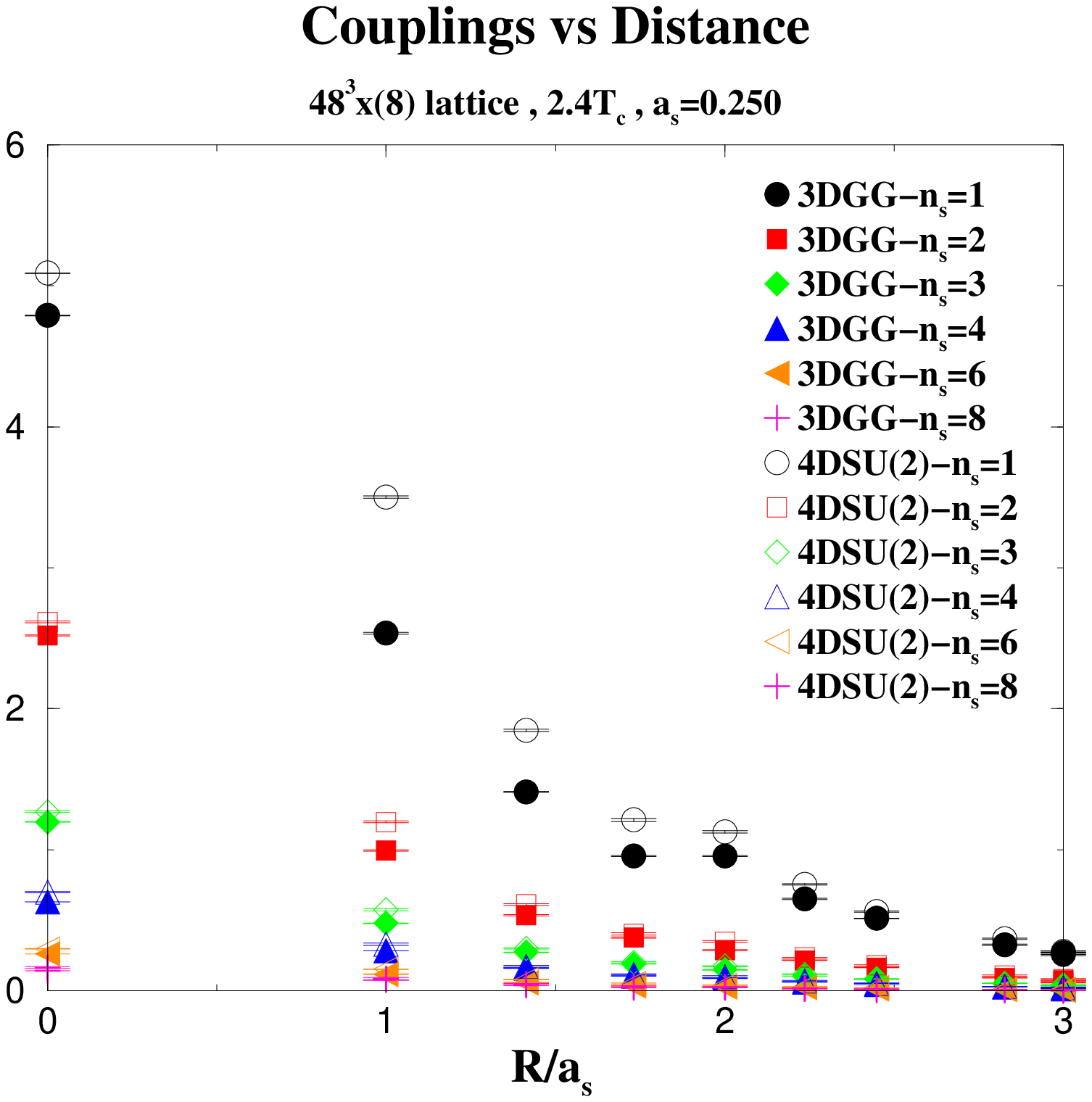, width=9cm, height=9cm}
\caption{The relation between the couplings and distance at $2.4T_c$.}
\label{distanceas25024tc}
}
However at high temperature the  
timelike monopoles are almost in wrapped monopole loops and the obtained 
actions at each time-slice are expected to be same. This is seen actually 
as shown in Fig.\ref{comparetimeaction}.
So in (SU(2))$_{4D}$ we may use the timelike monopoles 
after blockspin transformations completely in the time direction.
Here to perform the blockspin transformation means an averaging 
of the timelike monopoles at each time-slice.

Because there is no conservation law in the instanton case,  we use the original Swendsen's method~\cite{swendsen84} 
to determine instanton actions (see Appendix \ref{origswed}).
We assume that the instanton actions have 2-point interactions only and 
adopt 10 interactions within the distance 3 in unit of lattice spacing.

\clearpage

\FIGURE{
\epsfig{file=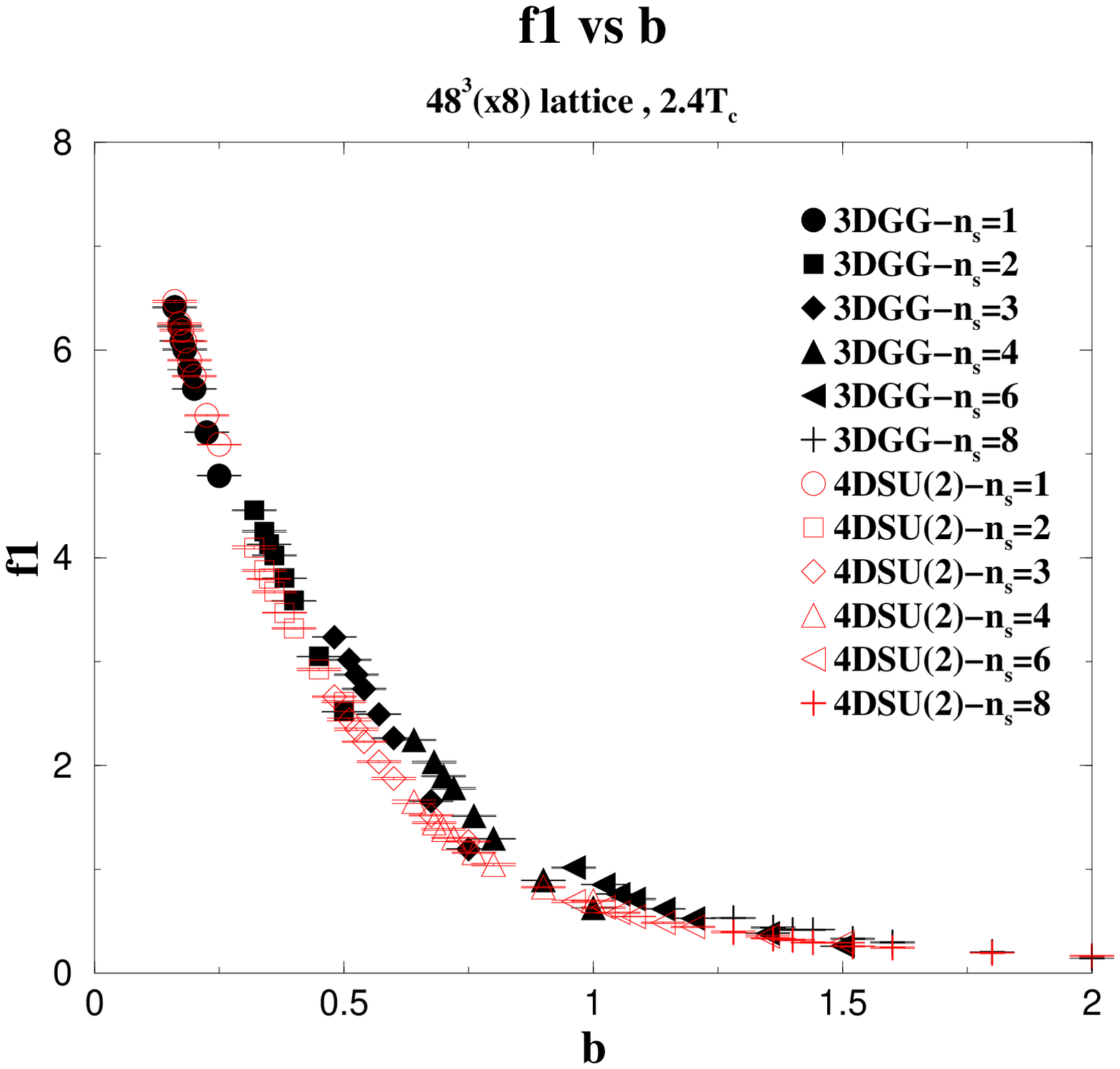, width=7cm, height=7cm}
\epsfig{file=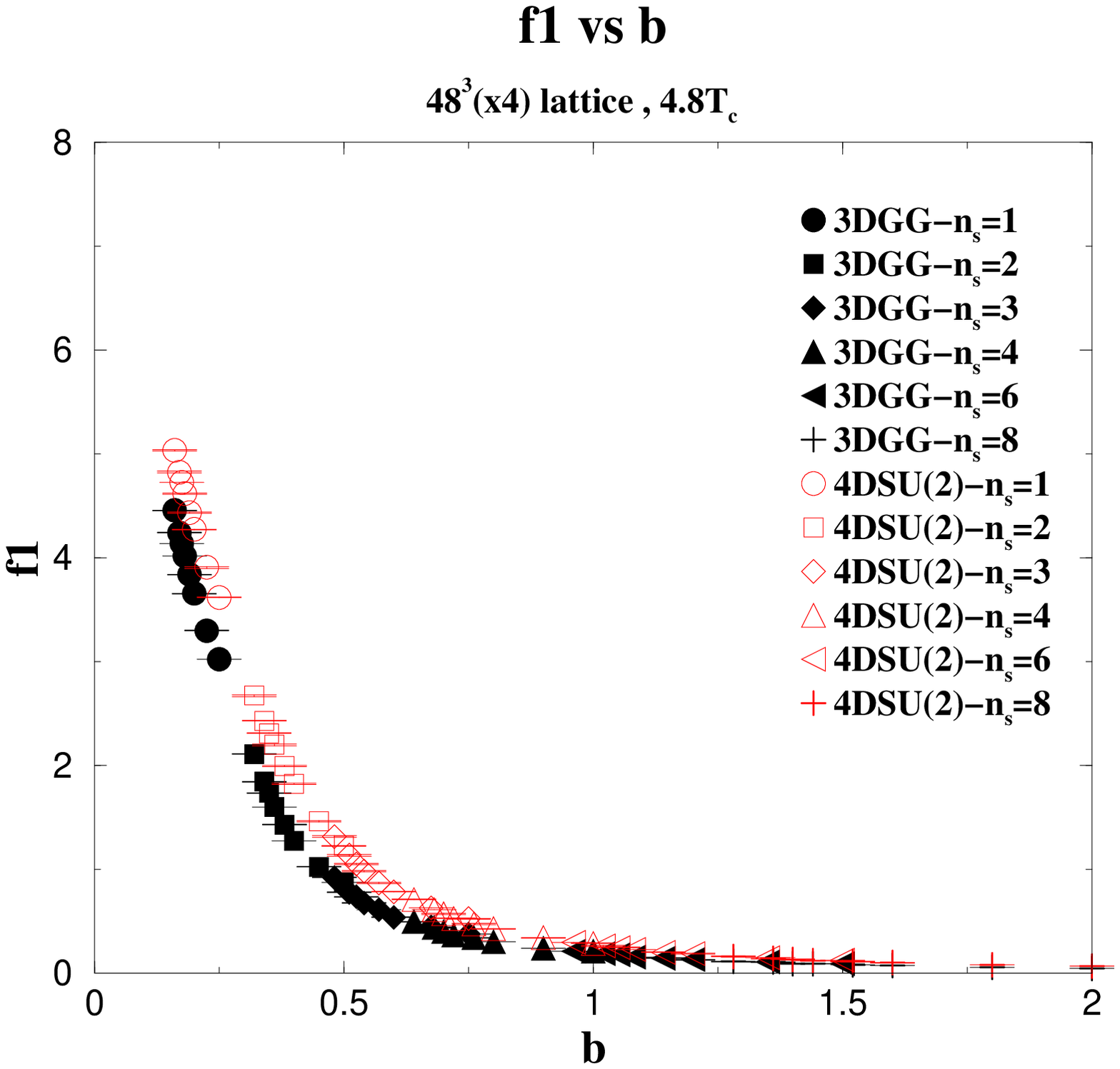, width=7cm, height=7cm}
\epsfig{file=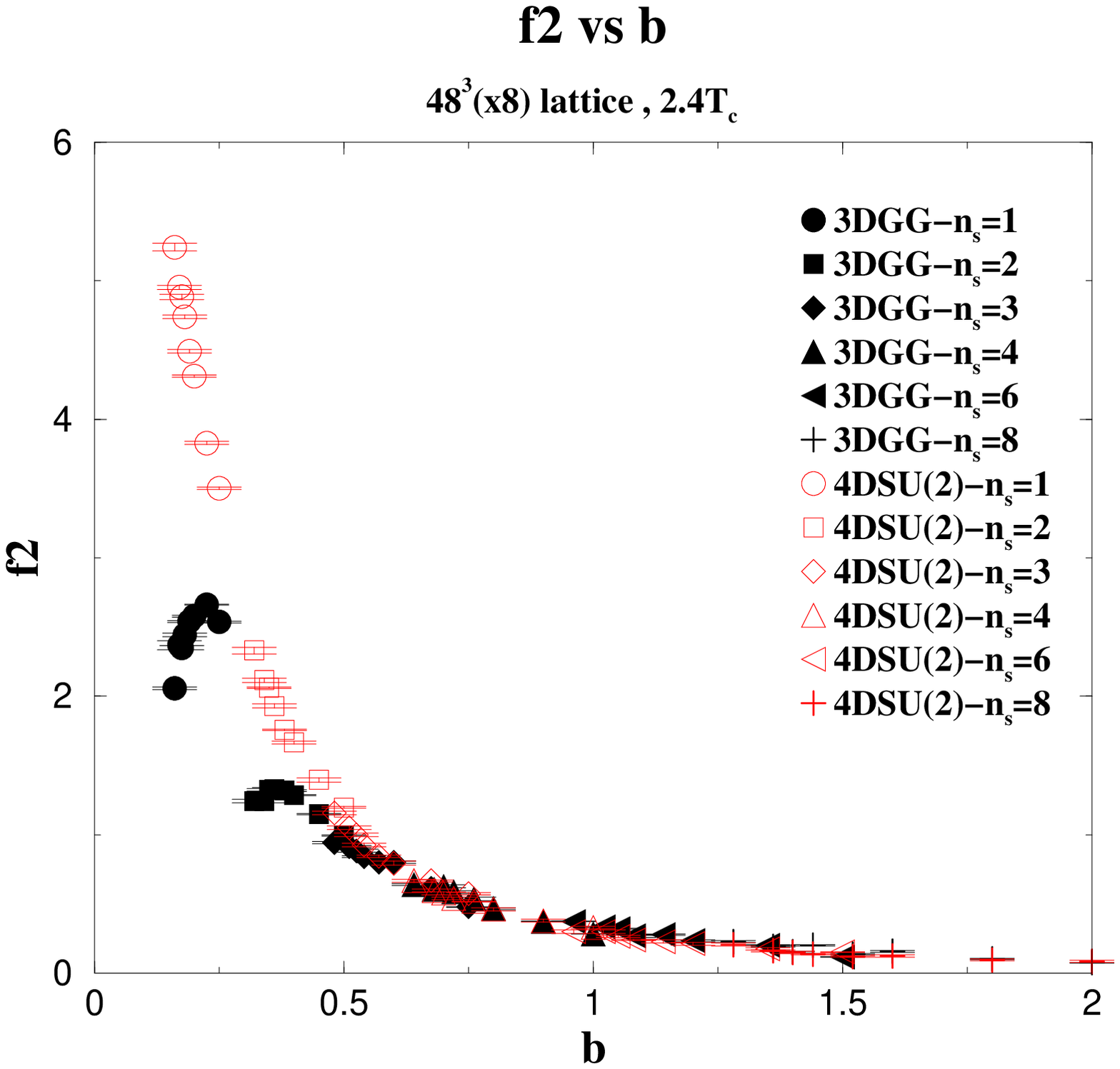, width=7cm, height=7cm}
\epsfig{file=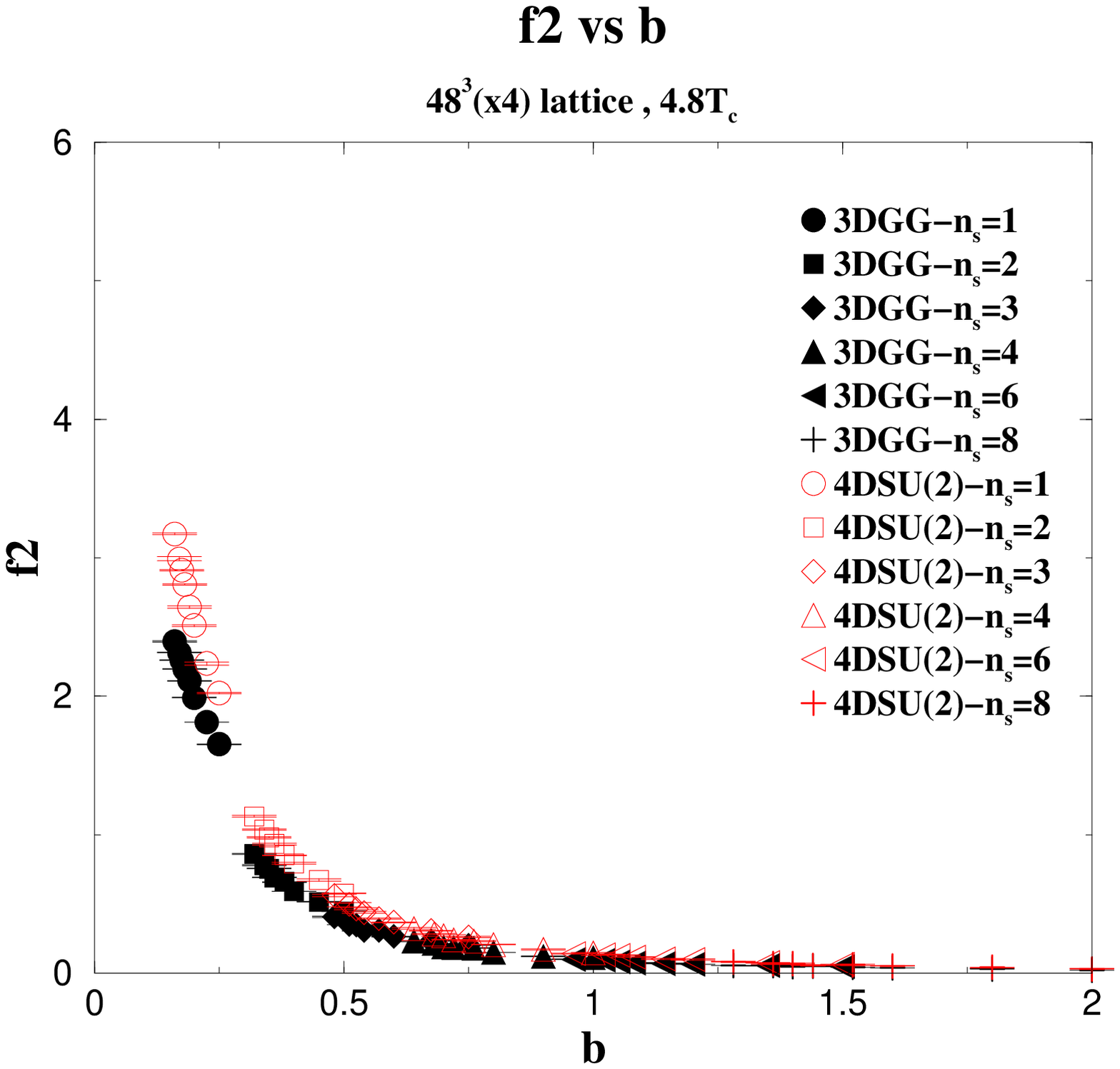, width=7cm, height=7cm}
\epsfig{file=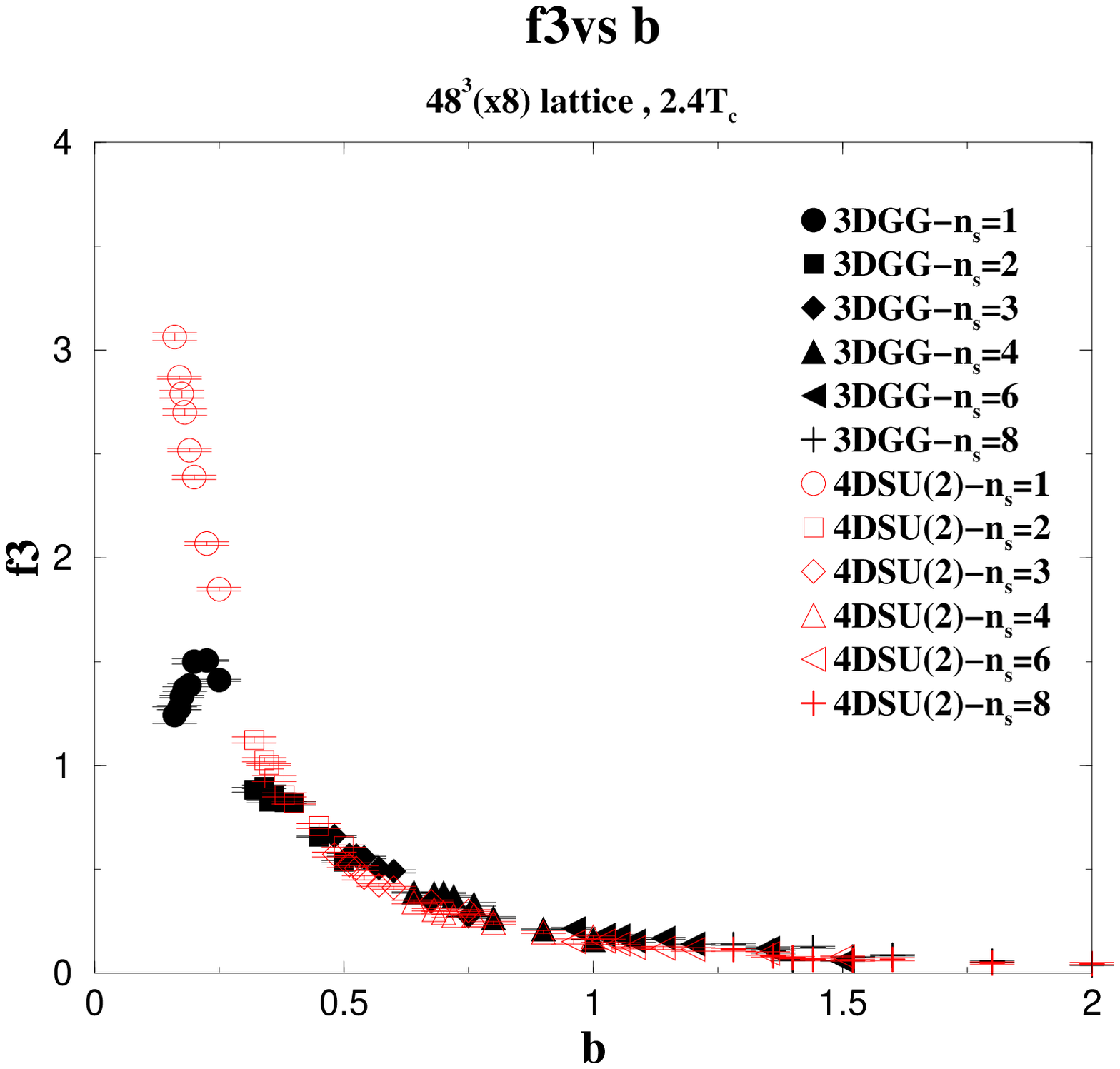, width=7cm, height=7cm}
\epsfig{file=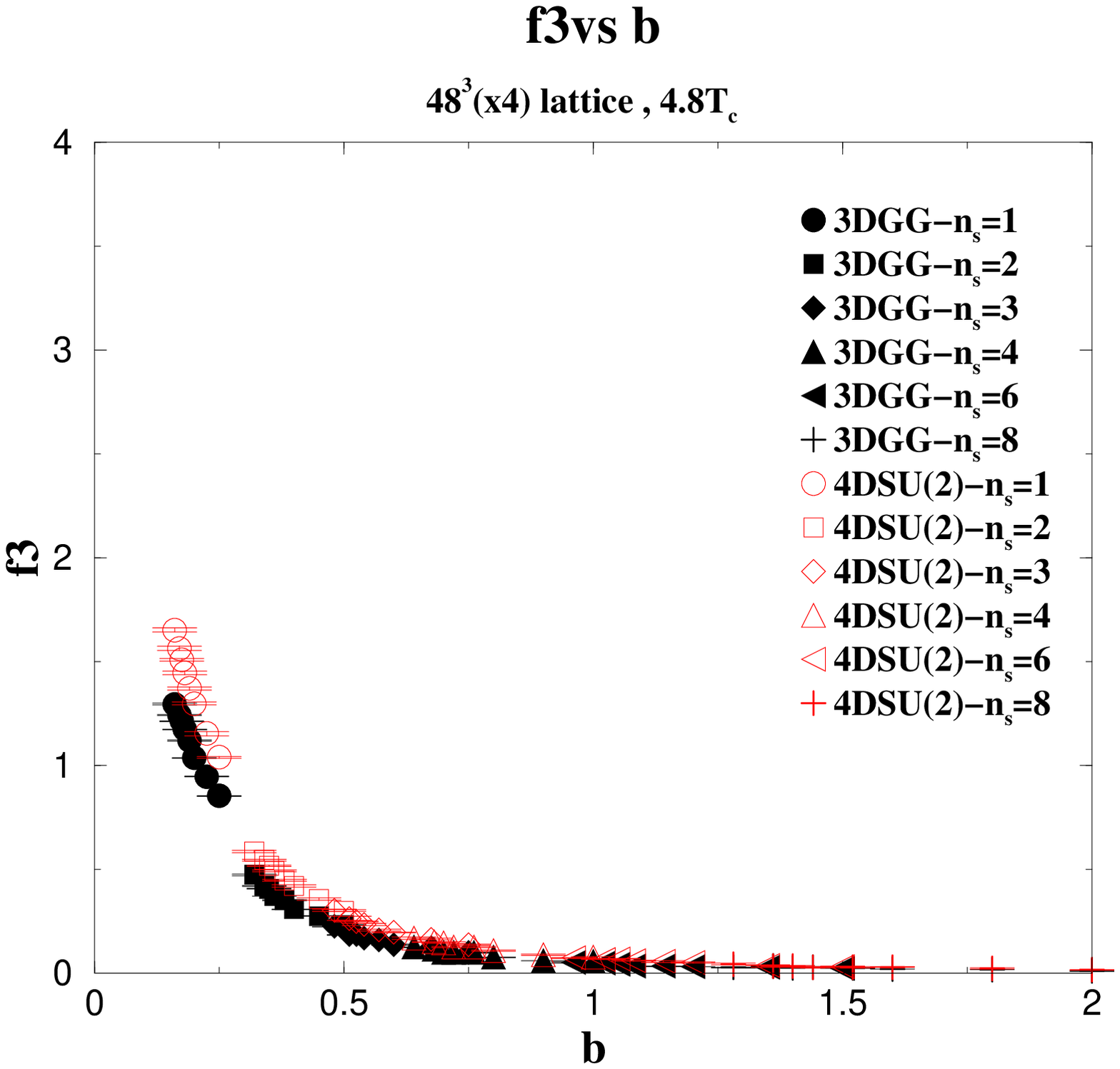, width=7cm, height=7cm}
\caption{Coupling $f_1$(top), $f_2$(middle) and $f_3$(bottom)
 at $2.4T_c$(left column) and $4.8T_c$(right column).}
\label{f123b19224tc}
}

\clearpage

\FIGURE{
\epsfig{file=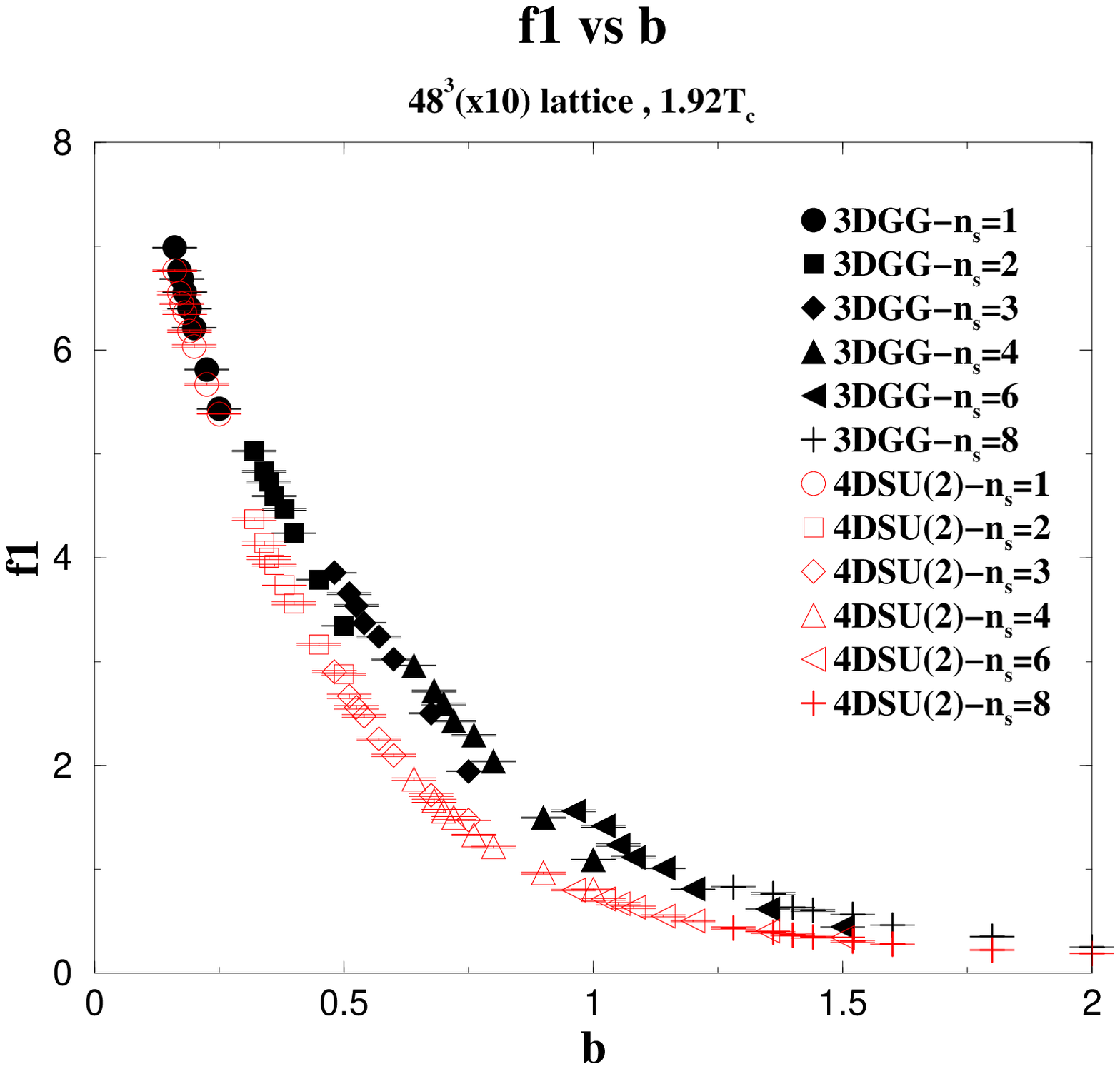, width=7cm, height=7cm}
\epsfig{file=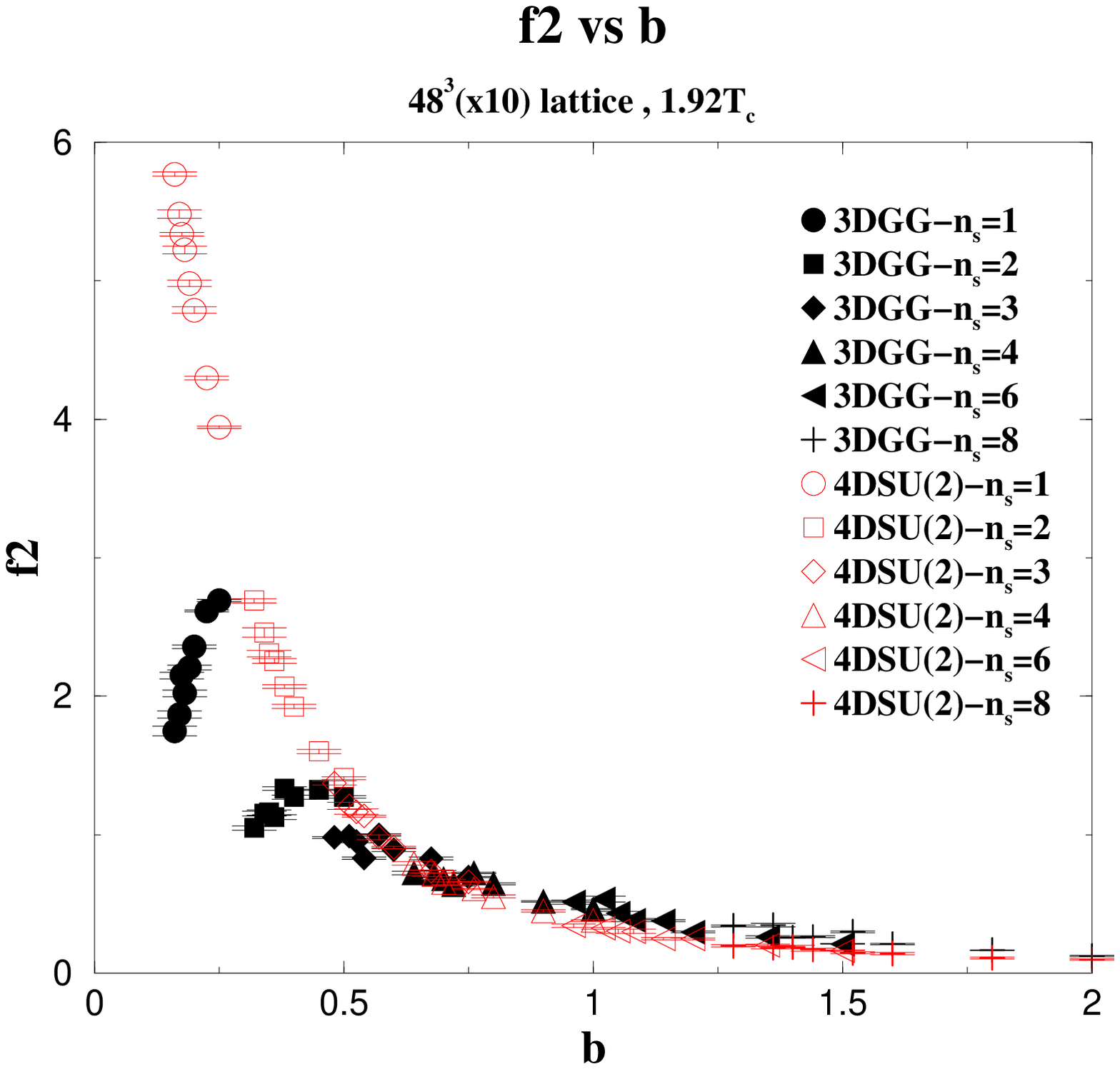, width=7cm, height=7cm}
\epsfig{file=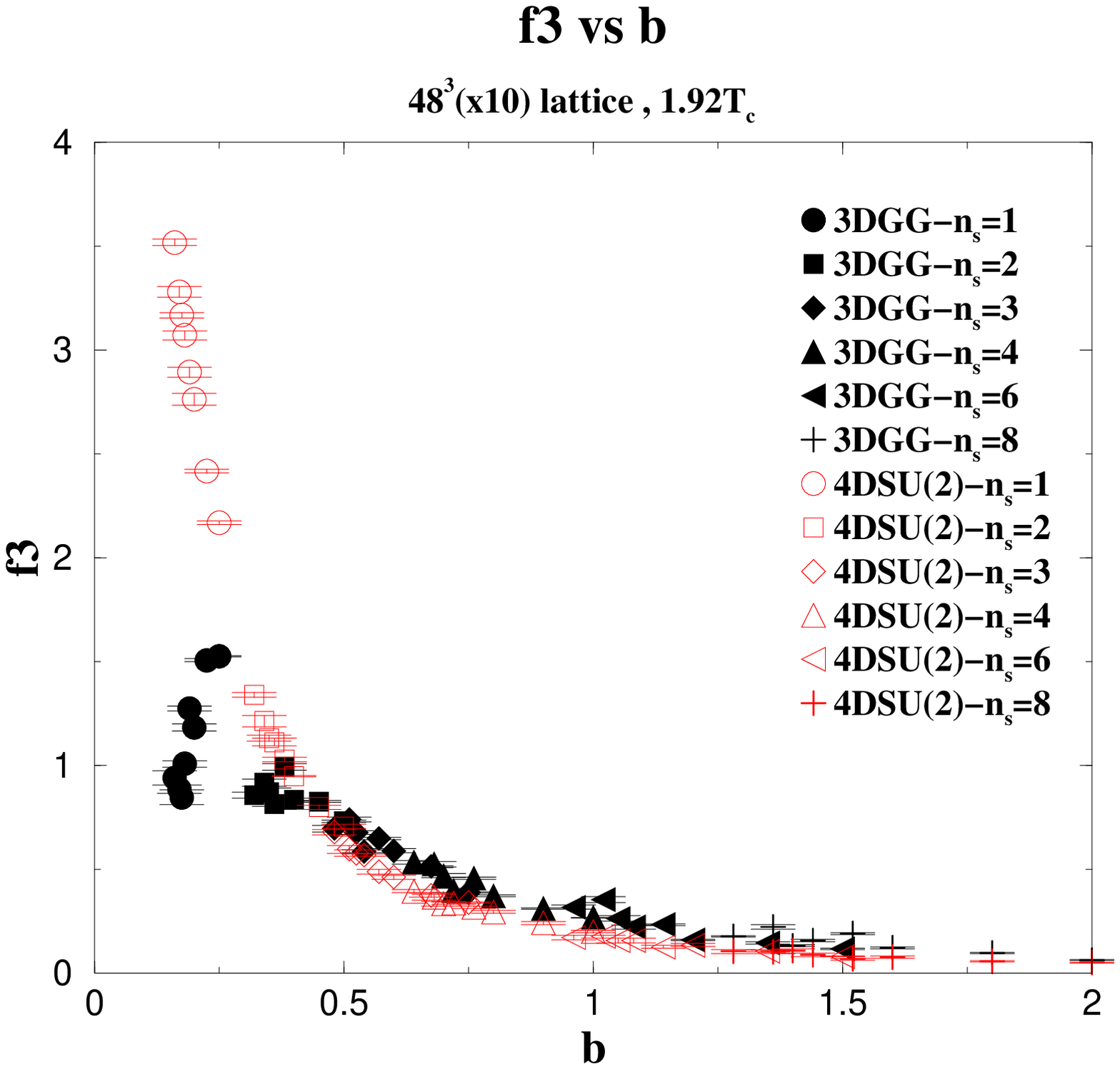, width=7cm, height=7cm}
\caption{Coupling $f_1$(left), $f_2$(right) and $f_3$(bottom)
 at $1.92T_c$.}
\label{f123b4896tc}
}

In Fig.~\ref{distanceas25024tc} we show the distance-dependence of 
the couplings at $b=$0.25,  0.50,  0.75,  1.00,  1.50 and 2.00 
for $T=2.4T_c$.
The couplings of the 3D instanton action are different from 
those of the 4D timelike monopole action
at small $b$ regions, especially in the case of the blockspin factor 
$n_s=1$.
However when we perform the blockspin transformation, 
both couplings tend to be the same.
To see the scaling behavior, we show the $n_s$-dependence of the couplings 
for both actions for different temperature in Fig.~\ref{f123b19224tc}, 
\ref{f123b4896tc}.
These figures show the good scaling behaviors for the couplings $f_1$, $f_2$ 
and $f_3$ in both actions, especially for $b > 0.4 (\sqrt{\sigma})^{-1}$.
From these figures it turns out that the couplings of the monopole actions 
originated from (SU(2))$_{4D}$ and those of the instanton actions 
in $(GG)_{3D}$ flow on the same 
renormalized trajectories in the large $b$ region at $T \ge 2.4T_c$.
In Fig.~\ref{f123b19224tc} we also show the case of $4.8T_c$.
The scaling behaviors look good  
and the agreement of both couplings is much better than 
that for $2.4T_c$.
On the other hand, the couplings at $1.92T_c$ are shown in 
Fig.~\ref{f123b4896tc}.
The figure shows that the couplings of both actions have a nice scaling 
at large $b$ region, but both actions do not coincide.
The temperature $T=1.92 T_c$ is so small 
that we can not apply the dimensional reduction.
The dimensional reduction works well 
at $T \ge 2.4T_c$ region also in the framework of the monopole (instanton)
action representing nonperturbative effects.

\FIGURE[t]{
\epsfig{file=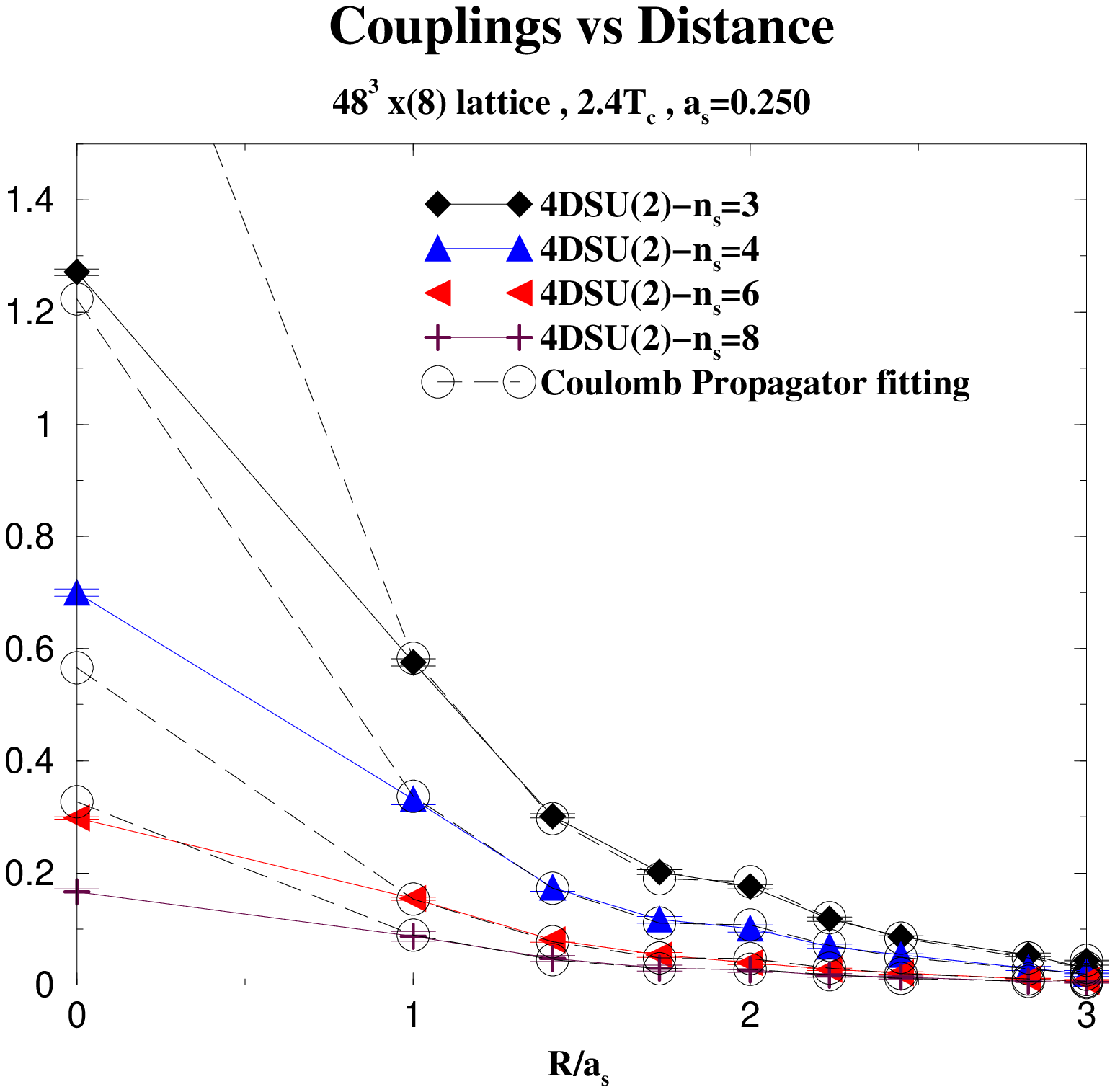, width=7cm, height=7cm}
\epsfig{file=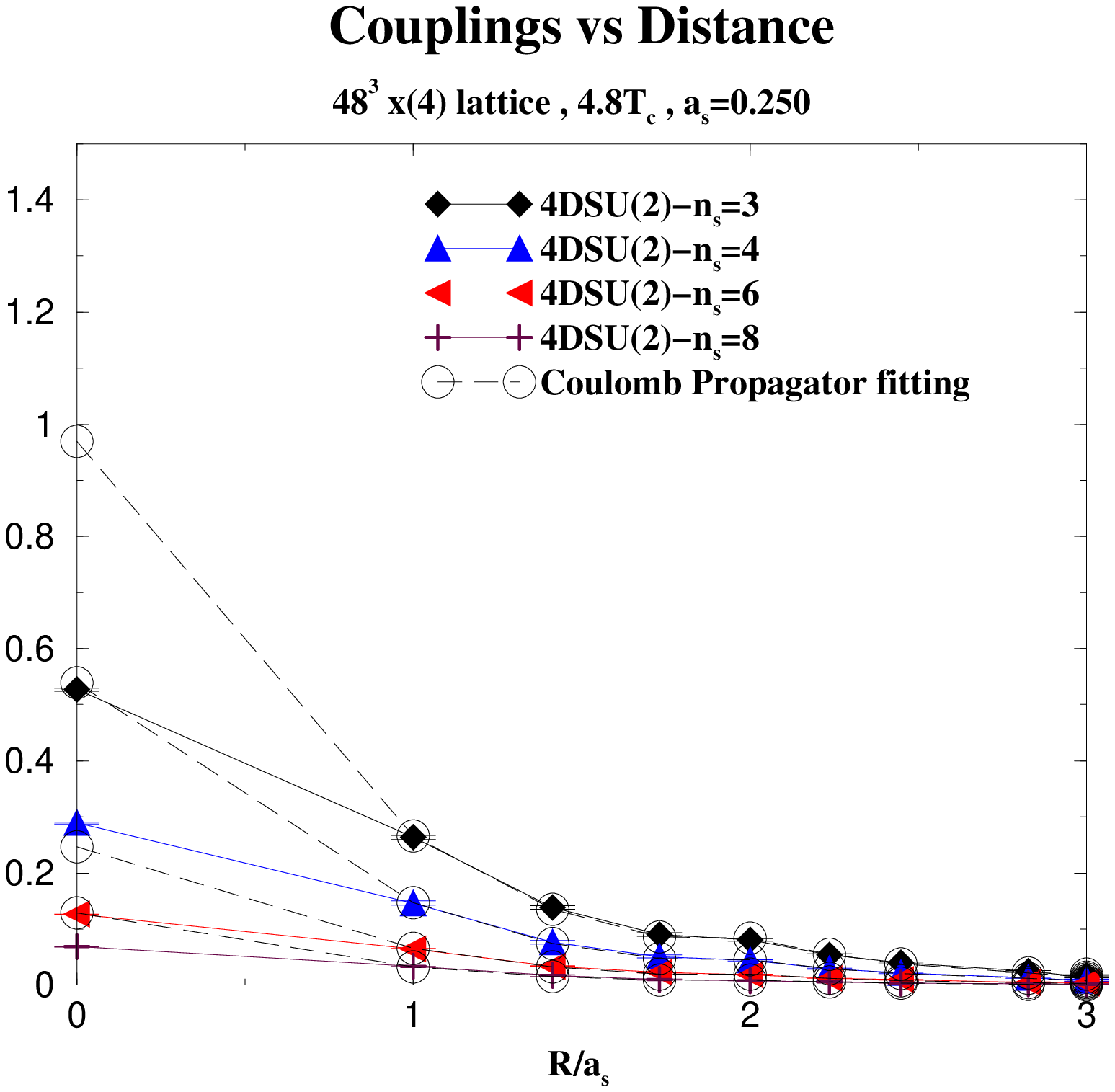, width=7cm, height=7cm}
\caption{The fitting of the 3-dimensional timelike monopole actions 
by the Coulomb propagator at $2.4T_c$ (left) and $4.8T_c$ (right).}
\label{coulomb}
}

Since we have obtained the monopole (instanton) action both in 
(SU(2))$_{4D}$ and in $(GG)_{3D}$, 
we consider the property of the actions.
As Polyakov showed in Ref.~\cite{polyakov77}, 
if instantons behave as a Coulomb gas, the string tension has a
non-zero finite value.
In order to explain the nonperturbative effect in the deconfinement 
phase such as the spatial string tension by instantons, 
we compare the obtained monopole (instanton) action with 
that of the Coulomb gas.
Using the method in Ref.~\cite{yazawa01}, we fit the timelike monopole action 
obtained from (SU(2))$_{4D}$ by the 3D lattice Coulomb 
propagator.
When we define the lattice Coulomb propagator as 
\\
\be
\Delta_{L}^{-1} (s-s') = C_1 \delta^3(s-s') 
                       + C_2 \sum_{i}\delta^3(s-(s'+\hat{i})) + ... , 
\ee
\\
we get a beautiful fit
\\
\be
f_i \sim Const. \times C_i (i\ne1)
\ee
\\
at $T=2.4T_c$ and $T=4.8T_c$ as shown in Fig.~\ref{coulomb}.
Here \{$f_i$\} are the couplings of the timelike monopole (instanton) action 
and the detail is shown in Appendix~\ref{3dcoupling}.
The results obtained here are very similar in Ref.~\cite{yazawa01}.
So we can conclude that the timelike monopoles (instantons) behave 
as a Coulomb gas.
This fact means that monopoles in the deconfinement phase form 
a Coulomb gas of the wrapped monopole loops and reproduce the spatial 
string tension.

\section{Concluding Remarks}

We have studied the effective monopole action at finite temperature in 
(SU(2))$_{4D}$.
(1) We have determined the anisotropy $\xi$ and the lattice spacings 
$a_s$ and $a_t$ for various $(\beta,  \gamma)$ on the anisotropic lattices 
in (SU(2))$_{4D}$.
Using the relations between the parameters $(\beta,  \gamma)$ and the 
lattice spacing $(a_s,  a_t)$,  the thermalized monopole current 
configurations are generated for various temperatures $(T \le 4.8T_c)$ 
in MA gauge.
After performing the blockspin transformations for space directions, 
we have obtained the almost perfect 4-dimensional effective monopole 
action under the assumption of two-point interactions alone.
The action depends only on the physical scale $b_s$ and the temperature $T$.
The temperature-dependence of the action appear with respect to the spacelike 
monopole couplings in the deconfinement phase,  whereas 
the timelike monopole couplings have no temperature-dependence.
(2) In $(GG)_{3D}$,  we have calculated the string tensions from the 
non-abelian,  abelian and instanton Wilson loops at the parameter 
regions obtained from the dimensional reduction of (SU(2))$_{4D}$.
The abelian dominance and the monopole dominance 
have been observed also.
Instantons play an important role for the 
infrared physics  in $(GG)_{3D}$.
(3) At high temperature (the deconfinement phase) in (SU(2))$_{4D}$,  
we have determined the 3-dimensional effective monopole action 
from $(GG)_{3D}$.
We compare the action with the timelike monopole action which is
obtained from (SU(2))$_{4D}$ at the same temperature.
The results show that both actions agree very well 
 at large $b$ region for $T \ge 2.4T_c$.
The dimensional reduction works well for the infrared physics also in the 
monopole-instanton  picture.
The timelike monopole (instanton) actions here obtained 
are fitted beautifully by the lattice Coulomb propagator.
The result means that in the deconfinement phase,  the mechanism 
reproducing the spatial string tension is the same as the one of 
$(GG)_{3D}$. 
Namely the Coulomb gas of the wrapped monopole loops induce 
the nonperturbative effects such as the spatial string tension.
Although the dimensional reduction 
works good  only for $T \ge 2.4T_c$,  the 4D timelike monopole 
actions for $T < 2.4T_c$ are 
very similar to the ones for $T \ge 2.4T_c$.
The nonperturbative effects in the deconfinement phase are given 
by the timelike monopoles in (SU(2))$_{4D}$.

The following subjects are very interesting to be studied.
(1) The exact mechanism of the confinement-deconfinement transition 
should be clarified. From the numerical study of critical exponents, 
spacelike monopoles play a key role in the mechanism. But we have not yet 
known what mechanism of spacelike monopoles is responsible for the transition.
Simple energy-entropy arguments may not be true,  since 
the energy of the system (which is well approximated by the self coupling 
of the monopole action) decreases monotonously 
as $b_s$ becomes larger even in the deconfinement phase.
If the entropy is governed by a kinematical factor 
which does not depend on $b_s$  
as in the zero-temperature case, 
energy-entropy arguments can not explain the transition.
(2) It is interesting to transform the obtained actions into those of different models 
like a dual abelian Higgs model or a string model.
We may get a different viewpoint with respect to the mechanism of the 
deconfinement transition.
(3) To study all nonperturbative effects such as Debye-screening mass and glueball
mass is also interesting. Are they all explained by monopoles?

\acknowledgments
The authors thank Shun-ichi Kitahara for fruitful discussions.
This work is supported by the Supercomputer Project of 
the Institute of Physical and Chemical Research (RIKEN).
T.S. acknowledges financial support 
from a JSPS Grant-in-aid for Scientific Research (B)  (No. 11695029).

\appendix
\section{Inverse Monte-Carlo Methods}\label{inversemonte}

\subsection{The Original Swendsen's Method}\label{origswed}

We apply the original Swendsen's method~\cite{swendsen84} 
to determine the 3D instanton
action from the thermalized instanton configurations.
The partition function of the theory described by the instantons is 
given by the following.
\be
Z = \Bigl( \prod_s \sum_{k(s)=-\infty}^{\infty} \Bigr) 
    \exp \bigl( - S[k] \bigr) , 
\ee
where $S[k]$ is an instanton action.
The action may be written as a linear combination of all independent 
operators which are summed over the whole lattice.
We denote each operator as $S_i [k]$.
Then the  action may be expressed as follows:
\\
\be
S[k] = \sum_i f_i S_i [k] , 
\ee
\\
where $f_i$ are coupling constants.
The expectation value of some operator $O[k]$ is 
written by 
\\
\be
\Bigl\langle O[k] \Bigr\rangle 
= \frac{1}{Z} \Bigl( \prod_s \sum_{k(s)=-\infty}^{\infty} \Bigr) 
              O[k] \exp \bigl( - S[k] \bigr) .
\ee
\\
Let us focus on a site $s^\prime$ and define $\hat{S}[k]$ which 
contains $k(s\prime)$. We get
\\
\be
\Bigl\langle O[k] \Bigr\rangle 
= \frac{1}{Z} \Bigl( {\prod_{s}}^{\prime} 
                     \sum_{k(s)=-\infty}^{\infty} \Bigr)
              \exp\bigl\{-(S[k]-\hat{S}[k])\bigr\}
              \sum_{k(s^\prime)=-\infty}^{\infty} 
              O[\hat{k}, \{k\}^\prime] \exp \bigl( - \hat{S}[k] \bigr) , 
\label{eq:oexp}
\ee
\\
where $\prod^{\prime}$ means the product except  the site $s^\prime$ 
and $\hat{k}=k(s^\prime)$ and $\{k\}^\prime$ is the coset of $\hat{k}$.
We rewrite Eq.(\ref{eq:oexp}) as 
\\
\be
\Bigl\langle O[k] \Bigr\rangle 
&=& \frac{1}{Z} \Bigl( \prod_{s} 
                     \sum_{k(s)=-\infty}^{\infty} \Bigr)
              \frac{ \sum_{\hat{k}=-\infty}^{\infty} 
                     O[\hat{k}, \{k\}^\prime]
              \exp\bigl\{-\hat{S}[\hat{k}, \{k\}^\prime]\bigr\} }
              { \sum_{\hat{k}=-\infty}^{\infty}
                \exp\bigl\{-\hat{S}[\hat{k}, \{k\}^\prime]\bigr\} } 
               \exp \bigl( - S[k] \bigr)
\\
&\equiv& \Bigl\langle \hat{O}[k] \Bigr\rangle , \label{eq:ohat1}
\ee
where
\be
\hat{O}[k] \equiv  \frac{ \sum_{\hat{k}=-\infty}^{\infty} 
                     O[\hat{k}, \{k\}^\prime]
              \exp\bigl\{-\hat{S}[\hat{k}, \{k\}^\prime]\bigr\} }
              { \sum_{\hat{k}=-\infty}^{\infty}
                \exp\bigl\{-\hat{S}[\hat{k}, \{k\}^\prime]\bigr\} } . 
\label{eq:ohatohat}
\ee
\\
When we use the definition of the instanton 
by DeGrand-Toussaint~\cite{degrand80}, 
the sum with respect to $\hat{k}$ change from $[-\infty, \infty]$ 
to $[-(3n^2-1), 3n^2-1]$ where n is a factor of blockspin transformation 
of instantons.
Using the identity Eq.(\ref{eq:ohat1}),  let us determine the instanton action iteratively.
Since we don't know the correct set of coupling constants $\{f_i\}$, 
we start from trial coupling constants $\{\tilde{f}_i\}$.
We define $\bar{O}$ in which the true coupling constants $\{f_i\}$ in 
Eq.(\ref{eq:ohatohat}) are replaced by the trial ones $\{\tilde{f}_i\}$ as 
\\
\be
\bar{O}[k] =  \frac{ \sum_{\hat{k}=k_{min}}^{k_{max}} 
                     O[\hat{k}, \{k\}^\prime]
              \exp\bigl\{-\sum_i \tilde{f}_i \hat{S}_i[\hat{k}, \{k\}^\prime]\bigr\} }
              { \sum_{\hat{k}=k_{min}}^{k_{max}}
                \exp\bigl\{-\sum_i \tilde{f}_i\hat{S}_i[\hat{k}, \{k\}^\prime]\bigr\} } . 
\ee
\\
When $f_i$ is not equal to $\tilde{f}_i$ for all $i$, 
$\Bigl\langle O[k] \Bigr\rangle \neq \Bigl\langle \bar{O}[k] \Bigr\rangle$.
But when $\{\tilde{f}_i\}$ are not far from $\{f_i\}$,  we get the 
following expansion:
\\
\be
\Bigl\langle O[k] - \bar{O}[k] \Bigr\rangle 
\sim \sum_i \Bigl\langle \bar{O}\bar{S}_i - \overline{OS}_i \Bigr\rangle  
\bigl( f_i - \tilde{f}_i \bigr) .
\ee
\\
In practice,  we use $S_i[k]$ as an operator $O[k]$ to get a 
good convergence.
Hence we get  linear equations for $\{f_i\}$ as 
\\
\be
\Bigl\langle S_i[k] - \bar{S}_i[k] \Bigr\rangle 
\sim \sum_j \Bigl\langle \bar{S}_i\bar{S}_j - \overline{S_iS_j} \Bigr\rangle  
\bigl( f_j - \tilde{f}_j \bigr) .\label{eq:iden}
\ee
\\
Starting from trial couplings $\{\tilde{f}_i\}$,  we first calculate 
the expectation value 
$\Bigl\langle S_i[k] - \bar{S}_i[k] \Bigr\rangle$ 
using the instanton configurations.
If the values of 
$\Bigl\langle S_i[k] - \bar{S}_i[k] \Bigr\rangle$ 
for all i are regarded as zero,  $\{\tilde{f}_i\}$ can be adopted 
as the true coupling constants.
If not,  we solve Eq.(\ref{eq:iden}) with respect to $\{f_i\}$ and adopt 
the solution $\{f_i\}$ as new trial couplings.
Repeating the above-mentioned procedure we can obtain the true 
coupling constants and determine the instanton action iteratively.

\subsection{The Modified Swendsen's Method}\label{modswed}

The modified Swendsen's Method~\cite{shiba95, kato98} 
is applied to determine the  
action of monopoles with current conservation law.
So we use the method to determine the 4-dimensional effective monopole 
action.

The partition function of the theory is written as
\\
\be
Z = \Bigl( \prod_{s, \mu} \sum_{k_\mu(s)=-\infty}^{\infty} \Bigr) 
    \Bigl( \prod_s \delta_{\partial_{\mu}^{\prime} k_\mu(s), 0} \Bigr)
    \exp \bigl( - S[k] \bigr) .\label{eq:zp}
\ee
\\
Using the expression of Eq.(\ref{eq:zp}),  we consider the expectation value 
of some operator $O[k]$ which is written by monopole currents
\\
\be
\Bigl\langle O[k] \Bigr\rangle 
= \frac{1}{Z} \Bigl( \prod_{s, \mu} \sum_{k_\mu(s)=-\infty}^{\infty} \Bigr) 
    \Bigl( \prod_s \delta_{\partial_{\mu}^{\prime} k_\mu(s), 0} \Bigr)
              O[k] \exp \bigl( - S[k] \bigr) .
\ee
\\
Because of the existence of the current conservation laws, 
we focus on a plaquette $(s^\prime, \hat{\mu}^\prime, \hat{\nu}^\prime)$ 
instead of a site $s^\prime$ and define $\hat{S}[k]$ as a part 
of $S[k]$ which contains currents along the plaquette, i.e. 
$\{ k_{\mu^\prime}(s^\prime), k_{\nu^\prime}(s^\prime+\hat{\mu}^\prime), 
k_{\mu^\prime}(s^\prime+\hat{\nu}^\prime), k_{\nu^\prime}(s^\prime) \}$.
Then we get 
\\
\be
\Bigl\langle O[k] \Bigr\rangle 
&=& \frac{1}{Z} \Bigl( {\prod_{s, \mu}}^{\prime} \sum_{k_\mu(s)=-\infty}^{\infty} \Bigr) 
    \Bigl( {\prod_{s}}^{\prime} \delta_{\partial_{\mu}^{\prime} k_\mu(s), 0} \Bigr)
              \exp\bigl\{-(S[k]-\hat{S}[k])\bigr\}  \nonumber \\
&\times&      \sum_{k_{\mu^\prime}(s^\prime)=-\infty}^{\infty}
              \sum_{k_{\nu^\prime}(s^\prime+\hat{\mu}^\prime)=-\infty}^{\infty}
              \sum_{k_{\mu^\prime}(s^\prime+\hat{\nu}^\prime)=-\infty}^{\infty}
              \sum_{k_{\nu^\prime}(s^\prime)=-\infty}^{\infty} \nonumber \\
&\times&      \delta_{\partial_{\mu}^{\prime} k_\mu(s^\prime), 0}
              \delta_{\partial_{\mu}^{\prime} k_\mu(s^\prime+\hat{\mu}^\prime), 0}
              \delta_{\partial_{\mu}^{\prime} k_\mu(s^\prime+\hat{\nu}^\prime), 0}
              \delta_{\partial_{\mu}^{\prime} k_\mu(s^\prime+\hat{\mu}^\prime+\hat{\nu}^\prime), 0} \nonumber \\
&\times&      O[k] \exp \bigl( - \hat{S}[k] \bigr) , \label{eq:oexp2}
\ee
\\
where ${\prod_{s, \mu}}^{\prime}$ and ${\prod_{s}}^{\prime}$ mean 
the product which excludes the links and the sites in the plaquette considered.
One of the $\delta$-functions on the four sites in the plaquette 
can be replaced by 
$\delta_{\partial_{\mu}^{\prime} k_\mu(s^\prime)+
         \partial_{\mu}^{\prime} k_\mu(s^\prime+\hat{\mu}^\prime)+
         \partial_{\mu}^{\prime} k_\mu(s^\prime+\hat{\nu}^\prime)+
         \partial_{\mu}^{\prime} k_\mu(s^\prime+\hat{\mu}^\prime+\hat{\nu}^\prime), 0}$ 
and this $\delta$-function does not contain any current in the plaquette.
Then Eq.(\ref{eq:oexp2}) is expressed as 
\\
\be
\Bigl\langle O[k] \Bigr\rangle 
&=& \frac{1}{Z} \Bigl( {\prod_{s, \mu}}^{\prime} \sum_{k_\mu(s)=-\infty}^{\infty} \Bigr) 
    \Bigl( {\prod_{s}}^{\prime} \delta_{\partial_{\mu}^{\prime} k_\mu(s), 0} \Bigr)
              \exp\bigl\{-(S[k]-\hat{S}[k])\bigr\} \nonumber \\
&\times& 
\delta_{\partial_{\mu}^{\prime} k_\mu(s^\prime)+
         \partial_{\mu}^{\prime} k_\mu(s^\prime+\hat{\mu}^\prime)+
         \partial_{\mu}^{\prime} k_\mu(s^\prime+\hat{\nu}^\prime)+
         \partial_{\mu}^{\prime} k_\mu(s^\prime+\hat{\mu}^\prime+\hat{\nu}^\prime), 0}  \nonumber \\
&\times&  \bigl( \sum \delta \bigr)_k 
          O[\hat{k}, \{k\}^\prime] \exp \bigl( - \hat{S}[k] \bigr) , 
\label{eq:oexp3}
\ee
\\
where $\{k\}^\prime$ does not contain the four currents in the plaquette 
considered and 
\be
\bigl( \sum \delta \bigr)_k 
&\equiv& \sum_{k_{\mu^\prime}(s^\prime)=-\infty}^{\infty}
         \sum_{k_{\nu^\prime}(s^\prime+\hat{\mu}^\prime)=-\infty}^{\infty}
         \sum_{k_{\mu^\prime}(s^\prime+\hat{\nu}^\prime)=-\infty}^{\infty}
         \sum_{k_{\nu^\prime}(s^\prime)=-\infty}^{\infty} \nonumber \\
&\times& \delta_{\partial_{\mu}^{\prime} k_\mu(s^\prime), 0}
         \delta_{\partial_{\mu}^{\prime} k_\mu(s^\prime+\hat{\mu}^\prime), 0}
         \delta_{\partial_{\mu}^{\prime} k_\mu(s^\prime+\hat{\nu}^\prime), 0} .
\ee
Defining the operator $\hat{O}[\hat{k}, \{k\}^\prime]$ as 
\\
\be
\hat{O}[\hat{k}, \{k\}^\prime] \equiv 
\frac{ \bigl( \sum \delta \bigr)_k O[\hat{k}, \{k\}^\prime] 
       \exp \bigl( - \hat{S}[k] \bigr) }
     { \bigl( \sum \delta \bigr)_k \exp \bigl( - \hat{S}[k] \bigr) } , 
\label{eq:ohatm1}
\ee
\\
then we can rewrite Eq.(\ref{eq:oexp3}) as 
\\
\be
\Bigl\langle O[k] \Bigr\rangle 
&=& \frac{1}{Z} \Bigl( \prod_{s, \mu} \sum_{k_\mu(s)=-\infty}^{\infty} \Bigr) 
    \Bigl( \prod_{s} \delta_{\partial_{\mu}^{\prime} k_\mu(s), 0} \Bigr)
    \hat{O}[\hat{k}, \{k\}^\prime]
              \exp\bigl\{-S[k]\bigr\} \\
&=& \Bigl\langle \hat{O}[k] \Bigr\rangle .
\ee
\\
From the three $\delta$-functions in $ \bigl( \sum \delta \bigr)_k$, 
there are three constraints for the four currents on the 
plaquette considered.
Namely only one current of the four is independent.
Define the independent variable $M$ and replace the current 
$\hat{k}_{\mu^\prime}(s^\prime)$ as 
\\
\be
\hat{k}_{\mu^\prime}(s^\prime) = k_{\mu^\prime}(s^\prime) + M .
\ee
\\
Using the three constraints for the four currents,  we get 
\be
\hat{k}_{\nu^\prime}(s^\prime) &=& k_{\nu^\prime}(s^\prime) - M ,  \\
\hat{k}_{\mu^\prime}(s^\prime+\hat{\nu}^\prime) &=& 
                  k_{\mu^\prime}(s^\prime+\hat{\nu}^\prime) - M ,  \\
\hat{k}_{\nu^\prime}(s^\prime+\hat{\mu}^\prime) &=& 
                  k_{\nu^\prime}(s^\prime+\hat{\mu}^\prime) + M .
\ee
Here we use the relation 
\be
\sum_{M=-\infty}^{\infty} \delta_{\hat{k}_{\mu^\prime}(s^\prime), 
                                  k_{\mu^\prime}(s^\prime) + M}
= 1 .
\ee
Then we can replace the sum with respect to $\hat{k}$ by the sum 
with respect to $M$.
When we use the DeGrand-Toussaint monopole definition,  the sum with 
respect to $M$ is restricted from $m_1$ to $m_2$ where 
\be
m_1 &=& -(3n^2-1)-min \{ k_{\mu^\prime}(s^\prime), 
                         k_{\nu^\prime}(s^\prime+\hat{\mu}^\prime), 
                        -k_{\mu^\prime}(s^\prime+\hat{\nu}^\prime), 
                        -k_{\nu^\prime}(s^\prime)
                      \} ,  \\
m_2 &=&  (3n^2-1)-max \{ k_{\mu^\prime}(s^\prime), 
                         k_{\nu^\prime}(s^\prime+\hat{\mu}^\prime), 
                        -k_{\mu^\prime}(s^\prime+\hat{\nu}^\prime), 
                        -k_{\nu^\prime}(s^\prime)
                      \} ,  
\ee
and $n$ is a number of blockspin transformations for all directions.
Hence we get another representation of Eq.(\ref{eq:ohatm1}) as 
\be
\hat{O}[k] = 
\frac{ \sum_{M=m_1}^{m_2} O[\bar{k}] 
       \exp \bigl( - \hat{S}[\bar{k}] \bigr) }
     { \sum_{M=m_1}^{m_2} \exp \bigl( - \hat{S}[\bar{k}] \bigr) } , 
\ee
where
\be
\bar{k}_\mu(s) = k_\mu(s) 
               + M (\delta_{s, s^\prime}\delta_{\mu, \mu^\prime}
                  + \delta_{s, s^\prime+\hat{\mu}^\prime}\delta_{\mu, \nu^\prime}
                  - \delta_{s, s^\prime+\hat{\nu}^\prime}\delta_{\mu, \mu^\prime}
                  - \delta_{s, s^\prime}\delta_{\mu, \nu^\prime} ) .
\ee
So we get the identity as follows:
\\
\be
\Bigl\langle O[k] \Bigr\rangle = \Bigl\langle \hat{O}[k] \Bigr\rangle .
\label{eq:iden2}
\ee
\\
Using Eq.(\ref{eq:iden2}) and the same procedure in Appendix A.1,  we can obtain 
the monopole action.

\section{The Quadratic Interactions Adopted}

\subsection{4D Effective Monopole Action}\label{4dcoupling}

Some comments on the 4D effective monopole action 
are in order.
(1) We have to distinguish spacelike monopoles from timelike monopoles.
(2) The current conservation laws exist at all sites. 
Using the conservation laws, we replace short-distance perpendicular 
interactions in terms of parallel interactions as many as possible as done in 
the $T=0$ case~\cite{shiba95}. 
(3) Monopole current configurations are generated on the anisotropic lattice.

We adopt 69 parallel- and 15 perpendicular-interactions 
in the following action :

\be
S[k] &=&  \sum_{i=1}^{84} f_i S_i . 
\ee
The interactions are summarized in Table~\ref{tb:84int}.

\begin{table}
\caption{The quadratic interactions used for the 4D effective monopole action.}
\label{tb:84int} 
\begin{tabular}{cllcll}
 $\mbra{f_i}$ & distance & \multicolumn{1}{c}{type} &
 $\mbra{f_i}$ & distance & \multicolumn{1}{c}{type} \\ 
\hline
$f_1$    & (0,0,0,0) & $k_i(s)k_i(s)$ &
$f_{43}$ & (0,2,1,1) & $k_i(s)k_i(s+2\hat{j}+\hat{l}+\hat{4})$ \\
$f_2$    & (0,0,0,0) & $k_4(s)k_4(s)$ &
$f_{44}$ & (0,2,1,1) & $k_i(s)k_i(s+\hat{j}+\hat{l}+2\hat{4})$ \\
$f_3$    & (1,0,0,0) & $k_i(s)k_i(s+\hat{i})$ &
$f_{45}$ & (0,2,1,1) & $k_4(s)k_4(s+2\hat{i}+\hat{j}+\hat{l})$ \\
$f_4$    & (1,0,0,0) & $k_4(s)k_4(s+\hat{4})$ &
$f_{46}$ & (2,1,1,1) & $k_i(s)k_i(s+2\hat{i}+\hat{j}+\hat{l}+\hat{4})$ \\
$f_5$    & (0,1,0,0) & $k_i(s)k_i(s+\hat{j})$ &
$f_{47}$ & (2,1,1,1) & $k_4(s)k_4(s+2\hat{4}+\hat{i}+\hat{j}+\hat{l})$ \\
$f_6$    & (0,1,0,0) & $k_i(s)k_i(s+\hat{4})$ &
$f_{48}$ & (1,2,1,1) & $k_i(s)k_i(s+\hat{i}+2\hat{j}+\hat{l}+\hat{4})$ \\
$f_7$    & (0,1,0,0) & $k_4(s)k_4(s+\hat{i})$ &
$f_{49}$ & (1,2,1,1) & $k_i(s)k_i(s+\hat{i}+\hat{j}+\hat{l}+2\hat{4})$ \\
$f_8$    & (1,1,0,0) & $k_i(s)k_i(s+\hat{i}+\hat{j})$ &
$f_{50}$ & (1,2,1,1) & $k_4(s)k_4(s+\hat{4}+2\hat{i}+\hat{j}+\hat{l})$ \\
$f_9$    & (1,1,0,0) & $k_i(s)k_i(s+\hat{i}+\hat{4})$ &
$f_{51}$ & (2,2,0,0) & $k_i(s)k_i(s+2\hat{i}+2\hat{j})$ \\
$f_{10}$ & (1,1,0,0) & $k_4(s)k_4(s+\hat{4}+\hat{i})$ &
$f_{52}$ & (2,2,0,0) & $k_i(s)k_i(s+2\hat{i}+2\hat{4})$ \\
$f_{11}$ & (0,1,1,0) & $k_i(s)k_i(s+\hat{j}+\hat{l})$ &
$f_{53}$ & (2,2,0,0) & $k_4(s)k_4(s+2\hat{4}+2\hat{i})$ \\
$f_{12}$ & (0,1,1,0) & $k_i(s)k_i(s+\hat{j}+\hat{4})$ &
$f_{54}$ & (0,2,2,0) & $k_i(s)k_i(s+2\hat{j}+2\hat{l})$ \\
$f_{13}$ & (0,1,1,0) & $k_4(s)k_4(s+\hat{i}+\hat{j})$ &
$f_{55}$ & (0,2,2,0) & $k_i(s)k_i(s+2\hat{j}+2\hat{4})$ \\
$f_{14}$ & (0,1,1,1) & $k_i(s)k_i(s+\hat{j}+\hat{l}+\hat{4})$ &
$f_{56}$ & (0,2,2,0) & $k_4(s)k_4(s+2\hat{i}+2\hat{j})$ \\
$f_{15}$ & (0,1,1,1) & $k_4(s)k_4(s+\hat{i}+\hat{j}+\hat{l})$ &
$f_{57}$ & (3,0,0,0) & $k_i(s)k_i(s+3\hat{i})$ \\
$f_{16}$ & (1,1,1,0) & $k_i(s)k_i(s+\hat{i}+\hat{j}+\hat{l})$ &
$f_{58}$ & (0,3,0,0) & $k_i(s)k_i(s+3\hat{j})$ \\
$f_{17}$ & (1,1,1,0) & $k_i(s)k_i(s+\hat{i}+\hat{j}+\hat{4})$ &
$f_{59}$ & (0,3,0,0) & $k_4(s)k_4(s+3\hat{i})$ \\
$f_{18}$ & (1,1,1,0) & $k_4(s)k_4(s+\hat{4}+\hat{i}+\hat{j})$ &
$f_{60}$ & (2,2,1,0) & $k_i(s)k_i(s+2\hat{i}+2\hat{j}+\hat{l})$ \\
$f_{19}$ & (2,0,0,0) & $k_i(s)k_i(s+2\hat{i})$ &
$f_{61}$ & (2,2,1,0) & $k_i(s)k_i(s+2\hat{i}+2\hat{j}+\hat{4})$ \\
$f_{20}$ & (2,0,0,0) & $k_4(s)k_4(s+2\hat{4})$ &
$f_{62}$ & (2,2,1,0) & $k_i(s)k_i(s+2\hat{i}+\hat{j}+2\hat{4})$ \\
$f_{21}$ & (0,2,0,0) & $k_i(s)k_i(s+2\hat{j})$ &
$f_{63}$ & (2,2,1,0) & $k_4(s)k_4(s+2\hat{4}+2\hat{i}+\hat{j})$ \\
$f_{22}$ & (0,2,0,0) & $k_i(s)k_i(s+2\hat{4})$ &
$f_{64}$ & (1,2,2,0) & $k_i(s)k_i(s+\hat{i}+2\hat{j}+2\hat{l})$ \\
$f_{23}$ & (0,2,0,0) & $k_4(s)k_4(s+2\hat{i})$ &
$f_{65}$ & (1,2,2,0) & $k_i(s)k_i(s+\hat{i}+2\hat{j}+2\hat{4})$ \\
$f_{24}$ & (1,1,1,1) & $k_i(s)k_i(s+\hat{i}+\hat{j}+\hat{l}+\hat{4})$ &
$f_{66}$ & (1,2,2,0) & $k_4(s)k_4(s+\hat{4}+2\hat{i}+2\hat{j})$ \\
$f_{25}$ & (1,1,1,1) & $k_4(s)k_4(s+\hat{4}+\hat{i}+\hat{j}+\hat{l})$ &
$f_{67}$ & (0,2,2,1) & $k_i(s)k_i(s+2\hat{j}+2\hat{l}+\hat{4})$ \\
$f_{26}$ & (2,1,0,0) & $k_i(s)k_i(s+2\hat{i}+\hat{j})$ &
$f_{68}$ & (0,2,2,1) & $k_i(s)k_i(s+2\hat{j}+\hat{l}+2\hat{4})$ \\
$f_{27}$ & (2,1,0,0) & $k_i(s)k_i(s+2\hat{i}+\hat{4})$ &
$f_{69}$ & (0,2,2,1) & $k_4(s)k_4(s+2\hat{i}+2\hat{j}+\hat{l})$ \\
$f_{28}$ & (2,1,0,0) & $k_4(s)k_4(s+2\hat{4}+\hat{i})$ &
$f_{70}$ & perpend.     & $k_i(s)k_j(s+\hat{4}+\hat{j})+...$ \\
$f_{29}$ & (1,2,0,0) & $k_i(s)k_i(s+\hat{i}+2\hat{j})$ &
$f_{71}$ & perpend.     & $k_i(s)k_j(s+\hat{l}+\hat{4})+...$ \\
$f_{30}$ & (1,2,0,0) & $k_i(s)k_i(s+\hat{i}+2\hat{4})$ &
$f_{72}$ & perpend.     & $k_i(s)k_4(s+\hat{j}+\hat{l})+...$ \\
$f_{31}$ & (1,2,0,0) & $k_4(s)k_4(s+\hat{4}+2\hat{i})$ &
$f_{73}$ & perpend.     & $k_i(s)k_j(s+\hat{j})+...$ \\
$f_{32}$ & (0,2,1,0) & $k_i(s)k_i(s+2\hat{j}+\hat{l})$ &
$f_{74}$ & perpend.     & $k_i(s)k_j(s+2\hat{4})+...$ \\
$f_{33}$ & (0,2,1,0) & $k_i(s)k_i(s+2\hat{j}+\hat{4})$ &
$f_{75}$ & perpend.     & $k_i(s)k_j(s+2\hat{4}+\hat{j})+...$ \\
$f_{34}$ & (0,2,1,0) & $k_i(s)k_i(s+\hat{j}+2\hat{4})$ &
$f_{76}$ & perpend.     & $k_i(s)k_j(s+2\hat{j}+\hat{l})+...$ \\
$f_{35}$ & (0,2,1,0) & $k_4(s)k_4(s+2\hat{i}+\hat{j})$ &
$f_{77}$ & perpend.     & $k_i(s)k_j(s+2\hat{j}+\hat{4})+...$ \\
$f_{36}$ & (2,1,1,0) & $k_i(s)k_i(s+2\hat{i}+\hat{j}+\hat{l})$ &
$f_{78}$ & perpend.     & $k_i(s)k_4(s+3\hat{i}+\hat{j})+...$ \\
$f_{37}$ & (2,1,1,0) & $k_i(s)k_i(s+2\hat{i}+\hat{j}+\hat{4})$ &
$f_{79}$ & perpend.     & $k_i(s)k_j(s+2\hat{l}+\hat{4})+...$ \\
$f_{38}$ & (2,1,1,0) & $k_4(s)k_4(s+2\hat{4}+\hat{i}+\hat{j})$ &
$f_{80}$ & perpend.     & $k_i(s)k_j(s+\hat{l}+2\hat{4})+...$ \\
$f_{39}$ & (1,2,1,0) & $k_i(s)k_i(s+\hat{i}+2\hat{j}+\hat{l})$ &
$f_{81}$ & perpend.     & $k_i(s)k_4(s+2\hat{j}+\hat{l})+...$ \\
$f_{40}$ & (1,2,1,0) & $k_i(s)k_i(s+\hat{i}+2\hat{j}+\hat{4})$ &
$f_{82}$ & perpend.     & $k_i(s)k_j(s+2\hat{i}+\hat{l}+2\hat{j})+...$ \\
$f_{41}$ & (1,2,1,0) & $k_i(s)k_i(s+\hat{i}+\hat{j}+2\hat{4})$ &
$f_{83}$ & perpend.     & $k_i(s)k_j(s+2\hat{i}+\hat{4}+2\hat{j})+...$ \\
$f_{42}$ & (1,2,1,0) & $k_4(s)k_4(s+\hat{4}+2\hat{i}+\hat{j})$ &
$f_{84}$ & perpend.     & $k_i(s)k_4(s+2\hat{i}+\hat{j}+2\hat{4})+...$ \\
         &           & \\
\end{tabular}
\end{table}

\clearpage

\subsection{3D Effective Monopole Action}\label{3dcoupling}

For 3D instanton action,  we adopt 10 interactions in the following action :
\\
\be
S[k] &=&  \sum_{i=1}^{10} f_i S_i 
\ee
\\
The interactions are summarized in Table~\ref{10int}.
\begin{table}
\caption{The quadratic interactions used for the 3D effective 
monopole (instanton) action.}
\label{10int} 
\begin{tabular}{cllcll}
coupling $\mbra{f_i}$ & distance & \multicolumn{1}{c}{type} &
coupling $\mbra{f_i}$ & distance & \multicolumn{1}{c}{type} \\ 
\hline
$f_1$    & \ (0,0,0) & $k(s)$ &
$f_6$    & \ (2,1,0) & $k(s+2\hat{i}+\hat{j})$ \\
$f_2$    & \ (1,0,0) & $k(s+\hat{i})$ &
$f_7$    & \ (2,1,1) & $k(s+2\hat{i}+\hat{j}+\hat{l})$ \\
$f_3$    & \ (1,1,0) & $k(s+\hat{i}+\hat{j})$ &
$f_8$    & \ (2,2,0) & $k(s+2\hat{i}+2\hat{j})$ \\
$f_4$    & \ (1,1,1) & $k(s+\hat{i}+\hat{j}+\hat{l})$ &
$f_9$    & \ (3,0,0) & $k(s+3\hat{i})$ \\
$f_5$    & \ (2,0,0) & $k(s+2\hat{i})$ &
$f_{10}$ & \ (2,2,1) & $k(s+2\hat{i}+2\hat{j}+\hat{l})$ \\
         &           & \\
\end{tabular}
\end{table}

\listoftables           
\listoffigures          

\end{document}